\RequirePackage{snapshot}
\documentclass[aps,pra,twocolumn,amsfonts,showpacs,longbibliography,superscriptaddress]{revtex4-2}
\usepackage[toc,page,titletoc]{appendix}
\usepackage{verbatim,epsfig,amsmath,amssymb,bm,epsf,graphicx,psfrag,bbold,amsthm,amsfonts}
\usepackage[bottom]{footmisc}
\usepackage[export]{adjustbox}
\usepackage{standalone}
\usepackage[caption=false]{subfig}
\usepackage{hyperref}
\hypersetup{
    colorlinks=true,
    linkcolor=black,
    citecolor=blue,
    filecolor=black,
    urlcolor=blue,
}
\usepackage[capitalize]{cleveref} 
\usepackage{enumitem}
\usepackage{grffile}
\usepackage{framed}
\usepackage{mathrsfs}
\usepackage{esint}
\setlist{nosep}
\usepackage{placeins}
\usepackage[all]{xy}
\usepackage{color}
\usepackage[utf8]{inputenc}
\usepackage{float}
\usepackage{multirow}
\usepackage[normalem]{ulem}
\usepackage{array}
\newcommand{\PreserveBackslash}[1]{\let\temp=\\#1\let\\=\temp}
\newcolumntype{C}[1]{>{\PreserveBackslash\centering}p{#1}}
\newcolumntype{R}[1]{>{\PreserveBackslash\raggedleft}p{#1}}
\newcolumntype{L}[1]{>{\PreserveBackslash\raggedright}p{#1}}

\usepackage{natbib}
\usepackage{tikz, mathtools}
\usetikzlibrary{external}
\usetikzlibrary{decorations.pathmorphing}
\usetikzlibrary{arrows.meta,arrows, fadings}
\tikzfading[name=fade top,
  top color=transparent!0, bottom color=transparent!100]
\usepackage{verbatim}
\usepackage{leftidx}

\usepackage{notes2bib}

\DeclarePairedDelimiter{\kett}{\lvert}{\rangle}

\newcommand{\moy}[1]{\langle #1 \rangle}
\newcommand{\ket}[1]{\mbox{$ | #1 \rangle $}}

\newcommand\A {{\cal A}}
\newcommand\C {{\cal C}}

\newcommand\cU {{\cal U}}
\newcommand\cV {{\cal V}}
\newcommand\cW {{\cal W}}

\newcommand\bZ {{\mathbb Z}}

\newcommand\beq {\begin{equation}}
	\newcommand\eeq {\end{equation}}
\newcommand\beqa {\begin{equatiobn}\begin{array}}
		\newcommand\eeqa {\end{array}\end{equation}}
\newcommand\bal {\begin{align}}
	\newcommand\eal {\end{align}}
\newcommand{\bea}{\begin{eqnarray}}
	\newcommand{\eea}{\end{eqnarray}}

\newcommand{\ztwo}{\mathbb{Z}_2}

\newcommand\SD[1]{[#1]}

\theoremstyle{plain}

\theoremstyle{definition}

\theoremstyle{remark}

\tikzset{snake it/.style={decorate,decoration={snake,segment length=2pt,amplitude=1pt}}}

\newcommand{\plaqy}{\tikz[baseline=.5ex, scale=.4]{
		\draw[line width=1.2pt](0,0)node[](A){}--(90:1)node[](B){};
		\draw[line width=1.2pt](1,0)node[](D){}--++(90:1)node[](C){};
		\shade[inner color=white, outer color = red] (A) circle (0.2);
		\shade[inner color=white, outer color = red] (B) circle (0.2);
		\shade[inner color=white, outer color = red] (C) circle (0.2);
		\shade[inner color=white, outer color = red] (D) circle (0.2);
	}
}

\newcommand{\plaqypm}{\tikz[baseline=.5ex, scale=.4]{
		\draw[line width=1.2pt](0,0)node[](A){}--(90:1)node[](B){};
		\draw[blue, snake it](1,0)node[](D){}--++(90:1)node[](C){};
		\shade[inner color=white, outer color = red] (A) circle (0.2);
		\shade[inner color=white, outer color = red] (B) circle (0.2);
		\shade[inner color=white, outer color = red] (C) circle (0.2);
		\shade[inner color=white, outer color = red] (D) circle (0.2);
	}
}

\newcommand{\plaqymp}{\tikz[baseline=.5ex, scale=.4]{
		\draw[blue,snake it](0,0)node[](A){}--(90:1)node[](B){};
		\draw[line width=1.2pt](1,0)node[](D){}--++(90:1)node[](C){};
		\shade[inner color=white, outer color = red] (A) circle (0.2);
		\shade[inner color=white, outer color = red] (B) circle (0.2);
		\shade[inner color=white, outer color = red] (C) circle (0.2);
		\shade[inner color=white, outer color = red] (D) circle (0.2);
	}
}

\newcommand{\plaqydp}{\tikz[baseline=.5ex, scale=.4]{
     \draw[thin,blue](-0.05,0)node[]{}--(-0.05,1)node[]{};
    \draw[thin,blue](0.05,0)node[]{}--(0.05,1)node[]{};
   \draw[line width=1.2pt](1,0)node[](D){}--++(90:1)node[](C){};
    \shade[inner color=white, outer color = red] (A) circle (0.2);
    \shade[inner color=white, outer color = red] (B) circle (0.2);
    \shade[inner color=white, outer color = red] (C) circle (0.2);
    \shade[inner color=white, outer color = red] (D) circle (0.2);
    }
}

\newcommand{\plaqypd}{\tikz[baseline=.5ex, scale=.4]{
     \draw[thin,blue](1-0.05,0)node[]{}--(1-0.05,1)node[]{};
    \draw[thin,blue](1+0.05,0)node[]{}--(1+0.05,1)node[]{};
   \draw[line width=1.2pt](0,0)node[](D){}--++(90:1)node[](C){};
    \shade[inner color=white, outer color = red] (0,0) circle (0.2);
    \shade[inner color=white, outer color = red] (0,1) circle (0.2);
    \shade[inner color=white, outer color = red] (1,0) circle (0.2);
    \shade[inner color=white, outer color = red] (1,1) circle (0.2);
    }
}

\newcommand{\plaqydm}{\tikz[baseline=.5ex, scale=.4]{
     \draw[thin,blue](-0.05,0)node[]{}--(-0.05,1)node[]{};
    \draw[thin,blue](0.05,0)node[]{}--(0.05,1)node[]{};
   \draw[blue,snake it](1,0)node[](D){}--++(90:1)node[](C){};
    \shade[inner color=white, outer color = red] (A) circle (0.2);
    \shade[inner color=white, outer color = red] (B) circle (0.2);
    \shade[inner color=white, outer color = red] (C) circle (0.2);
    \shade[inner color=white, outer color = red] (D) circle (0.2);
    }
}

\newcommand{\plaqymd}{\tikz[baseline=.5ex, scale=.4]{
     \draw[thin,blue](1-0.05,0)node[]{}--(1-0.05,1)node[]{};
    \draw[thin,blue](1+0.05,0)node[]{}--(1+0.05,1)node[]{};
   \draw[blue,snake it](0,0)node[](D){}--++(90:1)node[](C){};
    \shade[inner color=white, outer color = red] (0,0) circle (0.2);
    \shade[inner color=white, outer color = red] (0,1) circle (0.2);
    \shade[inner color=white, outer color = red] (1,0) circle (0.2);
    \shade[inner color=white, outer color = red] (1,1) circle (0.2);
    }
}

\newcommand{\gspi}{\tikz[baseline=.5ex, scale=.4]{
		\draw[line width=1.2pt,white](0,0)node[]{}--(-1,0)node[]{};
		\draw[line width=1.2pt,white](0,1)node[]{}--(-1,1)node[]{};
		\draw[line width=1.2pt,white](7,0)node[]{}--(8,0)node[]{};
		\draw[line width=1.2pt,white](7,1)node[]{}--(8,1)node[]{};
		\foreach \x in {0,2,4,6}
		{\fill[rounded corners,gray,opacity=0.2] (\x-0.4,-0.4) rectangle (\x+1.4,1.4);}
		\foreach \x in {0,1,2,3,4,5,6,7}
		{\draw[line width=1.2pt](\x,0)node[]{}--(\x,1)node[]{};
			\shade[inner color=white, outer color = red] (\x,0) circle (0.2);
			\shade[inner color=white, outer color = red] (\x,1) circle (0.2);
	}}
}

\newcommand{\gszero}{\tikz[baseline=.5ex, scale=.4]{
		\draw[line width=1.2pt,white](0,0)node[]{}--(-1,0)node[]{};
		\draw[line width=1.2pt,white](0,1)node[]{}--(-1,1)node[]{};
		\draw[line width=1.2pt,white](7,0)node[]{}--(8,0)node[]{};
		\draw[line width=1.2pt,white](7,1)node[]{}--(8,1)node[]{};
		\foreach \x in {0,2,4,6}
		{\fill[rounded corners,gray,opacity=0.2] (\x-0.4,-0.4) rectangle (\x+1.4,1.4);}
		\foreach \x in {0,1,2,3,4,5,6,7}
		{\draw[blue,snake it](\x,0)node[]{}--(\x,1)node[]{};
			\shade[inner color=white, outer color = red] (\x,0) circle (0.2);
			\shade[inner color=white, outer color = red] (\x,1) circle (0.2);
	}}
}
\newcommand{\gsmpibytwo}{\tikz[baseline=.5ex, scale=.4]{
		\draw[line width=1.2pt,white](0,0)node[]{}--(-1,0)node[]{};
		\draw[line width=1.2pt,white](0,1)node[]{}--(-1,1)node[]{};
		\draw[line width=1.2pt,white](7,0)node[]{}--(8,0)node[]{};
		\draw[line width=1.2pt,white](7,1)node[]{}--(8,1)node[]{};
		\foreach \x in {0,2,4,6}
		{\fill[rounded corners,gray,opacity=0.2] (\x-0.4,-0.4) rectangle (\x+1.4,1.4);
			\draw[line width=1.2pt](\x,0)node[]{}--(\x+1,0)node[]{};
			\draw[line width=1.2pt](\x,1)node[]{}--(\x+1,1)node[]{};
			\draw[line width=1.2pt](\x,0)node[]{}--(\x,1)node[]{};
			\draw[line width=1.2pt](\x+1,0)node[]{}--(\x+1,1)node[]{};
			\shade[inner color=white, outer color = red] (\x,0) circle (0.2);
			\shade[inner color=white, outer color = red] (\x,1) circle (0.2);
			\shade[inner color=white, outer color = red] (\x+1,0) circle (0.2);
			\shade[inner color=white, outer color = red] (\x+1,1) circle (0.2);
		}
	}
}
\newcommand{\gspibytwo}{\tikz[baseline=.5ex, scale=.4]{
		\foreach \x in {0,2,4,6}
		{\fill[rounded corners,gray,opacity=0.2] (\x-0.4,-0.4) rectangle (\x+1.4,1.4);
			\draw[line width=1.2pt](\x+1,0)node[]{}--(\x+2,0)node[]{};
			\draw[line width=1.2pt](\x+1,1)node[]{}--(\x+2,1)node[]{};
			\draw[line width=1.2pt](\x,0)node[]{}--(\x,1)node[]{};
			\draw[line width=1.2pt](\x+1,0)node[]{}--(\x+1,1)node[]{};
			\shade[inner color=white, outer color = red] (\x,0) circle (0.2);
			\shade[inner color=white, outer color = red] (\x,1) circle (0.2);
			\shade[inner color=white, outer color = red] (\x+1,0) circle (0.2);
			\shade[inner color=white, outer color = red] (\x+1,1) circle (0.2);
		}
		\draw[line width=1.2pt](0,0)node[]{}--(-1,0)node[]{};
		\draw[line width=1.2pt](0,1)node[]{}--(-1,1)node[]{};
		\shade[inner color=white, outer color = red] (0,0) circle (0.2);
		\shade[inner color=white, outer color = red] (0,1) circle (0.2);
	}
}

\newcommand{\dimy}{\tikz[baseline=.5ex, scale=.4]{
		\draw[line width=1.3pt](0,0)node[](A){}--(90:1)node[](B){};
		\shade[inner color=white, outer color = red] (A) circle (0.2);
		\shade[inner color=white, outer color = red] (B) circle (0.2);
	}
}

\newcommand{\plaqyblue}{\tikz[baseline=.5ex, scale=.4]{
		\draw[blue, snake it](0,0)node[](A){}--(90:1)node[](B){};
		\draw[blue, snake it](1,0)node[](D){}--++(90:1)node[](C){};
		\shade[inner color=white, outer color = red] (A) circle (0.2);
		\shade[inner color=white, outer color = red] (B) circle (0.2);
		\shade[inner color=white, outer color = red] (C) circle (0.2);
		\shade[inner color=white, outer color = red] (D) circle (0.2);
	}
}

\newcommand{\dimyblue}{\tikz[baseline=.5ex, scale=.4]{
		\draw[blue, snake it](0,0)node[](A){}--(90:1)node[](B){};
		\shade[inner color=white, outer color = red] (A) circle (0.2);
		\shade[inner color=white, outer color = red] (B) circle (0.2);
	}
}

\newcommand{\dimydouble}{\tikz[baseline=.5ex, scale=.4]{
		\draw[thin,blue](-0.05,0)node[]{}--(-0.05,1)node[]{};
		\draw[thin,blue](0.05,0)node[]{}--(0.05,1)node[]{};
		\shade[inner color=white, outer color = red] (0,0) circle (0.2);
		\shade[inner color=white, outer color = red] (0,1) circle (0.2);
	}
}

\newcommand{\plaqx}{\tikz[baseline=.5ex, scale=.4]{
		\draw[line width=1.3pt](0,0)node[](A){}--(1,0)node[](B){};
		\draw[line width=1.3pt](0,1)node[](D){}--(1,1)node[](C){};
		\draw[line width=1.3pt] (0,0) -- (0,1);
		\draw[line width=1.3pt] (1,0) -- (1,1);
		\shade[inner color=white, outer color = red] (A) circle (0.2);
		\shade[inner color=white, outer color = red] (B) circle (0.2);
		\shade[inner color=white, outer color = red] (C) circle (0.2);
		\shade[inner color=white, outer color = red] (D) circle (0.2);
	}
}

\begin{document}
	
	\title{Charge pumps, boundary modes, and the necessity of unnecessary criticality}
	\author{Abhishodh Prakash}
	\email{abhishodhprakash@hri.res.in (he/him/his)}
	\affiliation{Rudolf Peierls Centre for Theoretical Physics, University of Oxford, United Kingdom}
	\affiliation{Harish-Chandra Research Institute, a CI of Homi Bhabha National Institute, Prayagraj (Allahabad), India}
	\author{S.A. Parameswaran}
	\email{sid.parameswaran@physics.ox.ac.uk}
	\affiliation{Rudolf Peierls Centre for Theoretical Physics, University of Oxford, United Kingdom}
	
	\begin{abstract}
		We link the presence of ``unnecessary'' quantum critical surfaces within a single gapped phase of matter to the non-trivial topology of {\it families} of gapped Hamiltonians that encircle the critical surface. We study a specific set of one-dimensional spin models where each such family forms a one-parameter loop in a two-dimensional phase diagram.  Foliating the non-critical region by such loops identifies ``radial'' and ``angular'' coordinates in the phase diagram that respectively parametrize different families and different members of a single family.  We show that each one-parameter family is a generalized Thouless charge pump, all with the same topological index, and hence the gapped phase undergoes one or more nontrivial boundary phase transitions as we vary the angular coordinate in a loop through members of one family. Tuning the radial coordinate generates loci of boundary critical points that terminate at endpoints of the bulk unnecessary critical line within the gapped phase. We discuss broader implications of our results and possible extensions to higher dimensions.  
	\end{abstract}
	
	\maketitle
	
	Understanding changes in dynamics and thermodynamics as the parameters of a system are varied is a central problem in many-body physics. For quantum systems near absolute zero, our interest here, these properties are encoded in the ground state and low-lying excitations, so a natural focus is on surfaces in parameter space on which the excitation gap closes. A much-studied example involves quantum criticality~\cite{Sachdev_book,SondhiQuantumPhaseTransitionsRevModPhys.69.315}, where the gap closes as a \textit{single} parameter is tuned ---  a situation intermediate between stable phases whose gaplessness is robust to all changes in parameters, and multicriticality, accessed by tuning multiple parameters. The most interesting scenario is when the gap vanishes continuously as the critical surface is approached. This signals a vanishing scale for fluctuations, validating a long-wavelength field theory description. This in turn captures the scaling behaviour of correlation functions, that can be studied within the renormalization group (RG) framework~\cite{WilsonRG_RevModPhys.55.583}.
	
	Conventional wisdom holds that quantum criticality is tied to a transition between different phases of matter. Such distinctions could be based on symmetry breaking in ground-state observables {\it à la} Landau, or could stem from topological features~\cite{Haldane_Nobel_RevModPhys.89.040502} of the ground state wavefunction that can be diagnosed using  non-local properties such as entanglement~~\cite{PollmanTurnerBergOshikawa_entanglementspectrum_PhysRevB.81.064439,PollmannTurnerSPTPhysRevB.86.125441}, as in several contemporary examples. However, recent work has identified instances where this linkage fails:  quantum criticality  emerges \textit{within a single phase of matter}~\cite{AnfusoRosch_PhysRevB.75.144420,BiSenthil_UC_PhysRevX.9.021034,JianXu_UC_PhysRevB.101.035118,APUCPhysRevLett.130.256401,Pollmann_Quotient,YuchiPhysRevLett.132.136501,ZhangSenthilUC2024Dirac,APClassicalDQC}. The resulting phase diagram resembles that of a liquid-gas transition ---  it involves a critical surface that terminates within a phase and can hence be avoided by a suitably chosen path in parameter space --- but with the difference that it is continuous rather than first-order. 	\begin{figure}[!t]
		\centering{
			\begin{tikzpicture}
		\draw[thick,->] (-0.6,-0.1) -- ++(7.,0.0) node[align=right,below] {$J_U$};
		\draw[thick,->] (-0.6,-0.1) -- ++(0.0,3.) node[align=left,above] {$\delta$};
		\fill[blue,opacity=0.05] (-0.5,0.) rectangle ++(6.8,2.8);

    \draw[dotted] (6.,2.8) to[out=-90,in=20] (5,1.8);
    \draw[dotted] (6,0) to[out=90,in=-20] (5,2.8-1.8);

   \draw[dotted] (-0.2,2.8) to[out=-90,in=160] (0.8,1.8);
    \draw[dotted] (-0.2,0) to[out=90,in=-160] (0.8,2.8-1.8);
    
            \draw[thick] (2,1.4) -- (5.8-2,2.8-1.4);
            \fill[red] (2,1.4) circle (0.05);
            \fill[red] (5.8-2,2.8-1.4) circle (0.05);
            \fill[blue,opacity=0.1] (1.6,2.8) to[out=-90,in = 140] (2,1.4) -- (5.8-2,1.4) to[out=40,in=-90]  (5.8-1.6,2.8) -- cycle;
            \draw[dashed] (1.6,2.8) to[out=-90,in = 140] (2,1.4);
            \draw[dashed]  (5.8-1.6,2.8)to[in=40,out=-90]  (5.8-2,1.4);

            \fill[green,opacity=0.2,rounded corners] (2,1.4) to[bend left] (1.3,2) to[bend right] (1.3,0.8) to[bend left] (2,1.4);
            \draw[rounded corners] (2,1.4) to[bend left] (1.3,2) to[bend right] (1.3,0.8) to[bend left] (2,1.4);
            \fill[yellow,opacity=0.2,rounded corners] (5.8-2,1.4) to[bend right] (5.8-1.3,2)  to[bend left] (5.8-1.3,0.8) to[bend right] (5.8-2,1.4)  ;
            \draw[rounded corners] (5.8-2,1.4) to[bend right] (5.8-1.3,2)  to[bend left] (5.8-1.3,0.8) to[bend right] (5.8-2,1.4) ;
            \node[] at(1.42,1.4) {\scriptsize XY$_1^*$};
            \node[] at(4.41,1.4) {\scriptsize XY$_1$};
            \node[] at(2.9,1.52) {\scriptsize XY$_2$};

            \node[] at(2.1,2.5) {\scriptsize T$_1$};
            \node[] at(0.1,2.5) {\scriptsize T$_1$};
            \node[] at(5.7,2.5) {\scriptsize T$_1$};

            \node[] at(2.1,0.3) {\scriptsize T$_3$};

            \node[] at(-0.0,1.8) {\scriptsize T$_2$};
            \node[] at(5.8,1.8) {\scriptsize T$_4$};

            \node[] at(2.9,-0.25) {0};
            \node[] at(-0.8,0.25) {-1};
            \node[] at(-0.8,1.4) {0};
            \node[] at(-0.8,2.7) {1};
              \node[] at(2.9,1) {\scriptsize $c=3/2$};
              \draw[->,thin] (2.4,1.05) -- (2.05,1.3);
              \draw[->,thin] (3.4,1.05) -- (3.75,1.3);
              
              \def\ya{.1};
              \def\yb{.23};
              \foreach   \x in {2.4,3.0} {\fill[rounded corners,gray,opacity=0.2] (\x,\ya) rectangle (\x+0.5,\ya+.5); \draw[thick]  (\x+0.15,\yb) rectangle(\x+0.35,\yb+0.23);
              \shade[inner color=white, outer color = red] (\x+0.15,\yb) circle (0.06);\shade[inner color=white, outer color = red] (\x+0.35,\yb) circle (0.06);\shade[inner color=white, outer color = red] (\x+0.15,\yb+0.23) circle (0.06);\shade[inner color=white, outer color = red] (\x+0.35,\yb+0.23) circle (0.06);};
              
              \def\ya{2.2};
              \def\yb{2.33};
              \foreach \x in {3.15} {\draw[thick]   (\x+0.45,\yb+0.23)--(\x+0.2,\yb+0.23)-- (\x+0.2,\yb)--(\x+0.45,\yb);}
              \foreach \x in {3.0} {\draw[thick]  (\x+0.15,\yb) rectangle (\x-0.25,\yb+0.23);}
              \foreach   \x in {2.4} {\draw[blue]  (\x+0.135,\yb+0.23)-- (\x+0.135,\yb);\draw[blue]  (\x+0.165,\yb+0.23)-- (\x+0.165,\yb);}
              \foreach   \x in {2.4,3.0} {\fill[rounded corners,gray,opacity=0.2] (\x,\ya) rectangle (\x+0.5,\ya+.5); 
              \shade[inner color=white, outer color = red] (\x+0.15,\yb) circle (0.06);\shade[inner color=white, outer color = red] (\x+0.35,\yb) circle (0.06);\shade[inner color=white, outer color = red] (\x+0.15,\yb+0.23) circle (0.06);\shade[inner color=white, outer color = red] (\x+0.35,\yb+0.23) circle (0.06);};

              \def\sh{-2.05};
              \foreach \x in {3.15+\sh} {\draw[thick] (\x+0.45,\yb+0.23)--(\x+0.2,\yb+0.23)-- (\x+0.2,\yb)--(\x+0.45,\yb);}
              \foreach \x in {3.0+\sh} {\draw[thick]  (\x+0.15,\yb) rectangle (\x-0.25,\yb+0.23);}
              \foreach   \x in {2.4+\sh} {\draw[thick]  (\x+0.15,\yb+0.23)-- (\x+0.15,\yb);}
              \foreach   \x in {2.4+\sh,3.0+\sh} {\fill[rounded corners,gray,opacity=0.2] (\x,\ya) rectangle (\x+0.5,\ya+.5); 
              \shade[inner color=white, outer color = red] (\x+0.15,\yb) circle (0.06);\shade[inner color=white, outer color = red] (\x+0.35,\yb) circle (0.06);\shade[inner color=white, outer color = red] (\x+0.15,\yb+0.23) circle (0.06);\shade[inner color=white, outer color = red] (\x+0.35,\yb+0.23) circle (0.06);};
            
            \def\sh{1.85};
              \foreach \x in {3.15+\sh} {\draw[thick] (\x+0.45,\yb+0.23)--(\x+0.2,\yb+0.23)-- (\x+0.2,\yb)--(\x+0.45,\yb);}
              \foreach \x in {3.0+\sh} {\draw[thick]  (\x+0.15,\yb) rectangle (\x-0.25,\yb+0.23);}
              \foreach   \x in {2.4+\sh} {\draw[blue, snake it]  (\x+0.15,\yb+0.23)-- (\x+0.15,\yb);}
              \foreach   \x in {2.4+\sh,3.0+\sh} {\fill[rounded corners,gray,opacity=0.2] (\x,\ya) rectangle (\x+0.5,\ya+.5); 
              \shade[inner color=white, outer color = red] (\x+0.15,\yb) circle (0.06);\shade[inner color=white, outer color = red] (\x+0.35,\yb) circle (0.06);\shade[inner color=white, outer color = red] (\x+0.15,\yb+0.23) circle (0.06);\shade[inner color=white, outer color = red] (\x+0.35,\yb+0.23) circle (0.06);};
              
              \def\sh{0.45};
            \def\ya{1.15};
              \def\yb{1.28};
           \foreach   \x in {4.75+\sh,5.3+\sh} {\fill[rounded corners,gray,opacity=0.2] (\x,\ya) rectangle (\x+0.5,\ya+.5); \draw[blue, snake it]  (\x+0.15,\yb)--(\x+0.15,\yb+0.23);\draw[blue, snake it]  (\x+0.35,\yb)--(\x+0.35,\yb+0.23); 
              \shade[inner color=white, outer color = red] (\x+0.15,\yb) circle (0.06);\shade[inner color=white, outer color = red] (\x+0.35,\yb) circle (0.06);\shade[inner color=white, outer color = red] (\x+0.15,\yb+0.23) circle (0.06);\shade[inner color=white, outer color = red] (\x+0.35,\yb+0.23) circle (0.06);}

           \def\sh{-0.45};
           \foreach   \x in {0.0+\sh,0.55+\sh} {\fill[rounded corners,gray,opacity=0.2] (\x,\ya) rectangle (\x+0.5,\ya+.5); \draw[thick]  (\x+0.15,\yb)--(\x+0.15,\yb+0.23);\draw[thick]  (\x+0.35,\yb)--(\x+0.35,\yb+0.23); 
              \shade[inner color=white, outer color = red] (\x+0.15,\yb) circle (0.06);\shade[inner color=white, outer color = red] (\x+0.35,\yb) circle (0.06);\shade[inner color=white, outer color = red] (\x+0.15,\yb+0.23) circle (0.06);\shade[inner color=white, outer color = red] (\x+0.35,\yb+0.23) circle (0.06);}              
              \node[] at(2.9,3.0) {\scriptsize  $(a)~J_M<0$};         
		\end{tikzpicture} 
   
			
			\vspace{-0.2cm}
			     \begin{tikzpicture}[scale=1]
		\draw[thick,->] (-0.35,-0.1) -- ++(2.5,0.0) node[align=right,below] {$J_U$};
		\draw[thick,->] (-0.35,-0.1) -- ++(0.0,3.) node[align=left,above] {$\delta$};
		\fill[blue,opacity=0.05] (-0.25,0.) rectangle ++(2.25,2.8);
            \draw[dashed,blue] (1.8/2,2.8) -- (1.8/2,1.4);

  \draw[dotted] (-0.1,2.8) to[out=-90,in=140] (.4,1.8);
  \draw[dotted] (-0.1,0) to[out=90,in=-140] (.4,1);

  \draw[dotted] (1.85,2.8) to[out=-90,in=40] (1.85-.5,1.8);
  \draw[dotted] (1.85,0) to[out=90,in=-40] (1.85-.5,1);

\node[] at(0.16,2.6) {\scriptsize T$_1$};
\node[] at(1.67,2.6) {\scriptsize T$_1$};
\node[] at(0.0,1.8) {\scriptsize T$_2$};
\node[] at(1.8,1.8) {\scriptsize T$_4$};
\node[] at(0.95,0.67) {\scriptsize T$_3$};

            \def\ya{2.25};
              \def\yb{2.38};
              \foreach   \x in {1} {\fill[rounded corners,gray,opacity=0.2] (\x,\ya) rectangle (\x+0.5,\ya+.5);\draw[blue, snake it]  (\x+0.15,\yb)--(\x+0.15,\yb+0.23); \draw[thick] (\x+0.35,\yb+0.23)-- (\x+0.35,\yb)--(\x+0.55,\yb);\draw[thick]  (\x+0.35,\yb+0.23)--(\x+0.55,\yb+0.23);
              \shade[inner color=white, outer color = red] (\x+0.15,\yb) circle (0.06);\shade[inner color=white, outer color = red] (\x+0.35,\yb) circle (0.06);\shade[inner color=white, outer color = red] (\x+0.15,\yb+0.23) circle (0.06);\shade[inner color=white, outer color = red] (\x+0.35,\yb+0.23) circle (0.06);};

              \foreach   \x in {.3} {\fill[rounded corners,gray,opacity=0.2] (\x,\ya) rectangle (\x+0.5,\ya+.5);\draw[thick]  (\x+0.15,\yb)--(\x+0.15,\yb+0.23); \draw[thick] (\x+0.35,\yb+0.23)--   (\x+0.35,\yb)--(\x+0.55,\yb);\draw[thick]  (\x+0.35,\yb+0.23)--(\x+0.55,\yb+0.23);
              \shade[inner color=white, outer color = red] (\x+0.15,\yb) circle (0.06);\shade[inner color=white, outer color = red] (\x+0.35,\yb) circle (0.06);\shade[inner color=white, outer color = red] (\x+0.15,\yb+0.23) circle (0.06);\shade[inner color=white, outer color = red] (\x+0.35,\yb+0.23) circle (0.06);};
            
            \def\ya{1.15};
              \def\yb{1.28};
           \foreach   \x in {-.2} {\fill[rounded corners,gray,opacity=0.2] (\x,\ya) rectangle (\x+0.5,\ya+.5); \draw[thick]  (\x+0.15,\yb)--(\x+0.15,\yb+0.23);\draw[thick]  (\x+0.35,\yb)--(\x+0.35,\yb+0.23); 
              \shade[inner color=white, outer color = red] (\x+0.15,\yb) circle (0.06);\shade[inner color=white, outer color = red] (\x+0.35,\yb) circle (0.06);\shade[inner color=white, outer color = red] (\x+0.15,\yb+0.23) circle (0.06);\shade[inner color=white, outer color = red] (\x+0.35,\yb+0.23) circle (0.06);};

           \foreach   \x in {1.45} {\fill[rounded corners,gray,opacity=0.2] (\x,\ya) rectangle (\x+0.5,\ya+.5); \draw[blue, snake it]  (\x+0.15,\yb)--(\x+0.15,\yb+0.23);\draw[blue, snake it]  (\x+0.35,\yb)--(\x+0.35,\yb+0.23); 
              \shade[inner color=white, outer color = red] (\x+0.15,\yb) circle (0.06);\shade[inner color=white, outer color = red] (\x+0.35,\yb) circle (0.06);\shade[inner color=white, outer color = red] (\x+0.15,\yb+0.23) circle (0.06);\shade[inner color=white, outer color = red] (\x+0.35,\yb+0.23) circle (0.06);};              

              \def\ya{.05};
              \def\yb{.18};
              \foreach   \x in {1.35/2} {\fill[rounded corners,gray,opacity=0.2] (\x,\ya) rectangle (\x+0.5,\ya+.5); \draw[thick]  (\x+0.15,\yb)rectangle (\x+0.35,\yb+0.23);
              \shade[inner color=white, outer color = red] (\x+0.15,\yb) circle (0.06);\shade[inner color=white, outer color = red] (\x+0.35,\yb) circle (0.06);\shade[inner color=white, outer color = red] (\x+0.15,\yb+0.23) circle (0.06);\shade[inner color=white, outer color = red] (\x+0.35,\yb+0.23) circle (0.06);};

            \fill[] (1.8/2,1.4) circle (.05);
            \node[] at(1.8/2,1.24) {\scriptsize XY$_3$};
            \node[] at(1.8/2,-0.28) {0};
            \node[] at(-0.55,0.25) {-1};
            \node[] at(-0.55,1.4) {0};
            \node[] at(-0.55,2.7) {1};
            \node[] at(0.9,3.0) {\scriptsize $(b)~J_M>0$};
		\end{tikzpicture} 
			\begin{tikzpicture}[scale = 1]

\node[] at (1.5,1.25) { $h_L^+$};
\node[] at (1.5,-0.23) { $h_L^+$};
   \node[] at (-0.8,1.9) { $(c)$ };
   \node[] at (1.8,-1.2) { $h_L^{\pm} \equiv (1\pm \delta) (S^x_{a}S^x_{b}+S^y_{a}S^y_{b}+ \Delta S^z_{a}S^z_{b} )$ };
   \node[] at (1.8,-1.8) { $h_R \equiv~~~ J_U\left(S^x_{a}S^x_{b}+S^y_{a}S^y_{b} \right)+ J_M S^z_{a}S^z_{b}$ };
   \node[] at (1.8,-2.4) { };
    \foreach \x in {0,2}
    {\fill[rounded corners,gray,opacity=0.2] (\x-0.4,-0.4) rectangle (\x+1.4,1.4);
    \node[] at (\x+.5,1.25) { $h_L^-$};
    \node[] at (\x+.5,-0.23) { $h_L^-$};
     \node[] at (\x,0.5) { $h_R$};
    \node[] at (\x+1,0.5) {$h_R$};
     \draw[line width=1.2pt,red,dashed](\x,0)node[]{}--(\x+1,0)node[]{};
     \draw[line width=1.2pt,red, dashed](\x,1)node[]{}--(\x+1,1)node[]{};
    \draw[line width=1.2pt](\x+1,0)node[]{}--(\x+2,0)node[]{};
    \draw[line width=1.2pt](\x+1,1)node[]{}--(\x+2,1)node[]{};
    \draw[line width=1.2pt,dotted, blue](\x,0)node[]{}--(\x,1)node[]{};
    \draw[line width=1.2pt,dotted, blue](\x+1,0)node[]{}--(\x+1,1)node[]{};
    \shade[inner color=white, outer color = red] (\x,0) circle (0.15);
    \shade[inner color=white, outer color = red] (\x,1) circle (0.15);
     \shade[inner color=white, outer color = red] (\x+1,0) circle (0.15);
    \shade[inner color=white, outer color = red] (\x+1,1) circle (0.15);
    }
    \draw[line width=1.2pt](0,0)node[]{}--(-1,0)node[]{};
    \draw[line width=1.2pt](0,1)node[]{}--(-1,1)node[]{};
    \shade[inner color=white, outer color = red] (0,0) circle (0.15);
    \shade[inner color=white, outer color = red] (0,1) circle (0.15);
\end{tikzpicture} 
		}\vspace{-0.5cm}
		\caption{\textbf{Unnecessary critical (a) and multicrital (b) phase diagrams in a spin ladder (c),} {for $J_M \approx \mp 5.2, \Delta \approx -0.05$}.
			(a), (b) show evolution of ground states near one end of an open ladder using graphical notation described in the text. Over the pumping cycle the ladder remains in a single bulk gapped phase, but its boundary transforms under distinct irreducible representations of  layer exchange ($\ztwo^L$) and spin-rotation ($O(2)$) symmetry for $\delta>0$ and $J_U <-|J_M|$ or $J_U > |J_M|$. These are separated by a single boundary transition (dashed line in (b)) or a non-trivial boundary phase involving a 2D boundary irrep of $O(2)$ (blue shaded region in (a)). Dotted lines in (a), (b) represent crossovers between `RG basins' of symmetric field theory vacua T$_{1,\ldots,4}$ of \cref{tab:vacuumsummary}.}
		\label{fig:phasediags}
		\vspace{-0.3cm}
	\end{figure}
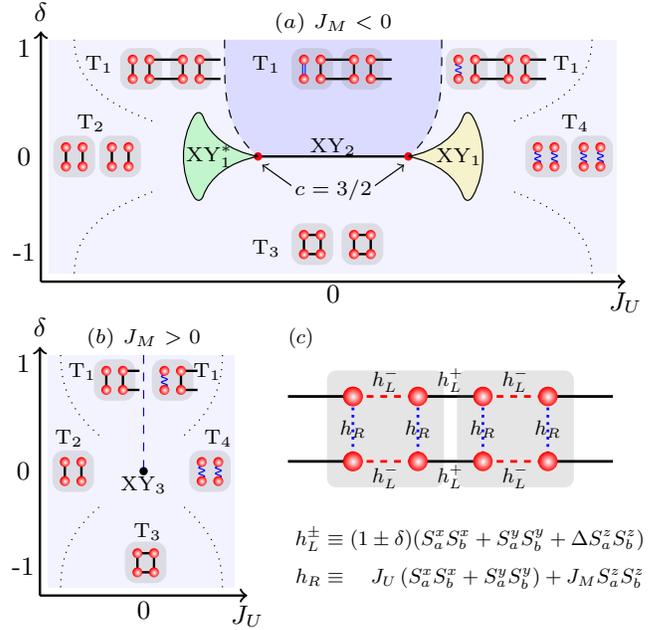     Such ``unnecessary critical points'' cannot be diagnosed by the usual tactic of studying gapped Hamiltonians straddling the critical surface: one can always find an adiabatic path between any two such points along which the bulk energy gap remains open, and so no sharp distinction can be drawn between Hamiltonians on the two ``sides''. There is as yet no general understanding of the criteria for the emergence and stability of unnecessary criticality.

	Here, we argue that ``unneccessary'' criticality  within a region of parameter space can be necessitated by the topological properties of the {\it families} of gapped Hamiltonians that enclose the region~\cite{kitaev2019}.  In the example we study here, this family can be viewed as a closed loop in a two-dimensional parameter space: while   each of its members taken in isolation  belongs to the same trivial phase, the one-parameter family of Hamiltonians over the closed loop represents a non-trivial pump~\cite{ThoulessChargePumpPhysRevB.27.6083,TeoKanePhysRevB.82.115120,YoshihitoYasuhiro_SPTPump_PhysRevResearch.2.042024,ShiozakiPumpFermionPhysRevB.106.165115,DeRoeckbachmann2023classificationgchargethoulesspumps,KapustinSpodyneiko2020higherdimensionalgeneralizationsthoulesscharge}  that transports symmetry charge across the system. It has been previously argued~\cite{ThorngrenDiabolicalPhysRevB.102.245113} that this guarantees that the bulk gap closes {somewhere} within the region enclosed by the pump. However,  this picture does not explain how a critical line rather than a multicritical point might arise, or discuss other aspects of the phase diagram ---  questions we address in this work, using a strategy we now outline. We first show that a class of spin chains previously introduced by us~\cite{APUCPhysRevLett.130.256401}  exemplifies the link between charge pumps and unnecessary criticality.  We explain how distinct variants of the pumping cycle (extracted through exact calculations on a one-parameter family  of zero-correlation-length Hamiltonians) can lead to distinct unnecessary critical scenarios in its interior (which is amenable to a field theoretic analysis). In doing so, we  prove a conjectured link~\cite{APUCPhysRevLett.130.256401} between {stable boundary modes} and bulk unnecessary criticality in our specific setting, and use this along with the bulk-boundary correspondence for Hamiltonian families~\cite{Seiberg_AnomaliesCouplingOne10.21468/SciPostPhys.8.1.001,Seiberg_AnomaliesCouplingTwo10.21468/SciPostPhys.8.1.002,HermeleGappedfamiliesPhysRevB.108.125147,ThorngrenDiabolicalPhysRevB.102.245113} to sharpen the conjecture in the general case. We close with a discussion of the broader implications of our work, and possible extensions of these ideas to higher dimensions.
	
	\textit{Model, Symmetries, and Phase Diagrams.---} We consider the Hamiltonian~\cite{APUCPhysRevLett.130.256401}  $H = H_{L,1} + H_{L,2} + H_R$, where 
	\begin{align}
		H_{L,\alpha} &= \sum_{j=1}^L  J_j(\delta) ( S^x_{\alpha j} S^x_{\alpha j+1}  +  S^y_{\alpha j} S^y_{\alpha j+1} + \Delta  S^z_{\alpha j} S^z_{\alpha j+1}),\nonumber\\
		H_R &=      \sum_{j} \left[ J_MS^z_{1 j} S^z_{2 j} + J_U \left(S^x_{1 j} S^x_{2 j}+S^y_{1 j} S^y_{2 j} \right)\right], \label{eq:H}
	\end{align}
	with $\vec{S} = \frac{1}{2}\vec{\sigma}$ in terms of the Pauli matrices $\sigma^\mu$. \cref{eq:H} describes a spin ladder (\cref{fig:phasediags}(c)) each of whose legs is an XXZ spin chain with a staggered coupling $J_j(\delta) = 1+ (-1)^j\delta$, and rung couplings $J_U, J_M$, and is invariant under the following symmetries: (i)  $U(1)$ spin rotations around the $S^z$-axis, under which $S^x_{\alpha j} \pm i S^{y}_{\alpha j} \equiv S^\pm_{\alpha j}  \mapsto e^{\pm i\chi} S^\pm_{\alpha j}$; (ii) $\ztwo^R$ spin reflections, $S^\pm_{\alpha j}  \mapsto S^\mp_{\alpha j}$, $S^z_{\alpha j} \mapsto -S^z_{\alpha j}$; and (iii) $\ztwo^L$ `leg exchange', $S^\mu_{\alpha j} \mapsto S^\mu_{\bar{\alpha} j}$, with $\bar{1}=2, \bar{2}=1$. Together, these symmetries form the non-Abelian group $\left( U(1) \rtimes \ztwo^R \right) \times \ztwo^L \cong O(2) \times \ztwo^L$~\cite{GroupTheoryTung1985}.
	
	Varying  the bond dimerization $\delta$ and rung coupling $J_U$ yields various possible phase diagrams shown in \cref{fig:phasediags}, depending on the (fixed) values of $\Delta, J_M$.  \cref{fig:phasediags}(a) shows the unnecessary critical behaviour that was a focus  of Ref.~\onlinecite{APUCPhysRevLett.130.256401}, where a  second-order critical surface accessible by tuning a single parameter is stable to arbitrary symmetry-preserving perturbations, and terminates in extended regions of gapless phase. Criticality can be avoided by a suitable path in parameter space that, crucially, remains within a single gapped bulk phase. However, the phase diagram has a striking feature~\cite{APUCPhysRevLett.130.256401}: the presence of robust edge modes  (shaded region in \cref{fig:phasediags}(a)) that do not enjoy any symmetry protection.  These edge modes disappear via the bulk unnecessary critical transition or a boundary transition. Adopting the unconventional view of treating bulk and boundary phase structures on the same footing would give a compelling rationalization for the presence of the critical line, leading  to the conjecture~\cite{APUCPhysRevLett.130.256401} that the presence of edge modes and the merger of bulk and boundary transitions is a generic feature of  unnecessary critical phase diagrams, consistent with several other examples~\cite{MoudgalayaPollmann_UC_PhysRevB.91.155128,Pollmann_Quotient,AnfusoRosch_PhysRevB.75.144420,BiSenthil_UC_PhysRevX.9.021034,JianXu_UC_PhysRevB.101.035118}.
	
	We demonstrate the validity of this conjecture for  the models of \cref{eq:H}, both for the unnecessary critical line in  \cref{fig:phasediags}(a)  and its deformation to  a single ``unnecessary multicritical'' point [\cref{fig:phasediags}(b)] 
	(Ref.~\cite{SupMat} discusses other cases involving spontaneous symmetry breaking.)
	Along the way, we make contact with progress in the general program  of classifying topological phases parametrized over any manifold and in arbitrary dimension~\cite{kitaev2019,HermeleGappedfamiliesPhysRevB.108.125147,KapustinSpodyneiko2020higherdimensionalgeneralizationsthoulesscharge,ShiozakiPumpPhysRevB.106.125108,ShiozakiPumpFermionPhysRevB.106.165115,DeRoeckbachmann2023classificationgchargethoulesspumps,Marvin2023chartingspacegroundstates,Hermele_HomotopicalFoundationdoi:10.1142/S0129055X24600031,Seiberg_AnomaliesCouplingOne10.21468/SciPostPhys.8.1.001,Seiberg_AnomaliesCouplingTwo10.21468/SciPostPhys.8.1.002,ThorngrenDiabolicalPhysRevB.102.245113,ZhenghanWang_modulispaceofGapped10.1063/5.0136906,HastingsAasenAdiabaticPathsPhysRevB.106.085122,KapusticSpodyneiko_HigherBerry_PhysRevB.101.235130,RyuHigherMPSPhysRevB.109.115152,RyuohyamaHigheMPS2024higherberryconnectionmatrix,RyuohyamaHigherPEPS2024higherberryphaseprojected,Ashvinsommer2024higherberrycurvaturewaveMPS,Ashvinsommer2024higherberrycurvaturewavePEPS}.  This classification depends on the dimensions of both space $d$ and  parameter space $d_P$; our case corresponds to $d=1$, $d_P=2$, but we will set the stage more generally. Let us take the region of bulk gap  closing to lie within some compact domain around the origin of parameter space. Each $d_P-1$ surface that encloses the critical region without intersecting it constitutes a $d_P-1$ dimensional family of  $d$-dimensional gapped Hamiltonians; for  obvious reasons, we refer to the $d_P-1$ parameters that label Hamiltonians within a family as ``angular'', and the single parameter labelling different families as ``radial''.  The topological classification of gapped families generalizes Thouless's notion of a $d=1, d_P=2$ pump of $U(1)$ charge~\cite{ThoulessChargePumpPhysRevB.27.6083} to higher dimensions, more parameters, and other symmetries.  ~\footnote{Formally, such pumps are deformation classes of Hamiltonians parametrized over the $d_P-1$ ``angular''  variables, just as gapped topological phases are deformation classes of gapped Hamiltonians over a point.} One result  of the classification is that a parameter-space region enclosed by a nontrivial pump must host some loci of ``diabolical points''~\cite{ThorngrenDiabolicalPhysRevB.102.245113} where the bulk gap closes,  necessitating the superficially ``unneccessary'' criticality. Another is a generalized bulk-boundary correspondence, wherein a gap on the $d-1$ dimensional boundary closes for one or more members of each family, forming loci of boundary transitions { within the same {\it bulk} phase} as we move ``radially inward'' between families. These terminate at endpoints of the unnecessary critical surface, consistent with the conjecture of Ref.~\cite{APUCPhysRevLett.130.256401}.
	
	Despite its intuitive appeal, it is unclear if this picture holds for \cref{eq:H} and (here and more generally) whether it offers any insight on different possible criticality patterns in the interior of the phase diagram. To that end,  we explore these questions in an exactly solvable limit of \cref{eq:H}, and then show how aspects of the pump and the boundary physics are imprinted on the long-wavelength Luttinger liquid field theory in the near-critical regime.

	\textit{Charge Pump and Boundary Criticality.---} We begin by focusing on the periphery of the phase diagrams in \cref{fig:phasediags}, where $|J_U| \rightarrow \infty$ or $|\delta| = 1$; the `angular' parameter is chosen to evolve through this loop. For these values, $H$ decomposes into a sum of decoupled four-spin problems: thus, the phase diagram is encircled by an exactly solvable ``fixed-point family'' of Hamiltonians.  It is convenient to introduce a graphical notation for four distinct rung states: the two Bell pairs (in the $S^z$-basis)

	$
	\kett[\Big]{\dimy}= \frac{1}{\sqrt{2}} \left(\biggl|\begin{tabular}{c}
		$\uparrow$    \\
		$\downarrow$   
	\end{tabular}\biggr> + \biggl|\begin{tabular}{c}
		$\downarrow$    \\
		$\uparrow$   
	\end{tabular}\biggr>  \right), ~    \kett[\Big]{\dimyblue} = \frac{1}{\sqrt{2}} \left(\biggl|\begin{tabular}{c}
		$\uparrow$    \\
		$\downarrow$   
	\end{tabular}\biggr> - \biggl|\begin{tabular}{c}
		$\downarrow$    \\
		$\uparrow$   
	\end{tabular}\biggr>  \right)$, 
	and the `doublet' of states 
	$\kett[\Big]{\dimydouble} = \biggl|\begin{tabular}{c}
		$\uparrow$    \\
		$\uparrow$   
	\end{tabular}\biggr>,~\biggl|\begin{tabular}{c}
		$\downarrow$    \\
		$\downarrow$   
	\end{tabular}\biggr>$. 
	These transform in different irreduciple representations (irreps) of $O(2)\cong \left( U(1) \rtimes \ztwo^R \right)$, distinguished, e.g. by the action of $\ztwo^R$: the Bell pairs form one-dimensional irreps with $\ztwo^R$ eigenvalue $\pm 1$ respectively, while the states in 
	the doublet are exchanged by $\ztwo^R$, forming a two-dimensional irrep. Meanwhile, the antisymmetric Bell pair changes sign under $\ztwo^L$ leg exchange, which leaves the other rung states invariant. Finally, we also introduce the `plaquette state'  
	$ \kett[\Big]{\plaqx} = \alpha \kett[\Big]{\plaqy} + \beta \kett[\Big]{\plaqyblue} + \gamma \left( \biggl|\begin{tabular}{cc}
		$\uparrow$    & $\downarrow$  \\
		$\uparrow$   & $\downarrow$ 
	\end{tabular}\biggr> + \biggl|\begin{tabular}{cc}
		$\downarrow$    & $\uparrow$  \\
		$\downarrow$   & $\uparrow$ 
	\end{tabular}\biggr>\right)$, 
	where $\alpha,\beta,\gamma$ depend on $\Delta,J_M,J_U$ in a complicated way.  This is invariant under all symmetries of $H$ and reduces to $\kett[\Big]{\plaqyblue}$ and $\kett[\Big]{\plaqy}$ for $J_U \rightarrow \pm \infty$ respectively. 
	
	For $ 4\Delta > |J_M| - \sqrt{J_M^2+16}$ and periodic boundary conditions, the ground state $\ket{\textrm{GS}(J_U,\delta)}$  is unique~\cite{SupMat}, and smoothly evolves between four representative points,
	\begin{subequations}
		\begin{align}
			\ket{\text{GS}(\infty,\delta)} &= \cdots \gszero \cdots \label{eq:GS0} , \\
			\ket{\text{GS}(0,1)} &=\cdots \gspibytwo \cdots, \label{eq:GSpiby2}\\
			\ket{\text{GS}(-\infty,\delta)} &= \cdots \gspi \cdots, \label{eq:GSpi} \\
			\ket{\text{GS}(0,-1)} &=  \cdots \gsmpibytwo \cdots \label{eq:GSmpiby2},
		\end{align}
	\end{subequations}
	while the bulk gap remains open: as advertised, we can avoid criticality while remaining in a single bulk phase.
	
	However, as we now show, in an {\it open} ladder, the boundary undergoes a phase transition (properly, a level crossing in $d=0$) that can be understood in terms of symmetry. A more formal {argument that \cref{eq:GS0,eq:GSpiby2,eq:GSpi,eq:GSmpiby2} describes a non-trivial family}  may be possible, e.g. by using  `suspension isomorphism' to construct generalized charge pumps~\cite{HermeleGappedfamiliesPhysRevB.108.125147}, but the argument here is more intuitive and relates directly to one of our objectives. 
	
	To proceed, we consider a semi-infinite chain and fix a fiducial four-site unit cell, indicated by the gray boxes in \cref{fig:phasediags}.
	Examining the four limiting cases \cref{eq:GS0}-\cref{eq:GSmpiby2}, we see that introducing a boundary does not affect the fixed-point family ground states except along the $\delta=1$ line. Here, the two spins at the boundary of the ladder decouple from the bulk and are governed by the effective Hamiltonian
	\begin{equation}
		H_\partial(J_M, J_U) = J_U\left(S^x_1 S^x_2 + S^y_1 S^y_2 \right) + J_M S^z_1 S^z_2.\label{eq:Hboundary}
	\end{equation}
	The ground state of \cref{eq:Hboundary} is the $\ztwo^{L,R}$-even state $\kett[\Big]{\dimy}$  for  $J_U< -|J_M|$,  and  the $\ztwo^{L, R}$-odd state $\kett[\Big]{\dimyblue}$ for  $J_U> |J_M|$. Since the bulk is a product of symmetric plaquettes, as we cycle through the fixed-point family, the ground state of the semi-infinite ladder must undergo a level crossing somewhere along the $\delta=1$, $J_U \in (-|J_M|, |J_M|)$ line  between distinct sectors of either $\ztwo^L$  or $O(2)$ (the latter follows from the change in $\ztwo^R$ eigenvalue). This is a purely boundary phenomenon, absent in the periodic case, and captures the fact that the cycle transfers symmetry charge across the system: the family is a nontrivial pump of both $\ztwo^L$ and $\ztwo^R$.  \footnote{The boundary charge returns to its original value at the end of a cycle without a second boundary level crossing, as charge is pumped through the bulk. Since this is impossible in a purely $d=0$ system, the boundary {family} is ``anomalous''~\cite{Seiberg_AnomaliesCouplingOne10.21468/SciPostPhys.8.1.001,Seiberg_AnomaliesCouplingTwo10.21468/SciPostPhys.8.1.002,ThorngrenDiabolicalPhysRevB.102.245113}.} {Althought the non-trivial pump only requires preserving either $\ztwo^L$ or $\ztwo^R$, the larger symmetry group is needed to stabilize the unnecessary critical line~\cite{APUCPhysRevLett.130.256401,SupMat}}.
	
	There are two possible ways in which the boundary charge can change along $\delta=1$. For $J_M>0$, we get a single direct level crossing (a boundary phase transition) at $J_U$ between the two boundary charge sectors described above, at which the ground state is twofold degenerate. This can be rationalised in terms of either symmetry. However, the non-Abelian nature of $O(2)$ admits a second possibility for $J_M<0$. In this case, there is a region of finite extent $J_U< |J_M|$  (a `boundary phase') where the ground state of \cref{eq:Hboundary} transforms in the $O(2)$ doublet irrep $\kett[\Big]{\dimydouble}$, i.e. there is a twofold degenerate boundary mode in the open system.  The semi-infinite ladder thus undergoes two distinct level-crossings at $J_U = \pm |J_M|$, at which its ground state is threefold degenerate. If we now tune parameters ``radially inwards'' towards the critical region, the symmetry argument above continues to hold even though the model is no  longer exactly solvable~\cite{SupMat}.

	\textit{Field theories near criticality.---} We now explore how the charge pump emerges near the ``unneccessary critical'' region, which admits a bosonized description~\cite{Giamarchi,SupMat}. The resulting theory is a two-component `Luttinger liquid', $H= H[\theta_1, \phi_1] + H[\theta_2, \phi_2] + H_{g}$, with 
	\begin{align}
		H[\theta, \phi] &= \frac{v}{2 \pi} \int dx \left(\frac{1}{4K} \left(\partial_x \phi\right)^2 + K \left(\partial_x \theta\right)^2\right),  \label{eq:compactboson}\\
		\text{and~} H_g& = \int dx\left( \frac{J_M}{4\pi^2}  \partial_x \phi_1 \partial_x\phi_2 +  \sum_{\mathcal{O}} g_{\mathcal{O}}  \mathcal{O}(x) \right). \label{eq:Hg}
	\end{align}
	$H_g$ includes an exactly marginal  operator with  coupling $\propto J_M$ and various operators of the form $\mathcal{O} \propto \cos(\Theta)$ where $\Theta$ is some combination of fields $\phi_\alpha, \theta_\alpha$. 
	To eliminate spurious symmetries relative to \cref{eq:H} requires  several such $\mathcal{O}$s~\cite{APUCPhysRevLett.130.256401,SupMat}, but the bulk phases in \cref{fig:phasediags} can be understood in terms of the RG flows of just four: $\mathcal{U} \equiv \sum_{\alpha}\cos(\phi_\alpha)$, $\mathcal{V}_\pm \equiv \cos (\phi_1\pm \phi_2)$, and $\mathcal{W}_-  \equiv \cos(\theta_1-\theta_2)$, with bare couplings $g_{\cU} \propto \delta$, $g_{\cV_{\pm}}\propto \mp J_M$, and $g_{\cW_-} \propto J_U$.   
	The scaling dimensions of these operators (denoted $[\cdots]$) are related: $\SD{\mathcal{W}_-} = 1/\SD{\mathcal{V}_-}$, and $\SD{\mathcal{U}} = (\SD{\mathcal{V}_+}+\SD{\mathcal{V}_-})/4$, and depend on {microscopic couplings}~\cite{SupMat}.   
    As $J_M$ is tuned, modifying $\SD{\mathcal{V}_\pm}$, different sets of operators become relevant (i.e., $[\mathcal{O}]<2$) and flow to strong coupling, pinning different field combinations with or without opening a gap.  
	Each gapless case is described by 
	a single-component Luttinger liquid $H[\theta, \phi]$, but the  long-wavelength  degrees of freedom, and hence the action of symmetries on $(\theta, \phi)$, differs depending on the combination of operators `left behind' as others flow to strong coupling~\cite{APUCPhysRevLett.130.256401,SupMat}. This ensures that they cannot be connected without changing  universality class or encountering an intermediate phase~\cite{VerresenThorngrenSECPhysRevX.11.041059,APSECTLLPhysRevB.108.245135}. Of these, XY$_{2}$ describes the unnecessary critical line in \cref{fig:phasediags}(a) and XY$_{3}$ the unnecessary multicritical point in \cref{fig:phasediags}(b), respectively reached by tuning one or two parameters, whereas the phases  XY$_1$ and XY$_1^*$  require no fine-tuning.  
	
	The trivial gapped phase emerges when both sectors of the Luttinger liquid are gapped  while preserving symmetry. This can happen in four distinct ways (T$_{1,\ldots,4}$; \cref{tab:vacuumsummary}); while these `vacua'  represent distinct basins of attraction under RG, they can all be adiabatically connected without any bulk phase transitions~\cite{SupMat,LECHEMINANT_SelfDualSineGordon_2002502}.  	
	\begin{table}[]
		\resizebox{\columnwidth}{!}{%
			\begin{tabular}{|c|c|c|c|}
				\hline
				\textbf{Vacuum} & \textbf{Scaling Dimensions}                                                                      & \textbf{Parameters} & \textbf{Pinned Fields}                                                         \\ \hline
				T$_1$           & $[\mathcal{U}] < [\mathcal{W}_-]$                                                                 & $\delta>0$          & $\langle\phi_1 \rangle = \langle \phi_2\rangle = \pi$                          \\ \hline
				T$_2$           & $[\mathcal{U}] > [\mathcal{W}_-]$, $[\mathcal{V}_+]<2$                                            & $J_U<0$             & $\langle\phi_1 +\phi_2 \rangle = 0$, $\langle \theta_1 - \theta_2\rangle =0$   \\ \hline
				T$_3$           & $[\mathcal{U}] < [\mathcal{W}_-]$                                                                 & $\delta<0$          & $\langle\phi_1 \rangle = \langle \phi_2\rangle = 0$                            \\ \hline
				T$_4$           & $[\mathcal{U}] > [\mathcal{W}_-]$, $[\mathcal{V}_+]<2$ & $J_U>0$             & $\langle\phi_1 +\phi_2 \rangle = 0$, $\langle \theta_1 - \theta_2\rangle =\pi$ \\ \hline
			\end{tabular}%
		}
		\caption{\label{tab:vacuumsummary}\textbf{Distinct  Luttinger liquid vacua corresponding to the same bulk gapped phase.} Depending on the operator  scaling dimensions and parameter choices, the flow to strong coupling pins different combinations of fields.}
	\end{table}

	However, the  boundary physics changes nontrivially as the gapped family evolves through the T$_1$ vacuum where the pinned fields take values $\langle \phi_\alpha\rangle = \pi$. To see this, we first observe that we can model a semi-infinite chain in  the T$_1$ vacuum for $x>0$ with open boundary conditions at $x=0$ as an infinite system where $g_{\cU} \propto \delta$ is smoothly modulated in space with $g_{\cU} (x\to\pm\infty) \lessgtr 0$,  such that the system lies  in the T$_3$ vacuum as $x\to -\infty$ and in the T$_1$ vacuum as $x\to \infty$~\cite{KeselmanBerg_PhysRevB.91.235309}   \footnote{This follows from the fact that  open boundary conditions on each XXZ leg corresponds to  $\phi_\alpha(x=0,t) = 0$ in a semi-infinite chain~\cite{EggertAffleck_boundaryXXZBosonizationPhysRevB.46.10866,Affleck1998edgeXXZ}.}.  This pins $\langle \phi_\alpha\rangle$ to their  asymptotic T$_{1,3}$ vacuum values of $0,\pi$ as $x\to \mp \infty$, but they will deviate from these values near the interface where $g_{\cU}$ vanishes. The interpolation between the vacua is controlled by the competition of the remaining cosines $\cV_+$, $\cV_-$ and $\cW_-$ near the interface. 
	
\begin{figure}
		\includegraphics[width=\columnwidth]{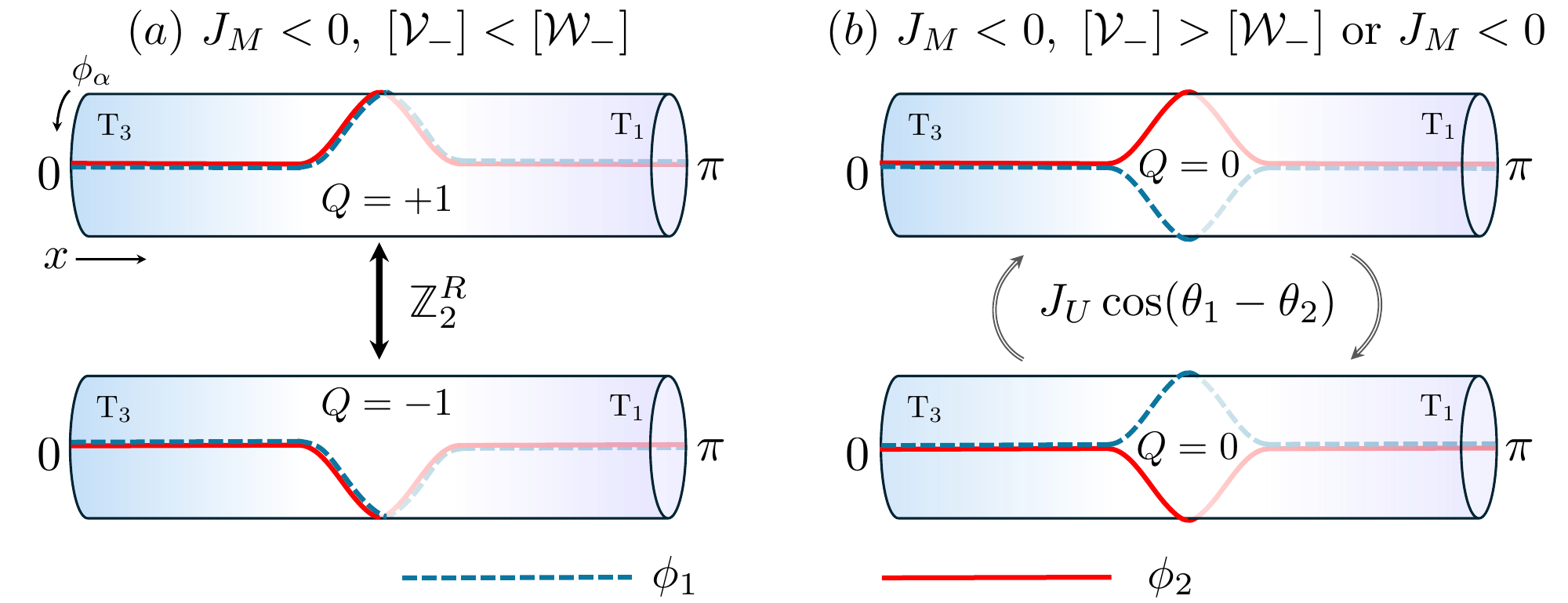}
		\caption{\label{fig:winding}\textbf{Interpolations at the T$_3$-T$_1$ interface.} For $J_M<0$ and $[\cV_-]<[\cW_-]$, $\phi_{1,2}$ wind in the same sense, so $\phi_1+\phi_2$ evolves smoothly from $0$ to $\pm2\pi$; this binds $U(1)$ charge $Q=\pm1$ to the interface, which is hence an $O(2)$ doublet. For  $J_M<0$ and $[\cV_-]>[\cW_-]$ or $J_M>0$,  $\phi_{1,2}$ wind oppositely; tunneling between the resulting $Q=0$ configurations via $J_U\cos(\theta_1-\theta_2)$ leads to a unique ground state whose $\ztwo^L$ charge is $-\text{sign}(J_U)$.}
	\end{figure}

	For $J_M<0$, there are two possibilities. When $\cV_- = \cos(\phi_1-\phi_2)$ dominates, since $g_{\cV_-}\propto J_M<0$, configurations where $\phi_1 = \phi_2$ (\cref{fig:winding}(a)) are favored, so that $\phi_1 + \phi_2$ evolves smoothly from $0$ to $\pm 2\pi$. Although the two vacua are invariant under all symmetries, the interpolations are exchanged by $\ztwo^R$ and involve a $U(1)$ charge $Q = \frac{1}{2\pi} \int_{-\infty}^{\infty} dx \partial_x(\phi_1 + \phi_2) = \pm 1$ localized near the interface.  We conclude that the interface transforms as a two-dimensional irrep of $O(2)$.   \footnote{Recall that  1d $O(2)$ irreps are $U(1)$ neutral, whereas 2d irreps consist of charge $\pm q$ states which are exchanged by $\ztwo^R$~\cite{GroupTheoryTung1985,SupMat}.} Since irreps are stable, we expect a finite window (a `phase') of boundary degeneracy, consistent with the physics for  $\delta\lesssim1$ and small $J_U$ in \cref{fig:phasediags}(a). In contrast, when $\cW_- = \cos(\theta_1-\theta_2)$ dominates,  $\theta_1-\theta_2$ is pinned to $0$ or $\pi$ depending on the sign of $g_{\cW_-}$. This forces its conjugate $\phi_1 - \phi_2$ to fluctuate at the interface, but allows  $\phi_1+\phi_2$ to have a definite value; although this could apparently be pinned to either $0$ or $\pi$ depending on the sign of $g_{\cV_+}$, the latter is frustrated by the asymptotic vacua~\cite{SupMat}. Since for $\phi_1+\phi_2=0$, $\phi_{1,2}$ wind in opposite senses, we conclude that the system tunnels back and forth between  $U(1)$-neutral configurations where $(\phi_1, \phi_2) = (\pm\pi/2, \mp\pi/2)$ near the interface (\cref{fig:winding}(b)). For $\theta_1-\theta_2=0$ ($g_{\cW_-} \propto J_U<0$), this leads to a unique $\ztwo^L$-even ground state, whereas for $\theta_1-\theta_2=\pi$ ($g_{\cW_-} \propto J_U>0$), it leads to a unique $\ztwo^L$-odd ground state, consistent with \cref{fig:phasediags}(a)  for $\delta \lesssim 1$ and large $|J_U|$. 
	
	The two possibilities above meet in a threefold degenerate crossing, since the two states in the $O(2)$ doublet and the $\ztwo^L$-even/odd  singlet are distinct on symmetry grounds and hence do not experience level repulsion. The crossings involving the even and odd singlets  respectively corresponds to the  line of boundary transitions in \cref{fig:phasediags}(a) for $J_U<0$ and $J_U>0$, which terminate at the bulk multicritical points (between the XY$_2$ line and  the XY$_{1}$/XY$_{1}^*$ phases resulting from the competition between the same operators $\cW_-,~\cV_-$ in the bulk). These are described by the central charge $c=3/2$ conformal field theories (CFTs) that the  two-component LL theory flows to when $[\cW_-]=[\cV_-],~[\cV_+]>2$ and $g_{\cU} = 0$~\cite{LECHEMINANT_SelfDualSineGordon_2002502,Schulz_Higherspinbosinization_PhysRevB.34.6372,SupMat,XueJia_C1p5_PhysRevA.109.062226}.

	For $J_M>0$ {and parameters consistent with \cref{fig:phasediags}(b)}, it is possible to show that  {$\cW_-$ dominates ($[\cW_-]<[\cV_-]$ near the multicritical point)} always, so generically we only have the first possibility above: the boundary is  $\ztwo^L$-even for $g_{\cW_-} \propto J_U<0$, and  $\ztwo^L$-odd for $g_{\cW_-} \propto J_U>0$. Evidently, tuning $J_U$ to zero leads to  a twofold degenerate boundary level crossing where the boundary $\ztwo^L$ charge changes; this will terminate at the XY$_3$ multicritical point as in \cref{fig:phasediags}(b).
	
	In the crossover region for large $J_U<0$ ($J_U>0$), $\theta_1-\theta_2$ is pinned at the edge of the T$_1$ vacuum to the same value as in the {\it bulk} of the adjacent T$_2$ (T$_4$) vacuum. Thus, by changing $\delta$ from $+1$ to $-1$ within T$_{2,4}$,  $\ztwo^L$ charge can be transferred continuously from the boundary into  the bulk. This allows us to smoothly go between T$_1$ and T$_3$, completing the pump cycle, and underscoring the essential role of the T$_{2,4}$ vacua. We have thus reproduced the essential features of the pump entirely within the field theory, as promised.

	\textit{Discussion.---} In this work, we have explicitly demonstrated a link between unnecessary criticality in the interior of a phase diagram and a nontrivial `pump' property of the family of gapped Hamiltonians that encircles the critical region, in the setting of a concrete one-dimensional model. Our exact microscopic arguments far from criticality were mirrored by  field theory calculations proximate to it. In both cases, we demonstrated the pumping of a $\ztwo$ symmetry charge by examining the changes at the boundaries of an open system. In the latter case, our analysis related these properties to the presence of multiple RG fixed points corresponding to the same bulk gapped phase --- a principle with possible utility beyond the specific questions studied here. These arguments naturally demonstrate a `bulk boundary correspondence' that argues that terminii of unnecessary critical surfaces source lines of boundary transitions in any gapped family of Hamiltonians that encircles them. 
	
	An outstanding puzzle is to identify a topological invariant that characterizes the gapped family in the bulk, without resorting to a boundary analysis, similar to the classification of Thouless pumps of $U(1)$ charge. While this is  a difficult problem in general, the exactly solvable limit may be especially amenable to matrix-product state methods~\cite{Ashvinsommer2024higherberrycurvaturewaveMPS,RyuHigherMPSPhysRevB.109.115152,RyuohyamaHigheMPS2024higherberryconnectionmatrix,Ashvinsommer2024higherberrycurvaturewavePEPS,RyuohyamaHigherPEPS2024higherberryphaseprojected} or the techniques of Ref.~\cite{HermeleGappedfamiliesPhysRevB.108.125147}.

 What broader lessons might our $d=1$ example teach us? Most obviously, we see that the presence of a nontrivial pump (or its higher-dimensional generalizations) at the periphery of the parameter space is at least a {\it sufficient} condition for a gapless point in the interior --- indeed, this is implicit in the notion that such a nontrivial gapped family defines a noncontractible loop in parameter space. However the presence of unnecessary criticality --- as opposed to  {\it multi}criticality --- cannot be deduced quite so readily. Here,  we show the bulk-boundary correspondence to plays a key role in one dimension: while in its basic incarnation it is a corollary of the pump property, our work sharpens the link between the boundary and bulk phase structures. To wit, the presence of a boundary {\it critical} point is matched to bulk unnecessary {\it multicriticality}, and a boundary {\it phase} to bulk {\it criticality}~\cite{SupMat}. We conjecture that this is a feature also in $d>1$, but defer its exploration to future work.

	\medskip 
	\noindent\textit{Acknowledgments.---}We thank Ryan Thorngren, Nick Jones, Senthil Todadri, Paul Fendley, Fabian Essler, Yuchi He, Gurkirat Singh, Chris Hooley, and Michele Fava for helpful discussions and correspondence, and Michele Fava for collaboration on related work~\cite{APUCPhysRevLett.130.256401}. We acknowledge support from the  European Research Council under the European Union Horizon 2020 Research and Innovation Programme, Grant Agreement No. 804213-TMCS, and from a Gutzwiller Fellowship at the Max Planck Institute for Complex Systems, where much of this work was completed (SAP).	
	

\onecolumngrid

  \renewcommand\appendixpagename{\centering \Large\uline{ Supplemental materials}}
  \newpage
  
\begin{appendices}
\renewcommand{\thesection}{\Roman{section}}
\section{Details of the exactly solvable family of Hamiltonians}
\subsection{Periodic boundary conditions}
\begin{figure}[!h]
	\centering
	\includegraphics[height=5cm]{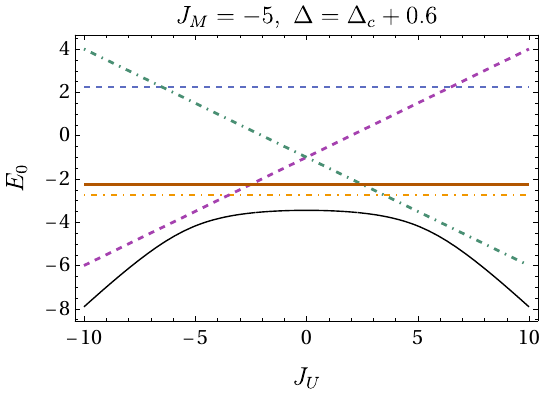}
	\includegraphics[height=5cm]{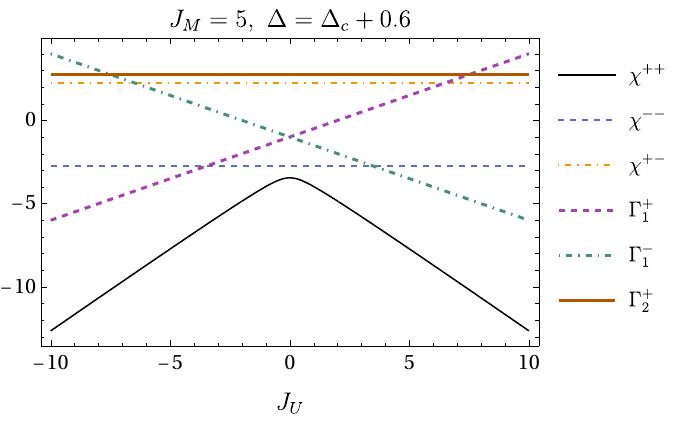}
	\includegraphics[height=5cm]{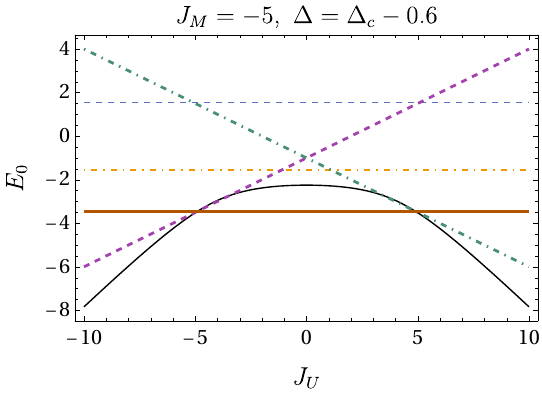}
	\includegraphics[height=5cm]{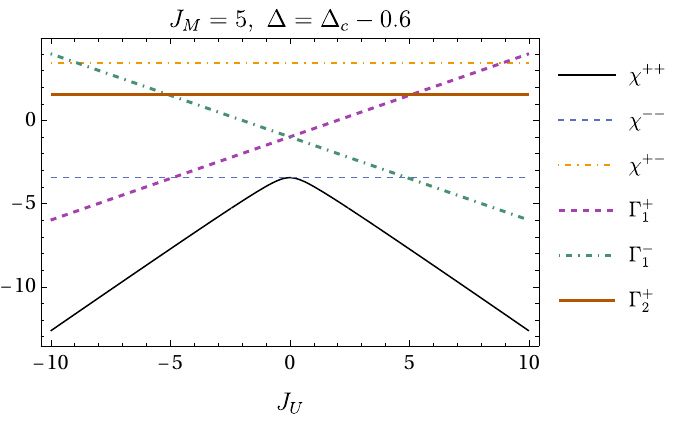}
	\caption{Minimum eigenvalue of each symmetry sector of the four qubit Hamiltonian in \cref{appeq:H_fourqubit} for $\Delta > \Delta_c$ (top row) which shows no level crossing and $\Delta < \Delta_c$ (bottom row) which does.  $\Delta_c \approx -0.35$ for $J_M = -5$, is the critical value of $\Delta$,defined in \cref{appeq:Deltacrit} above which the ground state is unique.\label{appfig:4qubit levels}}
	
\end{figure}
We provide more details on analyzing the Hamiltonian considered in the main text, 
\begin{equation}
	H = \sum_{j=1}^L  \left[\sum_{\alpha=1,2} (1+ (-1)^j \delta) ( S^x_{\alpha j} S^x_{\alpha j+1}  +  S^y_{\alpha j} S^y_{\alpha j+1} + \Delta  S^z_{\alpha j} S^z_{\alpha j+1}) +
	J_MS^z_{1 j} S^z_{2 j} + J_U \left(S^x_{1 j} S^x_{2 j}+S^y_{1 j} S^y_{2 j} \right) \right]. 
	\label{appeq:H}
\end{equation}
In the exactly solvable limit, when $|J_U| \rightarrow \infty$ or $|\delta| = 1$. For periodic boundary conditions, the system reduces to a disjoint collection of quantum mechanical systems of four qubits with the Hamiltonian
\begin{equation}
	H_{j,j+1} = \sum_{\alpha=1,2} 2 ( S^x_{\alpha j} S^x_{\alpha j+1}  +  S^y_{\alpha j} S^y_{\alpha j+1} + \Delta  S^z_{\alpha j} S^z_{\alpha j+1}) +\sum_{\beta=j,j+1}
	J_MS^z_{1 \beta} S^z_{2 \beta} + J_U \left(S^x_{1 \beta} S^x_{2 \beta}+S^y_{1 \beta} S^y_{2 \beta} \right).  \label{appeq:H_fourqubit}
\end{equation}
For $\delta>0$, $j$ is even whereas for $\delta<0$, $j$ is odd. Let us first derive the condition when there is no level crossing and \cref{appeq:H_fourqubit} has a unique ground state. To do this, observe that for $|J_U|\rightarrow \infty$, the ground state is always unique for any value of $\Delta,~J_M$,
\begin{equation}
	\ket{\text{GS}(\infty,\delta)} = \kett[\Big]{\plaqyblue}  , ~\ket{\text{GS}(-\infty,\delta)} =\kett[\Big]{\plaqy} \text{ with }
	\kett[\Big]{\dimy}= \frac{1}{\sqrt{2}} \left(\biggl|\begin{tabular}{c}
		$\uparrow$    \\
		$\downarrow$   
	\end{tabular}\biggr> + \biggl|\begin{tabular}{c}
		$\downarrow$    \\
		$\uparrow$   
	\end{tabular}\biggr>  \right), ~    \kett[\Big]{\dimyblue} = \frac{1}{\sqrt{2}} \left(\biggl|\begin{tabular}{c}
		$\uparrow$    \\
		$\downarrow$   
	\end{tabular}\biggr> - \biggl|\begin{tabular}{c}
		$\downarrow$    \\
		$\uparrow$   
	\end{tabular}\biggr>  \right). \label{appeq:boundary_Bell}
\end{equation}
If there were to be a level crossing, it would happen for small values of $J_U$. The condition for avoided level crossing is obtained at $J_U = 0$ where the eigenvalues of \cref{appeq:H_fourqubit} can easily be determined :
\begin{equation}
	\centering
	\begin{tabular}{c| c}
		\hline 
		E     & degeneracy  \\
		\hline 
		\hline
		$ \left(-2\Delta \pm \sqrt{16 + J_M^2} \right)/2$ & 1\\
		$ \left(\pm J_M - 2 \Delta \right)/2 $ & 1\\
		$\left(\pm J_M + 2 \Delta \right)/2$  & 2\\
		$\pm1$     & 4\\
		\hline 
	\end{tabular}
\end{equation}
Comparing the various eigenvalues, we see that $ \left(-2\Delta - \sqrt{16 + J_M^2} \right)/2$ is the unique ground state so long as
\begin{equation}
	\Delta> \Delta_c =  \frac{1}{4} \left(|J_M| -\sqrt{J_M^2+16} \right) \label{appeq:Deltacrit}
\end{equation}
as stated in the main text.  For $J_U \neq 0$, the Hamiltonian in \cref{appeq:H_fourqubit} can be block diagonalized into six different symmetry sectors. \cref{appfig:4qubit levels} shows the lowest eigenvalue in each of these. We see that level crossing is avoided for $\Delta > \Delta_c$ and is present otherwise.

We now provide some details on how the eigenvalues plotted in \cref{appfig:4qubit levels} were determined. Let us begin with a few comments about the irreducible representations (irreps) of the symmetry group $(U(1)\rtimes \ztwo^R) \times \ztwo^L \cong O(2)\times \ztwo^L$. There are four one-dimensional irreps $\chi^{++},\chi^{+-},\chi^{-+},\chi^{--}$, carrying no $U(1)$ charge but which transform as one of four non-trivial irreps of $\ztwo^R \times \ztwo^L$ and two distinct two-dimensional irreps, $\Gamma^{\pm}_q$ for each positive integer $q = 1,2,\ldots$ formed by a doublet of states with $U(1)$ charge $\pm q$ which are exchanged under $\ztwo^R$. The $\pm$ superscript denotes the charge under $\ztwo^L$ action by the irrep. These are summarized below 
\begin{equation}
	\centering
	\begin{tabular}{l|ccc}
		\hline
		Irrep  & $U(1)$ & $\ztwo^R$ & $\ztwo^L$   \\
		\hline
		\hline
		$\chi^{++}$  & +1 & +1 & +1 \\
		$\chi^{+-}$  & +1 & +1 & -1 \\
		$\chi^{-+}$  & +1 & -1 & +1 \\
		$\chi^{--}$  & +1 & -1 & -1 \\
		$\Gamma^+_q$ & $\begin{pmatrix}
			e^{iq\theta} & 0 \\
			0 & e^{-iq\theta}
		\end{pmatrix}$ & $\begin{pmatrix}
			0 & 1\\
			1 & 0
		\end{pmatrix}$ & $\begin{pmatrix}
			1 & 0 \\
			0 & 1
		\end{pmatrix}$\\
		$\Gamma^-_q$ & $\begin{pmatrix}
			e^{iq\theta} & 0 \\
			0 & e^{-iq\theta}
		\end{pmatrix}$ & $\begin{pmatrix}
			0 & 1\\
			1 & 0
		\end{pmatrix}$ & $-\begin{pmatrix}
			1 & 0 \\
			0 & 1
		\end{pmatrix}$\\
		\hline
	\end{tabular}
	\label{apptab:irrep}
\end{equation}

The Hilbert space of a single qubit transforms as a \emph{projective} representation of $O(2) \times \ztwo^L$. However, an even number of qubits can be decomposed into the linear irreps listed in \cref{apptab:irrep}. In particular, the 16 dimensional four-qubit Hilbert space of \cref{appeq:H_fourqubit} can be decomposed into the irreps shown in \cref{apptab:irrep} as 
\begin{equation}
	16 = 3 \chi^{++} \oplus 2  \chi^{--} \oplus \chi{-+} \oplus 2 \Gamma^{+}_1 \oplus 2 \Gamma^{-}_1 \oplus \Gamma^{+}_2. \label{appeq:16 irrep decomposition}
\end{equation}
\cref{appeq:H_fourqubit} can be block diagonalized into each of these sectors where matrix elements connect states within the degeneracy subspace of each irrep. The basis of states that achieve this block diagonalization is shown below along with the irrep and effective block Hamiltonian, $H_{\text{eff}}$ acting on the degeneracy subspace. 
\begin{equation}
	\begin{tabular}{|c|c|c|}
		\hline
		State(s) & Irrep & $H_{\text{eff}}$ \\
		\hline
		\hline
		$ \kett[\Big]{\plaqy}$  & \multirow{3}{*}{$\chi^{++}$}  & \multirow{3}{*}{$\begin{pmatrix}
				-\frac{J_M}{2}-J_U & \Delta  & -\sqrt{2} \\
				\Delta  & J_U-\frac{J_M}{2} & \sqrt{2} \\
				-\sqrt{2} & \sqrt{2} & \frac{1}{2} (J_M-2 \Delta ) \\
			\end{pmatrix}$} \\
		$ \kett[\Big]{\plaqyblue}$  &     &\\
		$\frac{1}{\sqrt{2}}\left( \biggl|\begin{tabular}{cc}
			$\uparrow$    & $\downarrow$  \\
			$\uparrow$   & $\downarrow$ 
		\end{tabular}\biggr> + \biggl|\begin{tabular}{cc}
			$\downarrow$    & $\uparrow$  \\
			$\downarrow$   & $\uparrow$ 
		\end{tabular}\biggr>\right)$  & & \\
		\hline
		\hline
		$ \kett[\Big]{\plaqypm}$ & \multirow{2}{*}{$\chi^{--}$}  & \multirow{2}{*}{$\begin{pmatrix}
				-\frac{J_M}{2} & \Delta  \\
				\Delta  & -\frac{J_M}{2} \\
			\end{pmatrix}$} \\
		$ \kett[\Big]{\plaqymp}$ &   &  \\
		\hline
		\hline   
		$\frac{1}{\sqrt{2}}\left( \biggl|\begin{tabular}{cc}
			$\uparrow$    & $\downarrow$  \\
			$\uparrow$   & $\downarrow$ 
		\end{tabular}\biggr> - \biggl|\begin{tabular}{cc}
			$\downarrow$    & $\uparrow$  \\
			$\downarrow$   & $\uparrow$ 
		\end{tabular}\biggr>\right)$  & $\chi^{-+}$ & $\frac{J_M}{2} - \Delta$ \\
		\hline
	\end{tabular}
	\hspace{1em}
	\begin{tabular}{|c|c|c|}
		\hline
		State(s) & Irrep & $H_{\text{eff}}$ \\
		\hline
		\hline
		$ \kett[\Big]{\plaqypd}$ & \multirow{2}{*}{$\Gamma^+_1$}  & \multirow{2}{*}{$\begin{pmatrix}
				\frac{J_U}{2} & 1  \\
				1  & \frac{J_U}{2} \\
			\end{pmatrix}$} \\
		$ \kett[\Big]{\plaqydp}$ &   &  \\
		\hline
		\hline   
		$ \kett[\Big]{\plaqymd}$ & \multirow{2}{*}{$\Gamma^-_1$}  & \multirow{2}{*}{$\begin{pmatrix}
				-\frac{J_U}{2} & 1  \\
				1  & -\frac{J_U}{2} \\
			\end{pmatrix}$} \\
		$ \kett[\Big]{\plaqydm}$ &   &  \\
		\hline
		\hline   
		$\biggl|\begin{tabular}{cc}
			$\uparrow$    & $\uparrow$  \\
			$\uparrow$   & $\uparrow$ 
		\end{tabular}\biggr>,\biggl|\begin{tabular}{cc}
			$\downarrow$    & $\downarrow$  \\
			$\downarrow$   & $\downarrow$ 
		\end{tabular}\biggr>$  & $\Gamma^+_2$ & $\frac{J_M}{2} + \Delta$ \\
		\hline
	\end{tabular}
	\label{apptab:decomposition 4 qubit}
\end{equation}
Diagonalizing these blocks allows us to determine the eigenvalues shown in \cref{appfig:4qubit levels}. In \cref{apptab:decomposition 4 qubit}, we have used the definition of Bell states in \cref{appeq:boundary_Bell} and also $ \kett[\Big]{\dimydouble} = \biggl|\begin{tabular}{c}
	$\uparrow$    \\
	$\uparrow$   
\end{tabular}\biggr>,~\biggl|\begin{tabular}{c}
	$\downarrow$    \\
	$\downarrow$   
\end{tabular}\biggr> $.

\subsection{Open boundary conditions}
\begin{figure}[!ht]
	\centering
	\includegraphics[height=4.5cm]{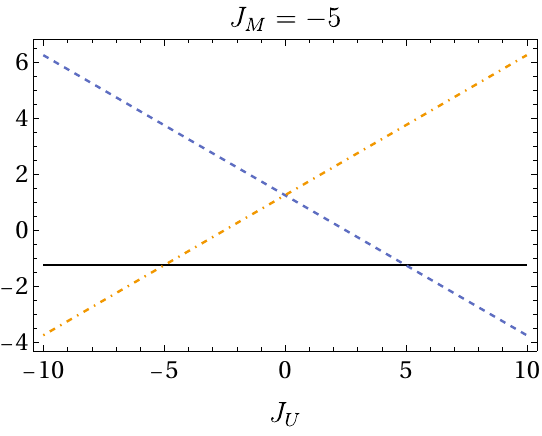}
	\includegraphics[height=4.5cm]{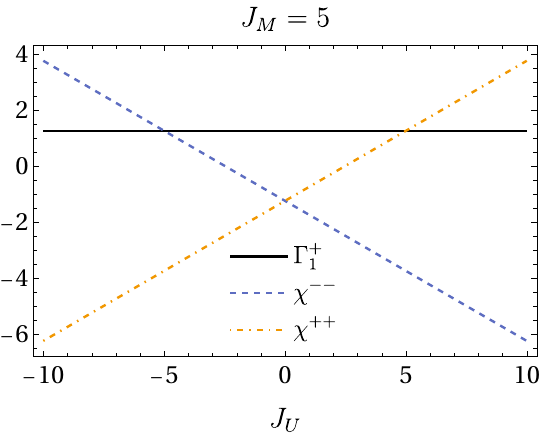}
	\includegraphics[height=4.7cm]{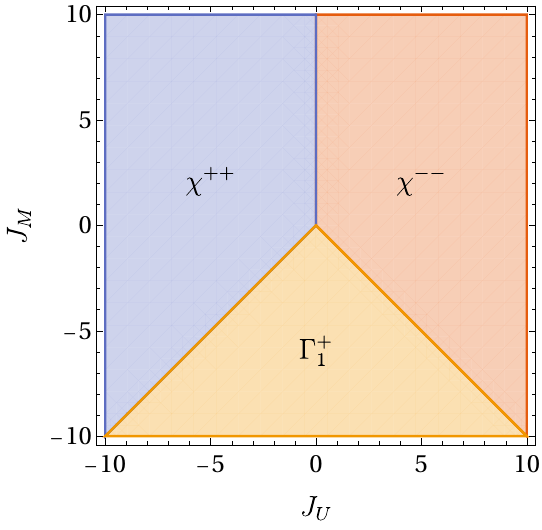}  
	\caption{Left, Middle: Plots of eigenvalues of the two-qubit boundary Hamiltonian in \cref{appeq:Hboundary}. Right: Graphical summary of the ground state irrep in various parameter regimes.}
	\label{appfig:2qubit levels}
\end{figure}
The effective two-qubit boundary Hamiltonian along $\delta = 1$ with open boundary conditions is 
\begin{equation}
	H_\partial(J_M, J_U) = J_U\left(S^x_1 S^x_2 + S^y_1 S^y_2 \right) + J_M S^z_1 S^z_2.\label{appeq:Hboundary}
\end{equation}
The four-dimensional Hilbert space can be decomposed into $O(2) \times \ztwo^R$ irreps as $2 = \chi^{++} \oplus \chi^{--} \oplus \Gamma^+_1$. Since none of the irreps has any degeneracies, \cref{appeq:Hboundary} is immediately diagonalized by going to the irrep basis, 
\begin{equation}
	\ket{\chi^{++}}=\kett[\Big]{\dimy},~\ket{\chi^{--}}=\kett[\Big]{\dimyblue},~\ket{\Gamma^+_1}=\kett[\Big]{\dimydouble},~~~E[\chi^{\pm \pm}] = -\frac{1}{4} \left(J_M \pm J_U \right),~E[\Gamma^+_1] = \frac{J_M}{4}. \label{appeq:boundary_eigenvalues}
\end{equation}
Which of these states has the lowest energy depends on the parameters. As seen in \cref{appfig:2qubit levels}, ground state is unique for $|J_U| > |J_M|$ and corresponds to $\ket{\chi^{\pm \pm}}$ for sign$(J_U) = \mp 1$ and undergoes a level crossing at $J_U=0$ for $J_M>0$. For $J_M<0$, the ground state is  $\ket{\Gamma^+_1}$ for $|J_U| < |J_M|$ and we have a level crossing with three-fold GSD at $|J_U| = |J_M|$.

\section{Stability of the boundary phase diagram away from the exactly solvable periphery}
In the main text, the boundary phase diagram was determined on the periphery of the phase diagram i.e. when $\delta = \pm 1$ or $J_U = \pm \infty$ where the system reduces to decoupled quantum mechanical systems. The nature of the phase diagram is preserved tuning `radially inwards', away from the exactly solvable periphery. To see this, note that once $\delta<1$  we introduce identical  coupling terms $V_\delta \propto 1-\delta$ between the  boundary spin pair and the bulk, and between adjacent plaquettes. Since the only nontrivial symmetry charge for $\delta=1$ is carried by the boundary spins, and $V_\delta$ (whose form can be deduced from \cref{appeq:H}) preserves all symmetries,  we can use perturbation theory to construct a new boundary charge operator  of the form $\mathcal{O}_\partial \propto \sum_r \mathcal{O}_r e^{-r/\xi}$, with $\xi = c_1/\ln[{(1-\delta)}/{\Delta}]$, where $\mathcal{O}_r$ acts only on unit cell $r$,  $\Delta$ is the bulk gap and $c_{1}$ is an $O(1)$ constant. $\langle \mathcal{O}_\partial\rangle$ now serves as the symmetry diagnostic for the level-crossings. Hence we expect the boundary critical point/phase to persist as long as the bulk gap remains open, forcing the charge to be localized near the boundary.

\section{Bosonization I: Characterizing the trivial phase}

\subsection{Bosonization preliminaries}

In this and following sections, we provide more details of how the phase diagrams in the main text, reproduced in \cref{appfig:gapless}, were obtained using bosonization. To start with, we consider the decoupled limit $J_U, J_M \rightarrow 0$ when the Hamiltonian reduces to two independent, indentical copies of XXZ spin chains. We bosonize each copy using standard techniques~\cite{Giamarchi,Haldane_LuttingerLiquid_1981} and then reintroduce $J_U,~J_M$ as perturbations to get a parent two-component Luttinger liquid with the following Hamiltonian
\begin{multline}
	H  \approx \frac{v}{2 \pi} \int dx \sum_{\alpha = 1}^{2} \left(\frac{1}{4K} \left(\partial_x \phi_\alpha\right)^2 + K \left(\partial_x \theta_\alpha\right)^2\right) +    \frac{J_M}{4 \pi^2} \int dx ~ \partial_x \phi_1 \partial_x \phi_2 + \frac{\mathcal{B}^2 J_M}{2}   \int dx\left( \cos(\phi_1 - \phi_2) - \cos(\phi_1 + \phi_2) \right) \\+2\mathcal{AC} \delta  \int dx~    \sum_{\alpha = 1}^{2}\cos \phi_\alpha +2 \mathcal{A}^2 J_U \int dx  \cos \left(\theta_1 - \theta_2\right) 
	+ \mathcal{C}^2 J_U \int dx \cos \left(\theta_1 - \theta_2\right) \cos \left(\phi_1 + \phi_2\right)  + \ldots. \label{appeq:H_bosonize}
\end{multline}

In the main text, Tthe above was written in a concise form as $H= H[\theta_1, \phi_1] + H[\theta_2, \phi_2] + H_{g}$, with 
\begin{align}
	H[\theta, \phi] &= \frac{v}{2 \pi} \int dx \left(\frac{1}{4K} \left(\partial_x \phi\right)^2 + K \left(\partial_x \theta\right)^2\right)  
	\text{ and~} H_g = \int dx\left( \frac{J_M}{4\pi^2}  \partial_x \phi_1 \partial_x\phi_2 +  \sum_{\mathcal{O}} g_{\mathcal{O}}  \mathcal{O}(x) \right). \label{appeq:Hg}
\end{align}
where $g_{\cU} = {2\mathcal{AC}} \delta$, $g_{\cV_{\pm}}= \mp \mathcal{B}^2 J_M$, and $g_{\cW_-} =2\mathcal{A}^2 J_U$. $\phi_\alpha \cong \phi_\alpha + 2 \pi $ and $\theta_\alpha \cong \theta_\alpha + 2 \pi $ are canonically conjugate compact boson fields with unit radius satisfying
\begin{equation}
	[\partial_x \phi_\alpha (x), \theta_\beta (x') ] = 2 \pi i \delta_{\alpha\beta} \delta(x-x'),
	\label{appeq:KacMoody_2component}
\end{equation}
The bosonized forms of the microscopic spin operators are~\cite{Giamarchi,Affleck1998edgeXXZ,EggertAffleck_boundaryXXZBosonizationPhysRevB.46.10866}
\begin{equation}
	S^{\pm}_{\alpha j} \approx  \exp{\left(\pm i \theta_\alpha\right)}\left(  (-1)^j \A~  + \C \cos \phi_\alpha + \ldots \right),~~
	S^z_{\alpha j} \approx \frac{1}{2 \pi} \partial_x \phi_\alpha + (-1)^j \mathcal{B} \sin \phi_\alpha + \ldots \label{appeq:Bosonization_spin}
\end{equation}
where $\mathcal{A},~\mathcal{B}$ and $\mathcal{C}$ are non-universal bosonization prefactors.  \cref{appeq:H_bosonize} denotes a conformal field theory (CFT) with central charge $c=2$ in the presence of various perturbations. Various qualitative aspects of the phase diagram shown in the main text (\cref{appfig:gapless}) can be understood by tracking the renormalization group (RG) flow of \cref{appeq:H_bosonize} by tracking the relevance (in the RG sense) of the primary operators~\cite{Ginsparg1988applied,Francesco2012conformal} shown in \cref{appeq:H_bosonize}
\begin{equation}
	\mathcal{W}_- \equiv \cos(\theta_1 - \theta_2),~\mathcal{V}_\pm \equiv \cos (\phi_1 \pm \phi_2),~\cU \equiv \sum_{\alpha=1,2}\cos \phi_\alpha,~\cV_+\cW_+ \equiv \cos(\theta_1 - \theta_2)\cos (\phi_1 + \phi_2). \label{appeq:UVdef}
\end{equation}
Recall that an operator is RG relevant if its scaling dimension is less than the space-time dimensions (2 in our case). The scaling dimensions of the operators in \cref{appeq:UVdef} are not all independent but related and can be expressed in terms of two independent ones, say $[\cV_\pm]$. For example $[\cW_-] = 1/[\cV_-]$ and  $[\cU] =([\cV_+] + [\cV_-])/4$. This has important consequences, for example it is impossible for both $\cU_-$ and $\cV_-$ to be simultaneously irrelevant. Finally, $\partial_x \phi_1 \partial_x \phi_2$ has scaling dimension $2$ and is exactly marginal. It plays an important role in changing scaling dimensions as the microscopic parameters of the Hamiltonian are varied~\cite{DijkgraafVerlindeVerlinde1988c}. While the exact values of scaling dimensions cannot be determined analytically, in the limit of small couplings $J_M,J_U$ and $\delta$, this can be determined perturbatively as~\cite{Giamarchi}
\begin{equation}
	[\cV_\pm] \approx 2K \left(1 \mp \frac{J_M K}{2 \pi v}\right), ~K = \frac{\pi}{2 \left(\pi-\arccos\Delta\right)},~v= \frac{K}{(2K-1)} \sin \left(\frac{\pi}{2K}\right).\label{appeq:Scaling_dim_perturbative}
\end{equation}
The Luttinger parameter $K$ and velocity $v$ are determined from the Bethe ansatz solution of the XXZ chain~\cite{Haldane_BetheLL_1981153}. Even though \cref{appeq:Scaling_dim_perturbative} is not reliable for large parameter values, it provides useful guidance to look for various phases which can be confirmed by other means, eg: numerical study. Using \cref{appeq:Bosonization_spin}, the symmetry action on spin operators can be translated to the fields as follows ($\tau^x$ is the standard Pauli-X operator)
\begin{equation}
	\centering
	\begin{tabular}{cccc}
		\hline
		&   $U(1)$   & $\ztwo^R$ & $\ztwo^L$  \\
		\hline
		\hline
		$\begin{pmatrix}
			S^\pm_{\alpha j} \\
			S^z_{\alpha j}
		\end{pmatrix} \mapsto$   & $\begin{pmatrix}
			e^{\pm i \chi} S^\pm_{\alpha j} \\
			S^z_{\alpha j}
		\end{pmatrix}$  & $\begin{pmatrix}
			S^\mp_{\alpha j} \\
			- S^z_{\alpha j}
		\end{pmatrix}$ &  $\tau^x_{\alpha \beta}\begin{pmatrix}
			S^\pm_{\beta j} \\
			S^z_{\beta j}
		\end{pmatrix}$ \\
		\hline
		$\begin{pmatrix}
			\theta_\alpha \\
			\phi_\alpha
		\end{pmatrix} \mapsto$ & $\begin{pmatrix}
			\theta_\alpha + \chi \\
			\phi_\alpha
		\end{pmatrix}$ & $-\begin{pmatrix}
			\theta_\alpha  \\
			\phi_\alpha
		\end{pmatrix}$ & $\tau^x_{\alpha \beta} \begin{pmatrix}
			\theta_\beta \\
			\phi_\beta
		\end{pmatrix}$ \\
		\hline
	\end{tabular}
	\label{apptab:Symmetry}
\end{equation}

\subsection{The trivial phase and absence of transitions within it}

The gapped phases shown in the phase diagrams of \cref{appeq:H} are produced when both independent sectors of the two-component Luttinger liquid in \cref{appeq:H_bosonize} are gapped out and a set of two linearly independent commuting field combinations is pinned to one or more values. When the pinned field combinations is unique and symmetry preserving, we get either a trivial or symmetry protected topological (SPT) phase~\cite{WenCompletePhysRevB.84.235128}. We list below the conditions for the former
\begin{equation}
	\begin{tabular}{c|c|c|c}
		\hline
		{Vacuum} & {Scaling Dimensions}                                                                      & {Parameters} & {Pinned Fields}                                                         \\ \hline \hline
		T$_1$           & $[\mathcal{U}] < [\mathcal{W}_-]$                                                                 & $\delta>0$          & $\langle\phi_1 \rangle = \langle \phi_2\rangle = \pi$                          \\ \hline
		T$_2$           & $[\mathcal{U}] > [\mathcal{W}_-]$, $[\mathcal{V}_+]<2$                                            & $J_U<0$             & $\langle\phi_1 +\phi_2 \rangle = 0$, $\langle \theta_1 - \theta_2\rangle =0$   \\ \hline
		T$_3$           & $[\mathcal{U}] < [\mathcal{W}_-]$                                                                 & $\delta<0$          & $\langle\phi_1 \rangle = \langle \phi_2\rangle = 0$                            \\ \hline
		T$_4$           & $[\mathcal{U}] > [\mathcal{W}_-]$, $[\mathcal{V}_+]<2$ & $J_U>0$             & $\langle\phi_1 +\phi_2 \rangle = 0$, $\langle \theta_1 - \theta_2\rangle =\pi$ \\ \hline
	\end{tabular}
\end{equation}
All four seemingly distinct regions T$_{1-4}$ are connected adiabatically without a bulk phase transition. 

Note: the pinning $\moy{\phi_1 + \phi_2} = \pi$ would corresponds to an SPT  phase~\cite{APSECTLLPhysRevB.108.245135,NAKAMURA_string} and is not found in our phase diagrams. While it is clear from \cref{appeq:H_bosonize} that for $J_M>0$, the coefficient of  $\cV_+$ is negative and favours pinning $\moy{\phi_1 + \phi_2} = 0$, we would naturally expect to get $\moy{\phi_1 + \phi_2} = \pi$ for $J_M<0$. However, this does not happen due to the presence of the operator $\cW_- \cV_+$. When $\cW_-$ dominates for large $J_U$, pinning $\moy{\theta_1 - \theta_2}$, this operator contributes an effective shift of the $\cV_+$ coefficient to $\mathcal{B}^2 |J_M|/2-\mathcal{C}^2 |J_U| $ which becomes negative and pins  $\moy{\phi_1 +\phi_2} = 0$. The pinning of  $\moy{\phi_1 + \phi_2} = \pi$ to produce an SPT can be driven by introducing additional diagonal couplings on the ladder~\cite{APSECTLLPhysRevB.108.245135}.

\begin{figure}[!ht]
	\centering
	\begin{tikzpicture}[scale=1.4]
		\draw[thick,->] (-0.35,-0.1) -- ++(4.5,0.0) node[align=right,below] {$J_U$};
		\draw[thick,->] (-0.35,-0.1) -- ++(0.0,3.) node[align=left,above] {$\delta$};
		\fill[blue,opacity=0.05] (-0.25,0.) rectangle ++(4.3,2.8);

		\draw[dotted] (3.8/4,2.75) -- (3.8/3,1.9);
		\draw[dotted] (3.8-3.8/4,2.75) -- (3.8-3.8/3,1.9);
		
		\draw[dotted] (3.8/4,0.05) -- (3.8/3,2.8-1.9);
		\draw[dotted] (3.8-3.8/4,0.05) -- (3.8-3.8/3,2.8-1.9);
		\node[] at(1.9,-0.28) {0};
		\node[] at(1.9,2.6) {\scriptsize T$_1$};
		\node[] at(1.9,2.3) {\scriptsize $\moy{\phi_\alpha} = \pi$};
		\node[] at(1.9,2.8-2.3) {\scriptsize T$_3$};
		\node[] at(1.9,2.8-2.6) {\scriptsize  $ \moy{\phi_\alpha} = 0$};
		\node[] at(0.4,1.8) {\scriptsize T$_2$};
		\node[] at(0.4,1.55) {\scriptsize  $\moy{\theta_1 - \theta_2} = 0$};
		\node[] at(0.4,2.8-1.55) {\scriptsize  $\moy{\phi_1 + \phi_2} = 0$};
		\node[] at(3.8-0.4,1.8) {\scriptsize T$_4$};
		\node[] at(3.8-0.4,1.55) {\scriptsize  $\moy{\theta_1 - \theta_2} = \pi$};
		\node[] at(3.8-0.4,2.8-1.55) {\scriptsize  $\moy{\phi_1 + \phi_2} = 0$};
	\end{tikzpicture}    
	\caption{A schematic representation of how the four pinned-field configurations leading to the trivial phase are generally placed in our phase diagrams. The dotted lines schematically indicate crossovers where the pinned values change without encountering singularities.}
	\label{appfig:trivial phase}
\end{figure}
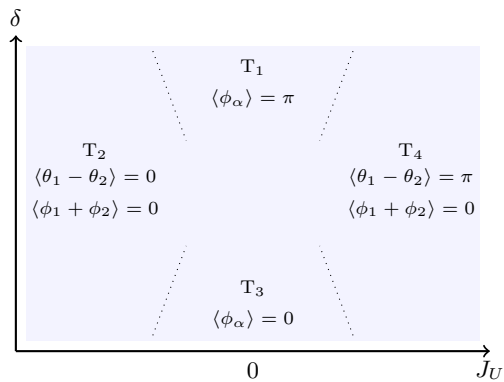

We may wonder whether the vacua T$_1$-T$_4$, schematically represented in \cref{appfig:trivial phase} which seemingly represent different basins of attraction under RG flow correspond to the same phase without encountering any phase transitions as we traverse between them (dotted lines in \cref{appfig:trivial phase}). To establish that this is indeed the case, we perform a standard change of basis on the compact bosons~\cite{Giamarchi} 
\begin{equation}
	\phi_\pm = \frac{\phi_1 \pm \phi_2}{2},~\theta_{\pm} = \theta_1 \pm \theta_2. \label{appeq:symmetric_antisymmetric}
\end{equation}
\cref{appeq:symmetric_antisymmetric} preserves the canonical commutation relations in \cref{appeq:KacMoody_2component} but halves compactification radius of $\phi_\pm$. For the purposes we will use this basis here and elsewhere, the compactification radii are not going to be important. We will use a different change of basis \cref{appeq:SL2Z XY1,appeq:SL2Z XY2} in the next subsection for when compactification radii are important, eg: in determining symmetry properties. The bosonized Hamiltonian in \cref{appeq:H_bosonize} can be rewritten in the new basis as 
\begin{multline}
	H \approx \sum_{\sigma = \pm}\frac{v_\sigma}{2 \pi} \int dx \left( \frac{1}{K_\sigma} (\partial_x\phi_\sigma)^2 + \frac{K_\sigma}{4} (\partial_x\theta_\sigma)^2   \right) 
	+ 2\mathcal{AC} \delta  \int dx~  ( \cos( \phi_+ + \phi_-) +  \cos(\phi_+ - \phi_-)) \\+2 \mathcal{A}^2 J_U \int dx  \cos\theta_- +  \frac{\mathcal{B}^2 J_M}{2} \int dx\left( \cos(2 \phi_-) - \cos(2 \phi_+) \right) 
	+ \mathcal{C}^2 J_U \int dx \cos \theta_- \cos (2 \phi_+)  + \ldots. \label{appeq:H_bosonize_symmetricAntisymmetric}
\end{multline}
The scaling dimensions of the various perturbing operators are 
\begin{equation}
	[\cos (2\phi_-)] =[\cos \theta_-]^{-1} = K_-,~ [\cos(\phi_+ \pm \phi_-)]= \frac{K_+ + K_-}{4},~[\cos\theta_- \cos (2 \phi_+)] =   \frac{1}{K_-} + K_+. \label{appeq:scalingdim_symmetricantisymmetric}
\end{equation}
Now, consider crossing from T$_1$ to T$_2$. The first thing to observe is that $\moy{\phi_1 + \phi_2} = 0$ along this journey. Thus, we can simplify the analysis of the trivial phase by replacing $\phi_+ \rightarrow \moy{\phi_+} \approx 0$ in \cref{appeq:H_bosonize_symmetricAntisymmetric} to get
\begin{multline}
	H \approx \frac{v_-}{2 \pi} \int dx \left( \frac{1}{K_-} (\partial_x\phi_-)^2 + \frac{K_-}{4} (\partial_x\theta_-)^2   \right) 
	+ g_1  \int dx~   \cos  \phi_-  + g_2 \int dx  \cos\theta_- + g_3 \int dx \cos(2 \phi_-)  
	+ \ldots. \label{appeq:H_bosonize_symmetricAntisymmetric_minus}
\end{multline}
where $g_1 \propto \delta,~g_2 \propto J_U$ and $g_3\propto J_M$. In other words, the physics of the change in vacuum can be analyzed using the $c=1$ fixed point theory of the single-component theory in \cref{appeq:H_bosonize_symmetricAntisymmetric_minus} rather than the full $c=2$ theory in \cref{appeq:H_bosonize,appeq:H_bosonize_symmetricAntisymmetric}. The change in the nature of the pinned fields comes from the competition between the lightest operators, $\cos \phi_-$ and $\cos \theta_-$. The subleading operator, $\cos (2\phi_-)$ can be safely ignored. As argued in \cite{LECHEMINANT_SelfDualSineGordon_2002502}, a putative transition occurs when the scaling dimensions of the two operators match, $[\cos \theta_-] = [\cos \phi_-] = 1/2$ when~$K_-=2$ giving us
\begin{equation}
	H \approx\frac{v_-}{4 \pi} \int dx \left( (\partial_x\phi_-)^2 + (\partial_x\theta_-)^2   \right) 
	+  g_1 \int dx~   \cos \phi_-   + g_2 \int dx~  \cos\theta_- . \label{appeq:SDSG_Q1_notselfdual}
\end{equation}
and if their coupling constants flows to identical values $g_{1,2} \rightarrow g$,
\begin{equation}
	H \approx\frac{v_-}{4 \pi} \int dx \left( (\partial_x\phi_-)^2 + (\partial_x\theta_-)^2   \right) 
	+  g \int dx~  \left( \cos \phi_-   +  \cos\theta_- \right). \label{appeq:SDSG_Q1}
\end{equation}
\cref{appeq:SDSG_Q1} is a self-dual Sine-Gordon model which was analyzed in \cite{LECHEMINANT_SelfDualSineGordon_2002502} (along with generalizations one of which we will encounter later)
where it was shown that it does not flow to any critical theory, but remains a trivial gapped state. This is because at this scaling dimension, there appears an enlarged $SU(2)$ symmetry~\cite{DijkgraafVerlindeVerlinde1988c} that allows us to rotate $\cos \theta_- \mapsto \sin \phi_-$ keeping all other terms in \cref{appeq:SDSG_Q1,appeq:SDSG_Q1_notselfdual} fixed. This changes \cref{appeq:SDSG_Q1_notselfdual} to
\begin{equation}
	H \approx\frac{v_-}{4 \pi} \int dx \left( (\partial_x\phi_-)^2 + (\partial_x\theta_-)^2   \right) 
	+   g_1 \int dx~   \cos \phi_-   + g_2 \int dx~  \sin\phi_- . \label{appeq:SDSG_Q1_rotated}
\end{equation}
This theory is the ordinary Sine-Gordon model which flows to a trivial massive phase pinning the field to the minimum of $\left( g_1    \cos \phi_-   + g_2   \sin\phi_- \right)$ thereby establishing that the dotted lines in \cref{appfig:trivial phase} do not represent any bulk phase transitions but merely crossovers. The authors of \cite{LECHEMINANT_SelfDualSineGordon_2002502} arrive at this conclusion using a microscopic mapping that should be of particular appeal to the working condensed matter theorist. The identical scaling dimension $[\cos \theta_-] = [\cos \phi_-]$ and the emergent $SU(2)$ symmetry also appears in the bosonized form of the spin-half Heisenberg chain. In fact, both perturbations in \cref{appeq:SDSG_Q1} can be identified with microscopic perturbations, bond-dimerization and staggered magnetic field in the X spin direction as follows
\begin{multline}
	H = \sum_j (1+ (-1)^j \delta) \vec{S}_j. \vec{S}_{j+1} + h \sum_j (-1)^j S^x_j \approx \frac{v}{4 \pi} \int dx \left( (\partial_x\phi)^2 + (\partial_x\theta)^2   \right) 
	+   2\mathcal{AC} \delta  \int dx \cos\phi   +   \mathcal{B} h  \int dx  \cos\theta \label{appeq:Heisenberg_perturbed}
\end{multline}
It is obvious that by a unitary transformation, the magnetic field can be oriented to the Z direction which gives us
\begin{multline}
	H = \sum_j (1+ (-1)^j \delta) \vec{S}_j. \vec{S}_{j+1} + h \sum_j (-1)^j S^z_j \approx \frac{v}{4 \pi} \int dx \left( (\partial_x\phi)^2 + (\partial_x\theta)^2   \right) 
	+   2\mathcal{AC} \delta  \int dx \cos\phi   +   \mathcal{B} h  \int dx  \sin\phi \label{appeq:Heisenberg_perturbed_rotated}
\end{multline}
Both the microscopic models and field theory in \cref{appeq:Heisenberg_perturbed,appeq:Heisenberg_perturbed_rotated} flows to a trivial gapped phase and thus, we see that T$_1$ and T$_2$ are smoothly connected by a crossover. The same argument can be applied replacing T$_1$ with T$_3$  and/or T$_2$ with T$_4$. However, the T$_{1,3}$ vacua cannot be connected directly nor can T$_{2,4}$. For generic parameter values the T$_{1,3}$ vacua are always interrupted within the trivial phase by crossover-compatible T$_{2,4}$ and vice-versa. For special parameter regimes however in the presence of accidental symmetries, a direct transition between T$_{1}$ and T$_{3}$ or T$_{2}$ and T$_{4}$ is forced and occurs via a bulk gap closure. We will discuss this next. 

\subsection{Special lines: enhanced symmetries and accidental SPTs}
There are two special parameter regimes of \cref{appeq:H} which illuminate important physics. First, if we eliminate bond dimerization by setting $\delta = 0$, the system is invariant under an enhanced translation symmetry by a single site. $\bZ: \vec{S}_{\alpha j} \mapsto\vec{S}_{\alpha j+1}$. The ground states for $J_U \rightarrow \pm \infty$,
\begin{equation}
	\ket{\text{GS}(\infty,\delta)} = \cdots \gszero \cdots,~
	\ket{\text{GS}(-\infty,\delta)} = \cdots \gspi \cdots \label{appeq:GSdelta0}
\end{equation}
are invariant under this enhanced symmetry. The dimers representing Bell pairs are defined in \cref{appeq:boundary_Bell}. We see that the two ground states, representing T$_2$ and T$_4$ carry distinct charges under $\ztwo^R$ and $\ztwo^L$ symmetries  on each rung of ladder making them distinct \emph{weak} SPT phases of matter~\cite{WenCompletePhysRevB.84.235128} that cannot be connected without explicitly breaking either translation or the two $\ztwo$ symmetries or going through a bulk gap closure. For $\delta \neq 0$, translation symmetry is explicitly broken and opens a path to connect the two ground states without a phase transition via intermediate crossovers to T$_1$ or T$_3$. 

For $J_U = 0$ on the other hand, a different symmetry emerges, allowing a separate $U(1)$ spin rotation on each layer. $S^{\pm}_{\alpha j} \mapsto e^{\pm i \chi_\alpha} S^{\pm}_{\alpha j}$ making the full symmetry $(U(1)\times U(1)) \rtimes \ztwo^R \times \ztwo^L$. As discussed in Ref.~\cite{APUCPhysRevLett.130.256401}, the ground states corresponding to T$_1$ and T$_3$ on this line represent distinct strong SPT phases protected by this enhanced on-site symmetry with T$_1$ being non-trivial with the choice of fiducial unit cell. This can be seen in two ways. First, T$_1$ can be detected using string operators with end points charged under $\ztwo^R$:
\begin{equation}
	C_\alpha = \prod_{j=1}^\infty \sigma^z_{2x+j} \sim \sin (\phi_\alpha/2) \label{appeq:string}
\end{equation}
which also implies the existence of edge modes that transform as projective representations of the symmetry group. A second way is to couple the system to two independent background $U(1)$ gauge fields $A^{1,2}_\mu$ corresponding to flux $F^{1,2} = \epsilon^{\mu \nu} \partial_\mu A^{1,2}_\nu$ to get a topological response~\cite{APUCPhysRevLett.130.256401} when we evaluate the Euclidean partition function on a two-dimensional spacetime $\mathcal{M}_2$,
\begin{equation}
	\frac{\mathcal{Z}[A^1,A^2]}{\mathcal{Z}[0,0]}= \exp \left(i \frac{\moy{\phi_\alpha}}{2 \pi} \int_{\mathcal{M}_2} dx dt( F^1 + F^2)\right) = e^{i \pi (\mathscr{I}_1 + \mathscr{I}_2)}. \label{appeq:topological response}
\end{equation}
$\mathscr{I}_{1,2}$ denote the number of $U(1)$ instantons in space time, and \cref{appeq:topological response} measures the total instanton number parity, a topological $\ztwo$ valued invariant. The accidental symmetries and accidental SPTs serve to illustrate two important point. The first, mentioned earlier is taken in pairs, T$_{1,2}$ cannot be connected without an intermediary appearance of T$_{3,4}$ and vice-versa since they are distinct phases under the emergent symmetries. The second is that the edge modes in our phase diagrams can be understood in certain parameter regions as arising from accidental SPTs. For general parameters, a more careful discussion is provided below.

{However, we emphasise that while the accidental symmetries help us understand the phase diagram, they are not necessary. Observe that while the edge modes along $J_U=0$ are justified from the accidental SPT, there exists a parameter regime for $J_U \neq 0$ when there is no non-trivial SPT, where the edge modes are nonetheless stable. Furthermore, along the $\delta = 0$ line, we can explicitly break enhanced translation invariance by adding the following perturbation
	\begin{equation}
		\delta H = \varepsilon \sum_j (-1)^j  S^x_{1j} S^x_{2j} \label{appeq:accidental_transalation_breaking}
	\end{equation}
	without changing the form of the phase diagram particularly the unnecessary critical line. This is because, at long distances, \cref{appeq:accidental_transalation_breaking} rapidly oscillates and vanishes as can be verified using the bosonization formulas \cref{appeq:Bosonization_spin} leaving \cref{appeq:H_bosonize} essentially  unchanged. Indeed, since \cref{appeq:H_bosonize} already consists of the most relevant symmetry-allowed operators, any symmetric weak perturbation merely renormalizes the couplings in \cref{appeq:H_bosonize} and, at most, shifts the locus of the unnecessary critical line.
}

\subsection{Boundary transitions and edge modes}
\label{appsec:Edge}
\begin{figure}[!ht]
	\includegraphics[width=\textwidth]{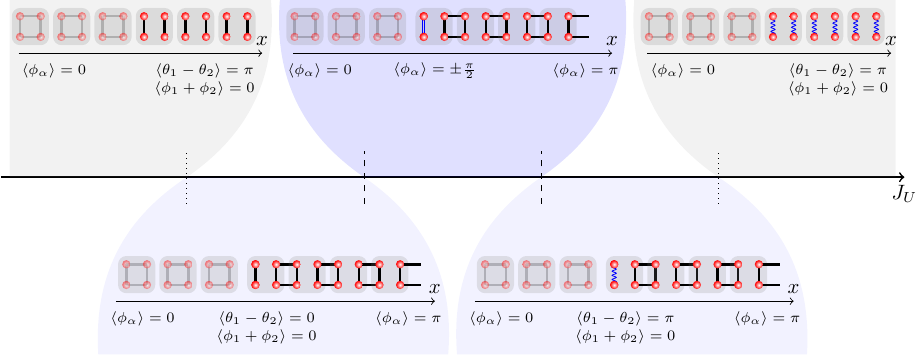}
	\caption{Schematic representation of the system with open boundaries in the $J_M<0$, $\delta>0$ region modeled as an interface to the fiducial T$_3$ vacuum. The microscopic ground states with open boundaries are placed next to a ghostly microscopic T$_3$ ground state to compare with the bosonization results. \label{appfig:interpolation}.}
\end{figure}

In order to characterize the boundary properties of the vacua T$_{1-4}$ of the trivial phase using the continuum bosonization formulation, we find it convenient to replace the boundary with an interface to a fiducial vacuum which we pick to be T$_3$ ($\moy{\phi_\alpha} = 0$)~\cite{Affleck1998edgeXXZ,KeselmanBerg_PhysRevB.91.235309}. This allows us to study an infinite system with spatially modulated couplings instead of a semi-infinite chain. Naturally, the T$_3$ vacuum itself is left without any non-trivial boundary physics and we are left to analyze T$_{1,2,4}$.

Let us first consider an interface that connects the T$_3$ and T$_2$ vacua. Dynamically, this corresponds to finding the ground state of the Hamiltonian in \cref{appeq:H_bosonize} with a spatially varying $\delta(x)$ where $\delta(x\to\pm\infty) \lessgtr 0$,  such that the system lies  in the T$_3$ vacuum as $x\to -\infty$ and in the T$_1$ vacuum as $x\to +\infty$. At the origin and its vicinity, when $\delta$ vanishes, the state of the system is determined by the competition between the other operators in \cref{appeq:H_bosonize}, 
\begin{multline}
	2 \mathcal{A}^2 J_U \int dx  \cos \left(\theta_1 - \theta_2\right) + \frac{\mathcal{B}^2}{2} J_M \int dx\left( \cos(\phi_1 - \phi_2) - \cos(\phi_1 + \phi_2) \right)
	+ \mathcal{C}^2 J_U \int dx \cos \left(\theta_1 - \theta_2\right) \cos \left(\phi_1 + \phi_2\right)   + \ldots \\ \equiv  \mathcal{A}^2 J_U \int dx ~\cW_- + \frac{\mathcal{B}^2}{2} J_M \int dx\left( \cV_- - \cV_+\right)
	+ \mathcal{C}^2 J_U \int dx~\cW_- \cV_+   + \ldots
	\label{appeq:T1T3interface}
\end{multline}
Let us first set $J_U =0$ in \cref{appeq:T1T3interface}. The lowest energy configuration is degenerate and corresponds to $\phi_1 = \phi_2 = \pm \frac{\pi}{2}$ ($J_M<0$) and $\phi_1 = -\phi_2 = \pm \frac{\pi}{2}$ ($J_M>0$). The two degenerate configurations are mapped to each other by $\ztwo^R$ and smoothly interpolate to the unique T$_{1,3}$ vacua as we go to $x \rightarrow \pm \infty$. The $J_M<0$ and $J_M>0$ cases are however distinct in their stability properties to switching on $J_U$. For $J_M<0$, the degenerate interpolations carries net $U(1)$ charges 
\begin{equation}
	Q = \frac{1}{2 \pi} \int dx ~\partial_x(\phi_1(x)+\phi_2(x)) = \pm 1.
\end{equation}
and thus form a two-dimensional irrep of $O(2)$ shown in \cref{apptab:irrep}. This is expected to be stable when we turn on $J_U$ up to some critical value $J_U^*$ and represents the region with boundary modes that shadow the unnecessary critical line in the phase diagram $J_M<0$ shown in \cref{appfig:gapless}(a). Beyond this, $\cW_-$ dominates at the interface, pinning $\moy{\theta_1 - \theta_2}$. 

The stability of the degeneracy at the junction can also be argued in a different way. Assume the converse, that is, for arbitrarily small $J_U$, $\cW_-$ dominates, pinning $\moy{\theta_1 - \theta_2} = 0/\pi$. Since $J_M<0$, the operator $-J_M \cV_+$ in \cref{appeq:T1T3interface} also simultaneously pins $\moy{\phi_1 + \phi_2} = \pi$. Both pinned fields are invariant under all symmetries, and the state contains no degeneracies. However, this configuration cannot be smoothly interpolated while preserving symmetries, to the T$_{1,3}$ vacua it straddles, which require $\moy{\phi_1 + \phi_2} = 2\pi \bZ$. We conclude that $\cW_-$ is suppressed for small $J_U$ until $|J^*_U| \approx\mathcal{B}^2|J_M|/(2 \mathcal{C}) $. Beyond this, if $\cW_-$ pins $\moy{\theta_1 - \theta_2} = 0/\pi$, the operator $\cV_+ \cW_-$ in \cref{appeq:T1T3interface} renormalizes the coefficient of $\cV_+$ to $ \mathcal{B}^2 |J_M|/2-\mathcal{C}^2 |J_U| $, which becomes negative, and pins $\moy{\phi_1 +\phi_2} = 0$. Thus, the degenerate 2d irrep at the junction is stable for some range $|J_U| < |J_U^*|$. As we go to $\delta \rightarrow 0$, The critical $J^*_U$, where $\cW_-$ and $\cV_-$ balance merges with the multicritical point separating the XY$_2$ unnecessary critical line from the XY$_1$ and XY$_1^*$ phases shown in \cref{appfig:gapless}(a). These gapless states arise from a competition between the same \emph{bulk} operators $\cW_-$ and $\cV_+$ nicely mirroring the interface story. The multicritical point that separates them occurs when $[\cW_-] = [\cV_-]$ and corresponds to a CFT with c=$3/2$ (see \cref{appsec:SEC}) . 

For $J_M>0$ on the other hand, the degenerate configurations $\moy{\phi_1} = -\moy{\phi_2} = \pm \frac{\pi}{2}$ do not carry any $U(1)$ charge and are not stable. It represents a level crossing between two 1d irreps which obtain by the introduction of arbitrarily small $J_U$ when $\cV_+$ pins $\moy{\phi_1 + \phi_2} = 0$ and $\cW_-$ pins $\moy{\theta_1 - \theta_2} = 0 /\pi$ for $J_U\lessgtr 0$. In the phase diagram with unnecessary multicriticality, \cref{appfig:gapless}(b), which obtains for $J_M>0$, the boundary transition merges with the bulk multicritical point where we have $[\cW_-] < [\cV_-]$ and the coupling constant of $\cW_-$ tuned to zero (see \cref{appsec:SEC}).

For both cases, the two pinned $\moy{\theta_1 - \theta_2}$ values at the interface correspond to even and odd 1d irreps localized on the interface. For large values of $|J_U|$ and/ or smaller values of $\delta$, the system crosses over to $T_{2,4}$ when the entire bulk for $x>0$ is characterized by pinned $\moy{\theta_1 - \theta_2} = 0/\pi$ and $\moy{\phi_1 + \phi_2} = 0$ indicating that the boundary charges have merged with the bulk by the charge pump.

\section{Bosonization II: Symmetry-enriched critical states}
\label{appsec:SEC}
\subsection{Symmetry enriched critical states}

\begin{figure}[!ht]
	\centering{
		\begin{tikzpicture}[scale = 1.5]
		\draw[thick,->] (-0.6,-0.1) -- ++(7.,0.0) node[align=right,below] {$J_U$};
		\draw[thick,->] (-0.6,-0.1) -- ++(0.0,3.) node[align=left,above] {$\delta$};
		\fill[blue,opacity=0.05] (-0.5,0.) rectangle ++(6.8,2.8);
 
    \draw[dotted] (6.,2.8) to[out=-90,in=20] (5,1.8);
    \draw[dotted] (6,0) to[out=90,in=-20] (5,2.8-1.8);

   \draw[dotted] (-0.2,2.8) to[out=-90,in=160] (0.8,1.8);
    \draw[dotted] (-0.2,0) to[out=90,in=-160] (0.8,2.8-1.8);
  
            \draw[thick] (2,1.4) -- (5.8-2,2.8-1.4);
            \fill[red] (2,1.4) circle (0.05);
            \fill[red] (5.8-2,2.8-1.4) circle (0.05);
            \fill[blue,opacity=0.1] (1.6,2.8) to[out=-90,in = 140] (2,1.4) -- (5.8-2,1.4) to[out=40,in=-90]  (5.8-1.6,2.8) -- cycle;
            \draw[dashed] (1.6,2.8) to[out=-90,in = 140] (2,1.4);
            \draw[dashed]  (5.8-1.6,2.8)to[in=40,out=-90]  (5.8-2,1.4);

            \fill[green,opacity=0.2,rounded corners] (2,1.4) to[bend left] (1.3,2) to[bend right] (1.3,0.8) to[bend left] (2,1.4);
            \draw[rounded corners] (2,1.4) to[bend left] (1.3,2) to[bend right] (1.3,0.8) to[bend left] (2,1.4);
            \fill[yellow,opacity=0.2,rounded corners] (5.8-2,1.4) to[bend right] (5.8-1.3,2)  to[bend left] (5.8-1.3,0.8) to[bend right] (5.8-2,1.4)  ;
            \draw[rounded corners] (5.8-2,1.4) to[bend right] (5.8-1.3,2)  to[bend left] (5.8-1.3,0.8) to[bend right] (5.8-2,1.4) ;
            \node[] at(1.42,1.4) {\scriptsize XY$_1^*$};
            \node[] at(4.41,1.4) {\scriptsize XY$_1$};
            \node[] at(2.9,1.52) {\scriptsize XY$_2$};

            \node[] at(2.2,2.5) {\scriptsize T$_1$};
            \node[] at(0.2,2.5) {\scriptsize T$_1$};
            \node[] at(5.6,2.5) {\scriptsize T$_1$};

            \node[] at(2.2,0.3) {\scriptsize T$_3$};

            \node[] at(0.0,1.8) {\scriptsize T$_2$};
            \node[] at(5.8,1.8) {\scriptsize T$_4$};

            \node[] at(2.9,-0.25) {0};
            \node[] at(-0.8,0.25) {-1};
            \node[] at(-0.8,1.4) {0};
            \node[] at(-0.8,2.7) {1};
              \node[] at(2.9,1) {\scriptsize $c=3/2$};
              \draw[->,thin] (2.4,1.05) -- (2.05,1.3);
              \draw[->,thin] (3.4,1.05) -- (3.75,1.3);
              
              \def\ya{.1};
              \def\yb{.23};
              \foreach   \x in {2.4,3.0} {\fill[rounded corners,gray,opacity=0.2] (\x,\ya) rectangle (\x+0.5,\ya+.5); \draw[thick]  (\x+0.15,\yb) rectangle(\x+0.35,\yb+0.23);
              \shade[inner color=white, outer color = red] (\x+0.15,\yb) circle (0.06);\shade[inner color=white, outer color = red] (\x+0.35,\yb) circle (0.06);\shade[inner color=white, outer color = red] (\x+0.15,\yb+0.23) circle (0.06);\shade[inner color=white, outer color = red] (\x+0.35,\yb+0.23) circle (0.06);};
              
              \def\ya{2.2};
              \def\yb{2.33};
              \foreach \x in {3.15} {\draw[thick]   (\x+0.45,\yb+0.23)--(\x+0.2,\yb+0.23)-- (\x+0.2,\yb)--(\x+0.45,\yb);}
              \foreach \x in {3.0} {\draw[thick]  (\x+0.15,\yb) rectangle (\x-0.25,\yb+0.23);}
              \foreach   \x in {2.4} {\draw[blue]  (\x+0.135,\yb+0.23)-- (\x+0.135,\yb);\draw[blue]  (\x+0.165,\yb+0.23)-- (\x+0.165,\yb);}
              \foreach   \x in {2.4,3.0} {\fill[rounded corners,gray,opacity=0.2] (\x,\ya) rectangle (\x+0.5,\ya+.5); 
              \shade[inner color=white, outer color = red] (\x+0.15,\yb) circle (0.06);\shade[inner color=white, outer color = red] (\x+0.35,\yb) circle (0.06);\shade[inner color=white, outer color = red] (\x+0.15,\yb+0.23) circle (0.06);\shade[inner color=white, outer color = red] (\x+0.35,\yb+0.23) circle (0.06);};

              \def\sh{-2.05};
              \foreach \x in {3.15+\sh} {\draw[thick] (\x+0.45,\yb+0.23)--(\x+0.2,\yb+0.23)-- (\x+0.2,\yb)--(\x+0.45,\yb);}
              \foreach \x in {3.0+\sh} {\draw[thick]  (\x+0.15,\yb) rectangle (\x-0.25,\yb+0.23);}
              \foreach   \x in {2.4+\sh} {\draw[thick]  (\x+0.15,\yb+0.23)-- (\x+0.15,\yb);}
              \foreach   \x in {2.4+\sh,3.0+\sh} {\fill[rounded corners,gray,opacity=0.2] (\x,\ya) rectangle (\x+0.5,\ya+.5); 
              \shade[inner color=white, outer color = red] (\x+0.15,\yb) circle (0.06);\shade[inner color=white, outer color = red] (\x+0.35,\yb) circle (0.06);\shade[inner color=white, outer color = red] (\x+0.15,\yb+0.23) circle (0.06);\shade[inner color=white, outer color = red] (\x+0.35,\yb+0.23) circle (0.06);};
            
            \def\sh{1.85};
              \foreach \x in {3.15+\sh} {\draw[thick] (\x+0.45,\yb+0.23)--(\x+0.2,\yb+0.23)-- (\x+0.2,\yb)--(\x+0.45,\yb);}
              \foreach \x in {3.0+\sh} {\draw[thick]  (\x+0.15,\yb) rectangle (\x-0.25,\yb+0.23);}
              \foreach   \x in {2.4+\sh} {\draw[blue, snake it]  (\x+0.15,\yb+0.23)-- (\x+0.15,\yb);}
              \foreach   \x in {2.4+\sh,3.0+\sh} {\fill[rounded corners,gray,opacity=0.2] (\x,\ya) rectangle (\x+0.5,\ya+.5); 
              \shade[inner color=white, outer color = red] (\x+0.15,\yb) circle (0.06);\shade[inner color=white, outer color = red] (\x+0.35,\yb) circle (0.06);\shade[inner color=white, outer color = red] (\x+0.15,\yb+0.23) circle (0.06);\shade[inner color=white, outer color = red] (\x+0.35,\yb+0.23) circle (0.06);};
              
              \def\sh{0.45};
            \def\ya{1.15};
              \def\yb{1.28};
           \foreach   \x in {4.75+\sh,5.3+\sh} {\fill[rounded corners,gray,opacity=0.2] (\x,\ya) rectangle (\x+0.5,\ya+.5); \draw[blue, snake it]  (\x+0.15,\yb)--(\x+0.15,\yb+0.23);\draw[blue, snake it]  (\x+0.35,\yb)--(\x+0.35,\yb+0.23); 
              \shade[inner color=white, outer color = red] (\x+0.15,\yb) circle (0.06);\shade[inner color=white, outer color = red] (\x+0.35,\yb) circle (0.06);\shade[inner color=white, outer color = red] (\x+0.15,\yb+0.23) circle (0.06);\shade[inner color=white, outer color = red] (\x+0.35,\yb+0.23) circle (0.06);}

           \def\sh{-0.45};
           \foreach   \x in {0.0+\sh,0.55+\sh} {\fill[rounded corners,gray,opacity=0.2] (\x,\ya) rectangle (\x+0.5,\ya+.5); \draw[thick]  (\x+0.15,\yb)--(\x+0.15,\yb+0.23);\draw[thick]  (\x+0.35,\yb)--(\x+0.35,\yb+0.23); 
              \shade[inner color=white, outer color = red] (\x+0.15,\yb) circle (0.06);\shade[inner color=white, outer color = red] (\x+0.35,\yb) circle (0.06);\shade[inner color=white, outer color = red] (\x+0.15,\yb+0.23) circle (0.06);\shade[inner color=white, outer color = red] (\x+0.35,\yb+0.23) circle (0.06);}              
              \node[] at(2.9,3.0) {\scriptsize (a) $J_M\approx -5.2,~\Delta\approx -0.05$};         
		\end{tikzpicture} 
		\includestandalone[width=.25\textwidth]{SupMatMulticriticalityPhasediagram}
	}\vspace{-0.0cm}
	\caption{Phase diagrams showing unnecessry criticality (a) and multicriticality (b). The rough parameter regimes around which we have found these phase diagrams are also indicated. The various symmetry enriched critical states XY$_1$, XY$_1^*$, XY$_2$ and XY$_3$ are all described by the compact boson CFT but distinguished by the symmetry charges carried by the low-energy fields. Crossover lines are schematic and exaggerated.  }
	\label{appfig:gapless}
	\vspace{-0.3cm}
\end{figure}
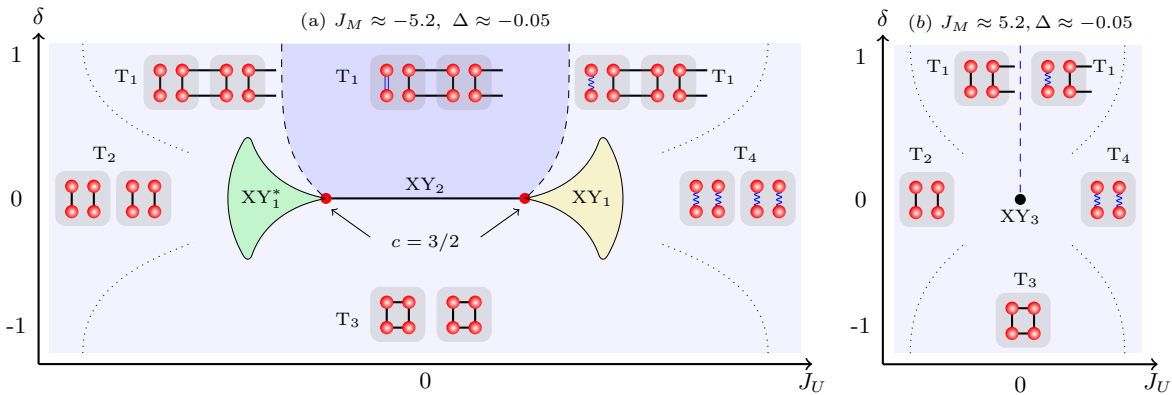

As we go to the origin of the phase diagrams, we see a cornucopia of `symmetry-enriched critical' states~\cite{VerresenThorngrenSECPhysRevX.11.041059,APSECTLLPhysRevB.108.245135,APUCPhysRevLett.130.256401}, which we label XY$_{1-3}$ and XY$_1^*$ in \cref{appfig:gapless} following the terminology of Ref.~\cite{Schulz_Higherspinbosinization_PhysRevB.34.6372,APUCPhysRevLett.130.256401,APSECTLLPhysRevB.108.245135}. These result when only one of the two Luttinger liquid components in \cref{appeq:H_bosonize} is gapped out.

All these critical states are described at long distances by a single compact-boson CFT with central charge $c=1$ described by the following effective Hamiltonian ($v_{\rm eff}$ sets the energy scale and $K_{\rm eff}$ determines the universal properties such as correlation exponents) 
\begin{equation}
	H_{\rm eff} \approx \frac{v_{\rm eff}}{2 \pi} \int dx \left(\frac{1}{4K_{\rm eff}} \left(\partial_x \phi\right)^2 + K_{\rm eff} \left(\partial_x \theta\right)^2\right).  \label{appeq:compactboson}
\end{equation}
They are however distinguished in the way microscopic symmetries of \cref{apptab:Symmetry} act on the surviving low-energy fields as a consequence of how fields are pinned in the gapped sector as summarized below 
\begin{equation}
	\centering
	\begin{tabular}{l|l|l|l|l|l}
		\hline
		State & Condition &Pinned field &  $U(1)$   & $\ztwo^R$ & $\ztwo^L$  \\
		\hline
		\hline
		XY$_1$ & $[\cW_-]<[\cV_-],~[\cV_+]>2,~J_U>0$ & $\moy{\theta_1 - \theta_2} = \pi$ &$\begin{pmatrix}
			\theta  \\
			\phi
		\end{pmatrix} \mapsto$ $\begin{pmatrix}
			\theta +\chi \\
			\phi
		\end{pmatrix}$&$\begin{pmatrix}
			\theta  \\
			\phi
		\end{pmatrix} \mapsto$ $-\begin{pmatrix}
			\theta  \\
			\phi
		\end{pmatrix}$ & $\begin{pmatrix}
			\theta  \\
			\phi
		\end{pmatrix} \mapsto$ $\begin{pmatrix}
			\theta  + \pi\\
			\phi
		\end{pmatrix}$\\
		\hline 
		XY$_1^*$& $[\cW_-]<[\cV_-],~[\cV_+]>2,~J_U<0$& $\moy{\theta_1 - \theta_2} = 0$ &$\begin{pmatrix}
			\theta  \\
			\phi
		\end{pmatrix} \mapsto$  $\begin{pmatrix}
			\theta +\chi \\
			\phi
		\end{pmatrix}$&$\begin{pmatrix}
			\theta  \\
			\phi
		\end{pmatrix} \mapsto$ $-\begin{pmatrix}
			\theta  \\
			\phi
		\end{pmatrix}$ &$\begin{pmatrix}
			\theta  \\
			\phi
		\end{pmatrix} \mapsto$ $\begin{pmatrix}
			\theta   \\
			\phi
		\end{pmatrix}$ \\
		\hline
		XY$_2$& $[\cW_-]>[\cV_-],~[\cV_+]>2,~\delta=0$ & $\moy{\phi_1 - \phi_2} = 0$ &$\begin{pmatrix}
			\theta  \\
			\phi
		\end{pmatrix} \mapsto$   $\begin{pmatrix}
			\theta +2\chi \\
			\phi
		\end{pmatrix}$&$\begin{pmatrix}
			\theta  \\
			\phi
		\end{pmatrix} \mapsto$ $-\begin{pmatrix}
			\theta  \\
			\phi
		\end{pmatrix}$ &$\begin{pmatrix}
			\theta  \\
			\phi
		\end{pmatrix} \mapsto$ $\begin{pmatrix}
			\theta   \\
			\phi
		\end{pmatrix}$ \\
		\hline
		XY$_3$& $[\cW_-]<[\cV_-],~[\cV_+]<2,~\delta=J_U=0$ & $\moy{\phi_1 + \phi_2} = 0$ &$\begin{pmatrix}
			\theta  \\
			\phi
		\end{pmatrix} \mapsto$  $\begin{pmatrix}
			\theta  \\
			\phi
		\end{pmatrix}$ &$\begin{pmatrix}
			\theta  \\
			\phi
		\end{pmatrix} \mapsto$ $-\begin{pmatrix}
			\theta  \\
			\phi
		\end{pmatrix}$ &$\begin{pmatrix}
			\theta  \\
			\phi
		\end{pmatrix} \mapsto$ $-\begin{pmatrix}
			\theta  \\
			\phi
		\end{pmatrix}$ \\
		\hline
	\end{tabular}
	\label{apptab:Symmetry_SEC}
\end{equation}
Gapless states with different subscripts are distinguished by the microscopic $U(1)$ charge carried by the low-energy fields whereas gapless states with different supercripts, XY$_1$ and XY$_1^*$ are distinguished by $\ztwo^L$ charges. Distinct symmetry charges assigned to the field, and therefore the underlying spectrum of operators classifying the CFT  serve as topological invariants and cannot be changed smoothly so long as the microscopic symmetries are present. This means that the gapless states cannot be connected smoothly, without either a change in universality class, or the appearance of other intermediary phases. To obtain the symmetry transformations shown in \cref{apptab:Symmetry_SEC}, we follow the strategy of Ref.~\cite{IGSPT_PhysRevB.104.075132} and use a convenient set of $SL(2,\bZ)$ transformations that preserve both the compatification radii and commutation relations in \cref{appeq:KacMoody_2component}. \cref{apptab:Symmetry_SEC} also specifies the parameter regime and field theoretic conditions leading to each gapless state. We see that XY$_{1}$ and XY$_{1}^*$ are phases requiring no fine-tuning of parameters whereas XY$_{2}$ and XY$_{3}$ require fine-tuning one and two parameters respectively. The $SL(2,\bZ)$ transformations will also help us understand this. 

\subsection{The XY$_2$ line and its vicinity} 

Let us begin with the case where $\phi_1 - \phi_2$ is pinned when $[\cV_-]<[\cW_-]$ and $[\cV_+]>2$ leading to the XY$_2$ line in \cref{appfig:gapless}(a). To analyze this region, the following transformation is suitable.
\begin{equation}
	\begin{pmatrix}
		\Phi \\
		\phi
	\end{pmatrix} = \begin{pmatrix}
		\phi_1 - \phi_2\\
		\phi_2
	\end{pmatrix},~\begin{pmatrix}
		\Theta\\
		\theta
	\end{pmatrix} = \begin{pmatrix}
		\theta_1\\
		\theta_1 + \theta_2
	\end{pmatrix}. \label{appeq:SL2Z XY2}
\end{equation}
The symmetry transformations in \cref{apptab:Symmetry} can be translated to the redefined fields in \cref{appeq:SL2Z XY2} as
\begin{equation}
	\begin{tabular}{c| c |c |c}
		\hline
		&  U(1)    &$\ztwo^R$ &$\ztwo^L$  \\
		\hline
		\hline
		$\begin{pmatrix}
			\Theta\\
			\theta
		\end{pmatrix} \mapsto$         & $\begin{pmatrix}
			\Theta+\chi\\
			\theta + 2\chi
		\end{pmatrix}$ & $- \begin{pmatrix}
			\Theta\\
			\theta
		\end{pmatrix}$ & $\begin{pmatrix}
			\theta-\Theta\\
			\theta
		\end{pmatrix}$ \\
		$\begin{pmatrix}
			\Phi\\
			\phi
		\end{pmatrix} \mapsto$ & $\begin{pmatrix}
			\Phi\\
			\phi
		\end{pmatrix} $ & $-\begin{pmatrix}
			\Phi\\
			\phi
		\end{pmatrix}$ & $\begin{pmatrix}
			- \Phi\\
			\Phi+\phi
		\end{pmatrix}$\\
		\hline
	\end{tabular} \label{appeq:XY2_symmetryaction}
\end{equation}
XY$_2$ is produced when $\moy{\Phi} = 0$. The low-energy physics of the system is described in terms of the conjugate pair $\theta,\phi$ with the effective Hamiltonian in \cref{appeq:compactboson}. The low-energy symmetry properties of these fields shown in \cref{apptab:Symmetry_SEC} is obtained by replacing the field $\Phi$ by its expectation value, $\Phi \approx \moy{\Phi} = 0$. The low-energy manifestation of various perturbations in \cref{appeq:H_bosonize} can be obtained as
\begin{align}
	\cU &\equiv \cos(\Phi+ \phi) + \cos \phi \approx \cos(\moy{\Phi}+ \phi) + \cos \phi = 2 \cos \phi, \nonumber\\
	\cW_- &\equiv \cos(2\Theta - \theta) \approx \moy{\cos(2\Theta - \theta)} = 0, \nonumber\\
	\cV_- &\equiv \cos \Phi \approx \cos \moy{\Phi} = \text{const}, \nonumber\\
	\cV_+ & \equiv \cos(\Phi + 2 \phi) \approx \cos(\moy{\Phi} + 2 \phi) = \cos(2 \phi). \label{appeq:XY2_effectiveperturbations}.
\end{align}
The phase diagram near the vicinity of the XY$_2$ line can be described by adding the perturbations in \cref{appeq:XY2_effectiveperturbations} to \cref{appeq:compactboson} to get
\begin{equation}
	H \approx \frac{v_{\rm eff}}{2 \pi} \int dx \left(\frac{1}{4K_{\rm eff}} \left(\partial_x \phi\right)^2 + K_{\rm eff} \left(\partial_x \theta\right)^2\right) + g_{\cU} \int dx \cos \phi - g_{\cV_+}\int dx \cos(2 \phi) \label{appeq:XY2proximate}
\end{equation}
where $g_{\cU} \propto \delta$ and $g_{\cV_+} \propto J_M$. The UC phase diagram obtains for  $\frac{1}{2}<K_{\rm eff}<2$ when $\cos(2\phi)$ is irrelevant whereas $\cos \phi$ is relevant. Thus for $g_{\cU} \neq 0$, the system gaps out to produce either T$_1$ or T$_3$ vacua.

\cref{appeq:XY2proximate} suggests other proximate phase diagrams: (i) when $K_{\rm eff}>2$, the XY$_2$ line is expected to get stabilized to a gapless phase~\cite{APSECTLLPhysRevB.108.245135} (ii) when $K_{\rm eff}<\frac{1}{2}$, the XY$_2$ line is expected to gap out to produce SSB phases which we will discuss in \cref{appsec:SSB}.

\subsection{XY$_1$, XY$_1^*$ phases and their vicinity} 
When $\theta_1 - \theta_2$ is pinned, a different transformation is suitable
\begin{equation}
	\begin{pmatrix}
		\Phi\\
		\phi
	\end{pmatrix} = 
	\begin{pmatrix}
		\phi_1\\
		\phi_1+\phi_2
	\end{pmatrix} ,\begin{pmatrix}
		\Theta\\
		\theta
	\end{pmatrix} = 
	\begin{pmatrix}
		\theta_1 - \theta_2\\
		\theta_2
	\end{pmatrix}, \label{appeq:SL2Z XY1}
\end{equation}
The symmetry action on the two fields shown in \cref{apptab:Symmetry} can be translated to the definitions in \cref{appeq:SL2Z XY1} as
\begin{equation}
	\begin{tabular}{c| c |c |c}
		\hline
		&  U(1)    &$\ztwo^R$ &$\ztwo^L$  \\
		\hline
		\hline
		$\begin{pmatrix}
			\Theta\\
			\theta
		\end{pmatrix} \mapsto$         & $\begin{pmatrix}
			\Theta \\
			\theta+ \chi
		\end{pmatrix}$ & $- \begin{pmatrix}
			\Theta\\
			\theta
		\end{pmatrix}$ & $\begin{pmatrix}
			-\Theta\\
			\theta + \Theta
		\end{pmatrix}$ \\
		$\begin{pmatrix}
			\Phi\\
			\phi
		\end{pmatrix} \mapsto$ & $\begin{pmatrix}
			\Phi\\
			\phi
		\end{pmatrix} $ & $-\begin{pmatrix}
			\Phi\\
			\phi
		\end{pmatrix}$ & $\begin{pmatrix}
			\phi-\Phi\\
			\phi
		\end{pmatrix} $\\
		\hline
	\end{tabular}
\end{equation}
The symmetry action in \cref{apptab:Symmetry_SEC} for XY$_1$ and XY$_1^*$ is obtained replacing the pinned field $\Theta \approx \moy{\Theta} = 0/\pi$ appropriately. The low-energy manifestation of various perturbations in \cref{appeq:H_bosonize} can be obtained as
\begin{align}
	\cU &\equiv \cos( \phi- \Phi) + \cos \Phi \approx \moy{\cos( \phi- \Phi)}  + \moy{\cos \Phi}=0 \nonumber\\
	\cW_- &\equiv \cos\Theta \approx \cos\moy{\Theta} = \text{const}, \nonumber\\
	\cV_- &\equiv \cos (2\Phi-\phi) \approx \moy{\cos (2\Phi-\phi)} = 0, \nonumber\\
	\cV_+ & \equiv \cos\phi. \label{appeq:XY1_effectiveperturbations}
\end{align} 
The phase diagram near the vicinity of the XY$_1$ or XY$_1^*$ can be written by adding the perturbations in \cref{appeq:XY1_effectiveperturbations} to \cref{appeq:compactboson} to get
\begin{equation}
	H \approx \frac{v_{\rm eff}}{2 \pi} \int dx \left(\frac{1}{4K_{\rm eff}} \left(\partial_x \phi\right)^2 + K_{\rm eff} \left(\partial_x \theta\right)^2\right) - g_{\cV_+}\int dx \cos\phi \label{appeq:XY1proximate}
\end{equation}
where $g_{\cV_+} \propto J_M$. It is clear from \cref{appeq:XY1proximate} that the gapless XY$_1$/ XY$_1^*$ phases obtain for $K_{\rm eff}>2$ and gap out to the trivial phase when $K_{\rm eff}<2$ via a BKT transition leading to either the T$_{2}$ or T$_{4}$ vacua. 

\subsection{The XY$_3$ point and its vicinity}

Finally, when $\phi_1 + \phi_2$ is pinned, the same $SL(2,\bZ)$ transformation in \cref{appeq:SL2Z XY1} can be used but with the pinned sector changed. To keep the notation of $\theta,\phi$ labeling the low-energy sector, let us rewrite \cref{appeq:SL2Z XY1} with the field labelling exchanged 
\begin{equation}
	\begin{pmatrix}
		\Phi\\
		\phi
	\end{pmatrix} = 
	\begin{pmatrix}
		\phi_1+\phi_2\\
		\phi_1
	\end{pmatrix} ,\begin{pmatrix}
		\Theta\\
		\theta
	\end{pmatrix} = 
	\begin{pmatrix}
		\theta_2\\
		\theta_1 - \theta_2
	\end{pmatrix}, \label{appeq:SL2Z XY3}
\end{equation}
The symmetry action on the two fields shown in \cref{apptab:Symmetry} can be translated to the definitions in \cref{appeq:SL2Z XY3} as
\begin{equation}
	\begin{tabular}{c| c |c |c}
		\hline
		&  U(1)    &$\ztwo^R$ &$\ztwo^L$  \\
		\hline
		\hline
		$\begin{pmatrix}
			\Theta\\
			\theta
		\end{pmatrix} \mapsto$         & $\begin{pmatrix}
			\Theta+ \chi\\
			\theta 
		\end{pmatrix}$ & $- \begin{pmatrix}
			\Theta\\
			\theta
		\end{pmatrix}$ & $\begin{pmatrix}
			\Theta + \theta\\
			-\theta
		\end{pmatrix}$ \\
		$\begin{pmatrix}
			\Phi\\
			\phi
		\end{pmatrix} \mapsto$ & $\begin{pmatrix}
			\Phi\\
			\phi
		\end{pmatrix} $ & $-\begin{pmatrix}
			\Phi\\
			\phi
		\end{pmatrix}$ & $\begin{pmatrix}
			\phi\\
			\Phi - \phi
		\end{pmatrix} $\\
		\hline
	\end{tabular}
\end{equation}

The symmetry action in \cref{apptab:Symmetry_SEC} for XY$_3$ is obtained by replacing the pinned field $\Phi\approx \moy{\Phi} = 0$.  The low-energy manifestation of various perturbations in \cref{appeq:H_bosonize} can be obtained as
\begin{align}
	\cU &\equiv \cos(\Phi- \phi) + \cos \phi \approx \cos(\moy{\Phi}- \phi) + \cos \phi = 2 \cos \phi, \nonumber\\
	\cW_- &\equiv \cos\theta, \nonumber\\
	\cV_- &\equiv \cos (2 \phi-\Phi) \approx \cos (2 \phi-\moy{\Phi}) = \cos(2\phi), \nonumber\\
	\cV_+ & \equiv \cos\Phi \approx \cos\moy{\Phi} = \text{const}. \label{appeq:XY3_effectiveperturbations}
\end{align}
The phase diagram near the vicinity of the XY$_3$ point can be written by adding the perturbations in \cref{appeq:XY3_effectiveperturbations} to \cref{appeq:compactboson} to get
\begin{equation}
	H \approx \frac{v_{\rm eff}}{2 \pi} \int dx \left(\frac{1}{4K_{\rm eff}} \left(\partial_x \phi\right)^2 + K_{\rm eff} \left(\partial_x \theta\right)^2\right) + g_{\cU} \int dx \cos \phi + g_{\cV_-}\int dx \cos(2 \phi) + g_{\cW_-} \int dx \cos \theta \label{appeq:XY3proximate}
\end{equation}
where $g_{\cU} \propto \delta$, $g_{\cV_-} \propto J_M$ and $g_{\cW_-} \propto J_U$. The phase diagram in \cref{appfig:gapless}(b) obtains when $\cV_-$ is irrelevant,  $\frac{1}{2}< K_{\rm eff} < 2$ which automatically leads to $[\cW_-]  < [\cV_-]$ and being the most relevant operator  where $[\cW_-]^{-1} = [\cV_-] =4 K_{\rm eff}$ and $[\cU] = K_{\rm eff}$. This means that near the XY$_3$ point, for $J_U \neq 0$, the system flows to a trivial phase with $T_{2,4}$ vacua. Fine-tuning $J_U =0$, the system flows to the $T_{1,3}$ vacua with the former exhibiting edge modes.

The dominance of bulk operators $\cW_-$ over $\cV_-$ near the  XY$_3$ point is shadowed by the fact that there exists no non-trivial boundary phase in this parameter regime and the edge modes represent a transition between different boundary charges, as discussed in \cref{appsec:Edge} and in the main text. Away from the XY$_3$ point, we expect the $T_{1,3}$ to persist over some narrow but finite range near the $J_U = 0$ line as shown in \cref{appfig:gapless}(b) and at large values of $J_U$, crossover to the $T_{2,4}$ vacua. \cref{appeq:XY3proximate} tells us that there should exist proximate phase diagrams when $\frac{1}{2}> K_{\rm eff}$ and the XY$_3$ point opens up to a SSB phase which we will discuss below. There is also seemingly a new route to unnecessary criticality when $K_{\rm eff} > 2$ along the $J_U = 0$ line. However, this is unstable to the addition of symmetry allowed operator, $\cos (2 \theta)$ to \cref{appeq:XY3proximate} which would be relevant. 

\section{Bosonization III: Symmetry breaking and Ising transitions}

\label{appsec:SSB}

\begin{figure}[!t]
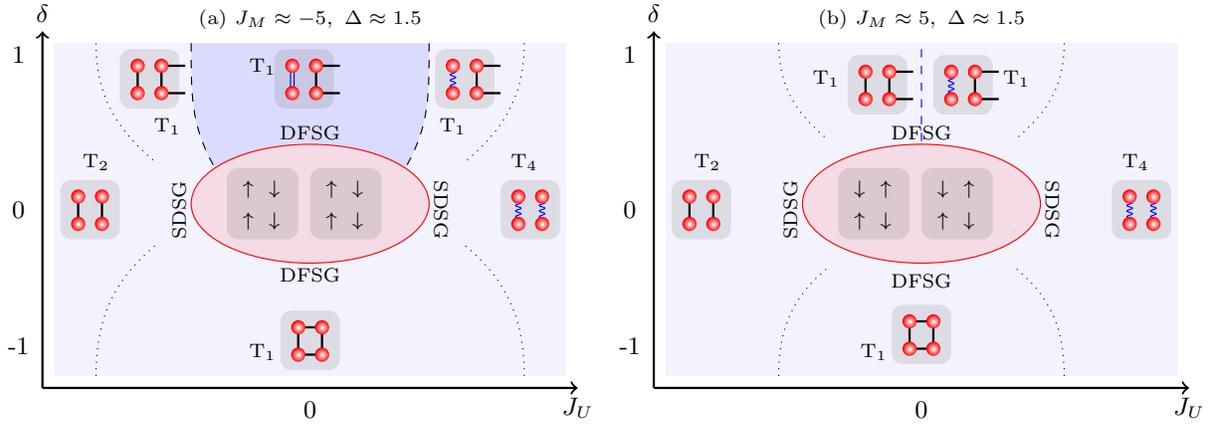

	\centering{
		\includestandalone[width=.9\textwidth]{SupMatSSBPhasediagram}
	}\vspace{-0.0cm}
	\caption{Phase diagrams exhibiting spontaneous symmetry breaking detected by order parameters in $\mathcal{O}_+$ (a) and  $\mathcal{O}_-$ (b) defined in \cref{appeq:order parameter Jm neg,appeq:order parameter Jm pos}. Rough parameter regimes where the phase diagrams have been confirmed are listed. Crossover lines are schematic and exaggerated. }
	\label{appfig:SSB}
	\vspace{-0.3cm}
\end{figure}

We discuss deformations of \cref{appfig:gapless} where the unnecessary critical and multicritcal theories are gapped out to produce symmetry breaking phases as shown in \cref{appfig:SSB} and the associated critical phenomena. This occurs when the $\cV_\pm$ operators in \cref{appeq:H_bosonize} are dominant and pin the fields $\phi_\alpha$ to degenerate values which transform into each other under broken symmetries. From \cref{appeq:H_bosonize}, we see that the nature of degenerate vacua and symmetry breaking depends on the sign of $J_M$ as indicated in \cref{appfig:SSB}. 

\subsection{Mean field analysis}
When $\cV_\pm$ and $\cU$ are relevant in \cref{appeq:H_bosonize}, we can use the following mean-field potential
\begin{equation}
	V(\phi_1,\phi_2) = g_{\cU} (\cos \phi_1 + \cos \phi_2) + g_{\cV} \sin \phi_1 \sin \phi_2. \label{appeq:VSSB}
\end{equation}
where $g_{\cU} \propto \delta$ and $g_{\cV} \propto J_M$ are renormalized couplings. Symmetry breaking is obtained for small values of $g_{\cU}$. For $g_{\cV}<0$, we get $\moy{\phi_1} = \moy{\phi_2} = \pm \Phi$ where $\Phi$ depends on the coupling constants. From \cref{apptab:Symmetry}, we see that the ground states are mapped to each other under spin reflections, $\ztwo^R$ but preserve layer exchange, $\ztwo^L$. The local charged order parameter that detects this phase is 
\begin{equation}
	\mathcal{O}_+ = S^z_1 + S^z_2 \sim (\sin\phi_1 + \sin \phi_2) \label{appeq:order parameter Jm neg}.
\end{equation}
For $J_M >0$ on the other hand, we have $\moy{\phi_1} = -\moy{\phi_2} = \pm \Phi$. From \cref{apptab:Symmetry}, we now see that the ground states are mapped to each other under both lattice reflections $\ztwo^R$ and layer exchange $\ztwo^L$ but the combined action leaves the vacua invariant. The local charged order parameter  that detects this phase is 
\begin{equation}
	\mathcal{O}_- = S^z_1 - S^z_2 \sim (\sin\phi_1 - \sin \phi_2) \label{appeq:order parameter Jm pos}.
\end{equation}
Caricatures of the ground state are shown in \cref{appfig:SSB}. As we tune other parameters in the theory, symmetry is restored via an Ising transiton. This can be understood in two differnt ways discussed below. 

\subsection{Double-frequency Sine-Gordon Ising criticality}
\begin{figure}[!ht]
	\centering
	\includegraphics[width=0.25\linewidth]{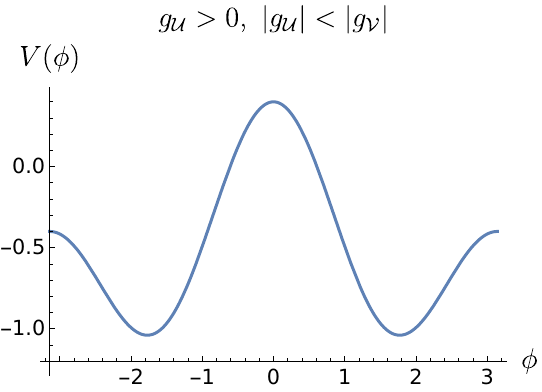}
	\includegraphics[width=0.25\linewidth]{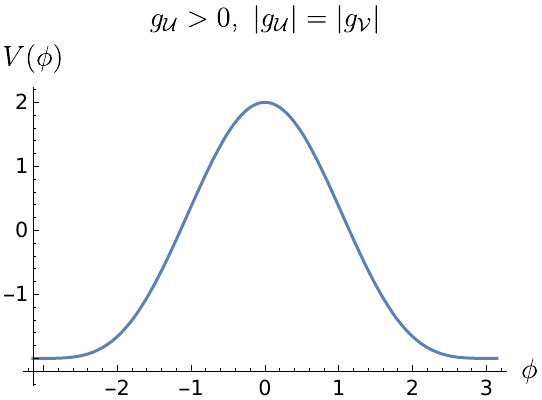}
	\includegraphics[width=0.25\linewidth]{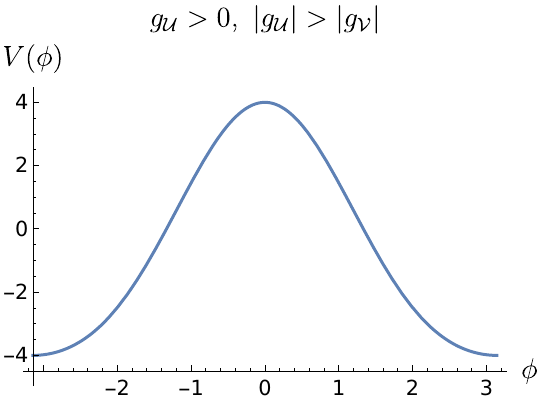}\\
	\vspace{.5em}
	\includegraphics[width=0.25\linewidth]{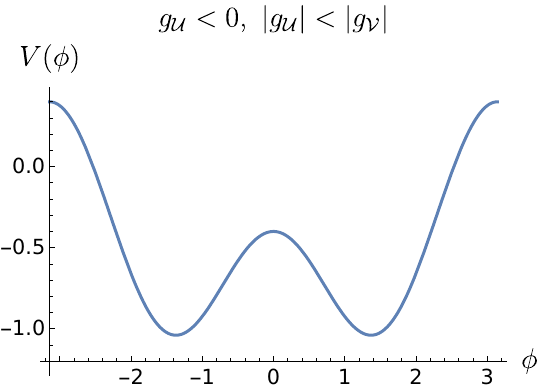}
	\includegraphics[width=0.25\linewidth]{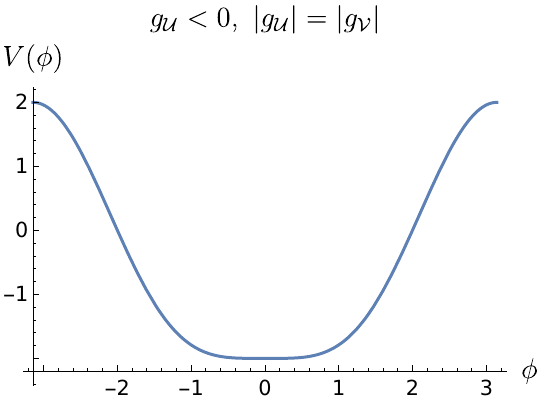}
	\includegraphics[width=0.25\linewidth]{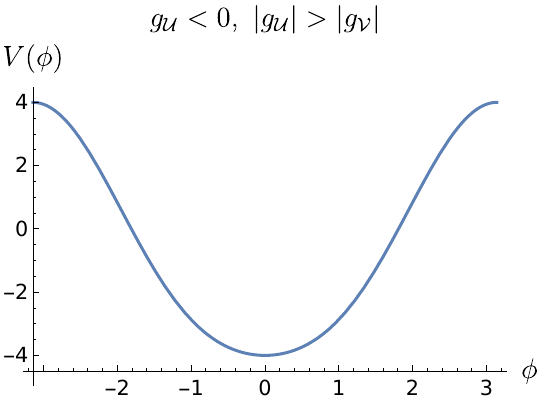}
	\caption{Symmetry breaking, restoration and the onset of Ising transition from the mean-field potential \cref{appeq:VSSB} projected to a single field as shown in \cref{appeq:VSSB_single} using $\moy{\phi_1} = \moy{\phi_2}$ for $J_M<0$ and $\moy{\phi_1} = -\moy{\phi_2}$ for $J_M>0$.}
	\label{appfig:Ising}
\end{figure}
The first route to symmetry restoration can be understood along the line $J_U = 0$ in \cref{appfig:SSB} within the mean field theory of \cref{appeq:VSSB}. As $|g_{\cU}|$ is increased beyond a critical value $|g_{\cU}| > |g_{\cV}|$, symmetry is restored by a continuous transition such that $\Phi \rightarrow 0$ for $g_{\cU} <0$ or $\Phi \rightarrow \pi$ for $g_{\cU} >0$ smoothly. 

For $g_{\cV} \propto J_M <0$, symmetry breaking and restoration preserves the ferromagnetic correlations between the legs of the ladder, which favors field configurations with $\moy{\phi_1} = \moy{\phi_2}$. Enforcing this in \cref{appeq:VSSB} gives us an effective potential in terms of a single field $\phi_1 = \phi_2 =\phi$
\begin{equation}
	V(\phi) = 2 g_{\cU} \cos \phi -  |g_{\cV}| \sin^2 \phi. \label{appeq:VSSB_single} 
\end{equation}
For $g_{\cU} \propto \delta = 0$, \cref{appeq:VSSB_single} has two degenerate vacua which merge at large values of $|\delta|$ at $\phi = 0$ for negative $\delta$ giving us the T$_3$ vacuum  and $\phi = \pi$ for positive $\delta$ giving us the T$_1$ vacuum  as shown in \cref{appfig:Ising}. Near the critical $g_{\cU}$, we can Taylor expand \cref{appeq:VSSB_single} where the compact nature of $\phi$ can be ignored and can be treated as a scalar field with a quartic interaction which is known to flow to the Ising universality class at criticality. 

A different way of saying this is by working in the language of compact bosons. Using the basis in \cref{appeq:SL2Z XY2} which is suitable for pinned $\moy{\phi_1 - \phi_2} =0$, the effective theory is as shown in \cref{appeq:XY2proximate} which is the so-called double frequency Sine-Gordon model ~\cite{DelfinoMussardo_1998675,FABRIZIO2000647} and is known to flow to the Ising CFT when tuned to criticality. 

For $g_{\cV} \propto J_M <0$, symmetry breaking and restoration takes place preserving an underlying antiferromagnetic correlation between the legs that favours $\moy{\phi_1} = - \moy{\phi_2} $. Using $\phi_1 = -\phi_2 = \phi$ in \cref{appeq:VSSB} gives the same potential as in \cref{appeq:VSSB_single} which exhibits an Ising transition in the mean field language. We can also reach the same conclusion by considering the basis in \cref{appeq:SL2Z XY3} to get the effective model in \cref{appeq:XY3proximate} which, setting $g_{\cW_-} =0$ corresponds to the DFSG model again and flows to Ising criticality.

\subsection{Self-dual Sine-Gordon Ising criticality}
We now consider starting within the symmetry broken phases in \cref{appfig:SSB} and going along the line $\delta = 0$. To analyze this transition, we will employ the basis in \cref{appeq:H_bosonize_symmetricAntisymmetric} with the $\theta_+,\phi_+$ sector kept gapped and focus on the sector $\theta_-, \phi_-$ shown in \cref{appeq:H_bosonize_symmetricAntisymmetric_minus} with $\delta = 0$  
\begin{equation}
	H \approx \frac{v_-}{2 \pi} \int dx \left( \frac{1}{K_-} (\partial_x\phi_-)^2 + \frac{K_-}{4} (\partial_x\theta_-)^2   \right)  + g_2 \int dx  \cos(\theta_-) + g_3 \int dx \cos(2 \phi_-) \label{appeq:H_bosonize_Antisymmetric_deltazero}
\end{equation}
where $g_{2} \propto J_U$ and $g_3 \propto J_M$. When $[\cos (2\phi_-)] < [\cos \theta_-]$, the system flows to the symmetry breaking phases. Symmetry is restored when $[\cos (2\phi_-)] > [\cos \theta_-]$ to give us the T$_2$ or T$_4$ vacua. The critical system obtains when $[\cos (2\phi_-)] = [\cos \theta_-]$ for $K_-=1$ and the coupling constants flow to the same value $g_{2,3} \rightarrow g$. In this regime, rescaling $\theta_- = 2 \vartheta_- $, \cref{appeq:H_bosonize_Antisymmetric_deltazero}   reduces to 
\begin{equation}
	H \approx \frac{v_-}{2 \pi} \int dx \left(  (\partial_x\phi_-)^2 +  (\partial_x\vartheta_-)^2   \right) + g \int dx \left( \cos(2\vartheta_-) +    \cos( 2\phi_-) . \right)\label{appeq:SDSG_Q2}
\end{equation}
\cref{appeq:SDSG_Q2} is also a self-dual Sine-Gordon (SDSG) model distinct from the one in \cref{appeq:SDSG_Q1}. Unlike the latter which flows to a trivial state, it was shown in Refs.~\cite{LECHEMINANT_SelfDualSineGordon_2002502,Schulz_Higherspinbosinization_PhysRevB.34.6372} that \cref{appeq:SDSG_Q2} flows to the Ising universality class.  

We have seen that the Ising universality class describing the transition out of the symmetry breaking phases to the trivial phase appears in two distinct ways, under the RG flow of the critical double-frequency Since-Gordon and Self-dual Sine Gordon models. When the Ising CFT is perturbed to flow to the symmetry-preserving disordered phase in each of these cases, we obtain the trivial phase as shown in \cref{appfig:SSB} in one of the four variants T$_{1-4}$. We are unable to determine how the DFSG and SDSG branches on the Ising critical locus in \cref{appfig:SSB} are connected. 

\subsection{The c = 3/2 points}
The critical points on the $\delta = 0$ line separating the XY$_2$ unnecessary critical line from the XY$_1$ and XY$_1^*$ lobes in \cref{appfig:gapless}(a) can be understood using the formulation in \cref{appeq:H_bosonize_symmetricAntisymmetric} when the $\theta_+,\phi_+$ sector is not gapped i.e. $\mathcal{V}_+$ is irrelevant. Here, the theory  in \cref{appeq:H_bosonize_symmetricAntisymmetric} reduces to 
\begin{equation}
	H \approx \frac{v_+}{2 \pi} \int dx \left( \frac{1}{K_+} (\partial_x\phi_+)^2 + \frac{K_+}{4} (\partial_x\theta_+)^2   \right) +\frac{v_-}{2 \pi} \int dx \left(  (\partial_x\phi_-)^2 +  (\partial_x\vartheta_-)^2   \right) + g \int dx \left( \cos(2\vartheta_-) +    \cos( 2\phi_-)  \right) .\label{appeq:SDSGplus LL}
\end{equation}
\cref{appeq:SDSGplus LL} represents $c=3/2$ conformal field theory corresponding to a stack of a $c=1$ compact boson CFT ($\theta_+,\phi_+$) with the $c=1/2$ Ising universality (SDSG) that has been identified as the critical theory separating symmetry-enriched critical states in Refs.~\cite{APUCPhysRevLett.130.256401,APSECTLLPhysRevB.108.245135,Schulz_Higherspinbosinization_PhysRevB.34.6372,XueJia_C1p5_PhysRevA.109.062226}.

\subsection{Explicit symmetry breaking}
Finally, let us comment on the instability of the phase diagram to explicitly breaking each of the symmetries individually. This can be done using the following operators and their bosonized forms
\begin{align}
	\text{Break } U(1)&:~\sum_j (-1)^j (S^x_{1 j}+S^x_{2 j})   \sim \int dx (\cos\theta_1 + \cos\theta_2) ~\text{ or } \sum_j (S^+_{1j}S^+_{2j}+S^-_{1j}S^-_{2j})\sim \int dx \cos(\theta_1 + \theta_2),\\
	\text{Break } \ztwo^R&:~\sum_j (-1)^j (S^z_{1 j}+S^z_{2 j})   \sim \int dx (\sin\phi_1 + \sin\phi_2),\\
	\text{Break } \ztwo^L&:~\sum_{\alpha = 1,2}\sum_j (-1)^{j+\alpha} (S^x_{\alpha j}S^x_{\alpha j+1}+S^y_{\alpha j}S^y_{\alpha j+1})   \sim \int dx (\cos\phi_1 - \cos\phi_2).
\end{align}
Breaking any of the three symmetries using the above operators gaps out both the unnecessary critical line and unnecessary multicritical point and renders the surrounding charge pump trivial. 

{
	\section{Pathways to higher dimensional examples}
	The main focus of this work was on a one-dimensional example, where we exploited the substantial analytical control afforded by bosonization. How do we generalize this study to higher dimensions? We can use our conjecture in the concluding part of the main text that the presence of stable boundary modes heralded an unnecessary critical surface and look for phase diagram containing a topological family with anomalously stable edge modes. 
	
	The notion of a charge pump in 1d generalises to an SPT pump in higher dimensions. More precisely, a non-contractible loop of states in $d$ spatial dimensions corresponds to a pump of a non-trivial $d-1$ dimensional SPT phase. For $d=1$, this recovers the charge pump since symmetry charges classify $0$ dimensional SPT phases. There are two complementary ways we can extend our insight to higher dimensional cases. First, there are several known examples of systems exhibiting unnecessary criticality in higher dimensions~\cite{BiSenthil_UC_PhysRevX.9.021034,JianXu_UC_PhysRevB.101.035118,ZhangSenthilUC2024Dirac}. We can check if the surrounding states form a non-trivial topological family. Second, we can construct non-trivial higher dimensional pumps and embed them in phase diagrams using the suspension isomorphism construction of Ref~\cite{HermeleGappedfamiliesPhysRevB.108.125147} using which we can look for the presence of unnecessary critical surfaces. We will briefly discuss both of these here to sketch a pathway for future investigations.
	
	\subsection{`Failed SPT' unnecessary critical surfaces}
	\begin{figure}[!h]
		\centering
		     \begin{tikzpicture}[scale=1.5]
		\draw[->] (-2,-1) --(2,-1);
        	\draw[->] (-2,-1) --(-2,1);
            \fill[blue,opacity=0.1] (-2+0.08,-1+0.08) rectangle (2-0.08,1-0.08);
            \fill[blue, opacity=0.1] (-1,0) to[bend left] (-1.5,1-0.08) -- (1.5,1-0.08) to[bend left] (1,0);
            \draw[dashed]  (-1.5,1-0.08)to[bend right](-1,0) ;
            \draw[dashed] (1.5,1-0.08) to[bend left] (1,0);
            \node[] at (2.1,-1) {$U$};
            \node[] at (-2,1.1) {$m$};
            \node[] at (-2.1,0) {0};
            \node[] at (0,-1.1) {0};
            \draw[thick,red] (-1,0) --(1,0);        
            \node[] at (-1.2,0) {?};
            \node[] at (1.2,0) {?};
		\end{tikzpicture}
		\caption{Schematic phase diagram for a `failed SPT' recipe to produce unnecessary criticality. The continuous red line represents a terminating critical theory described by a massless relativistic fermion theory. The shaded region represents edge modes, also corresponding to a massless relativistic fermion theory in a lower dimension. The question marks indicates that the nature of termination of the critical line is generally unknown.}
		\label{appfig:UC_failedSPT}
	\end{figure}
	A useful recipe to construct phase diagrams with unnecessary criticality is to consider a topological transition and  introduce a perturbation that removes the topological distinction between the two sides~\cite{BiSenthil_UC_PhysRevX.9.021034,JianXu_UC_PhysRevB.101.035118}. This occurs, for example, when a non-trivial free-fermion topological phase is trivialized by interactions~\cite{FidkowskiKitaev_PhysRevB.81.134509}. If there exists a direct, stable transition between such a system and the trivial phase in the free-fermion limit, by adding interactions, we can obtain a phase diagram with unnecessary criticality. Additional symmetries usually need to be imposed to ensure a direct transition rather than an intervening phase~\cite{JianXu_UC_PhysRevB.101.035118}.  To see this, recall that the vicinity of the free fermion critical theory is usually described by a relativistic free fermion with a \emph{single} symmetry-allowed mass term~\cite{QiXZhang_TITSC_review_RevModPhys.83.1057}. The two signs of the mass produce the free-fermion topological and trivial phases, respectively, and the critical theory is a relativistic massless fermion theory. For bulk spatial dimensions $d\ge2$, weak interactions are irrelevant for the massless fermion and the critical theory extends to an unnecessary critical line in the presence of interactions $U$. This is schematically shown in \cref{appfig:UC_failedSPT} using a solid line. The edge modes of the free-fermion topological phase, likewise,  are described by a massless relativistic fermion in one lower dimension~\cite{QiXZhang_TITSC_review_RevModPhys.83.1057}. Here too, weak interactions are either irrelevant (for bulk $d>2$) or marginal (for bulk $d=2$) and the edge modes are parametrically stable as indicated by the shaded region in \cref{appfig:UC_failedSPT}. For strong interactions, the unnecessary critical line terminates, as do the edge modes through a boundary transition leading us to the schematic phase diagram in \cref{appfig:UC_failedSPT}. Notice the similarity with our phase diagrams. This leads us to conjecture that the family of states surrounding the critical line form a non-trivial topological family. We leave a confirmation of our conjecture in known ``failed SPT'' models~\cite{BiSenthil_UC_PhysRevX.9.021034,JianXu_UC_PhysRevB.101.035118} and the determination of the nature of the topological family to future work.

	\subsection{Constructing examples using suspension isomorphism}
	\begin{figure}[!h]
		\centering
		\subfloat[$-1/\sqrt{2}\le \Delta \le 1$]{ \begin{tikzpicture}[scale=1]
\draw[->] (-4, -4) -- (4, -4) node[right] {$\delta_x$};
\draw[->] (-4,-4) -- (-4,4) node[above] {$\delta_y$};
\fill[blue,opacity=0.1] (-3.9,-3.9) rectangle (3.9,3.9);
\fill[red,opacity=0.2] (0,0) circle (1.);
\draw[thick,red] (0,0) circle (1.);
\draw[thick, dashed] (0,3.9) -- (0,1);
draw[dashed] ()
\def\sep{.5}
\newcommand{\drawafm}[2]{
  \begin{scope}[shift={(#1,#2)}]
    \foreach \i in {0,2} {
      \foreach \j in {0,2} {
        \draw[-{Stealth[open]}] (\i*\sep-0.25*\sep,\j*\sep-0.25*\sep) -- (\i*\sep+0.25*\sep,\j*\sep+0.25*\sep);
      }
    }
    \foreach \i in {1,3} {
      \foreach \j in {1,3} {
        \draw[-{Stealth[open]}] (\i*\sep-0.25*\sep,\j*\sep-0.25*\sep) -- (\i*\sep+0.25*\sep,\j*\sep+0.25*\sep);
      }
    }
        \foreach \i in {0,2} {
      \foreach \j in {1,3} {
        \draw[-{Stealth[open]}]  (\i*\sep+0.25*\sep,\j*\sep+0.25*\sep) -- (\i*\sep-0.25*\sep,\j*\sep-0.25*\sep) ;
      }
    }
      \foreach \i in {1,3} {
      \foreach \j in {0,2} {
        \draw[-{Stealth[open]}]  (\i*\sep+0.25*\sep,\j*\sep+0.25*\sep) -- (\i*\sep-0.25*\sep,\j*\sep-0.25*\sep) ;
      }
    }
  \end{scope}
}

\def\sep{0.4}      
\def\rad{0.1}     

\newcommand{\drawlatticexa}[2]{
  \begin{scope}[shift={(#1,#2)}]
    \foreach \i in {0,...,4} {
      \foreach \j in {0,...,4} {
        \shade[ball color=blue!80,opacity=1] (\i*\sep,\j*\sep) circle (\rad);
      }
    }
      \foreach \i in {0,2} {
      \foreach \j in {0,...,4} {
        \filldraw[red,rounded corners,opacity=.2] (\i*\sep-0.4*\sep,\j*\sep-0.4*\sep) rectangle (\i*\sep+1.4*\sep,\j*\sep+0.4*\sep);
      }
    }
    \foreach \i in {4} {
      \foreach \j in {0,...,4} {
        \filldraw[red,rounded corners,opacity=.2] (\i*\sep-0.4*\sep,\j*\sep-0.4*\sep) rectangle (\i*\sep+0.4*\sep,\j*\sep+0.4*\sep);
      }
    }
  \end{scope}
}
\newcommand{\drawlatticexb}[2]{
  \begin{scope}[shift={(#1,#2)}]
    \foreach \i in {-1,...,3} {
      \foreach \j in {0,...,4} {
        \shade[ball color=blue!80,opacity=1] (\i*\sep,\j*\sep) circle (\rad);
      }
    }
      \foreach \i in {0,2} {
      \foreach \j in {0,...,4} {
        \filldraw[red,rounded corners,opacity=.2] (\i*\sep-0.4*\sep,\j*\sep-0.4*\sep) rectangle (\i*\sep+1.4*\sep,\j*\sep+0.4*\sep);
      }
    }
    \foreach \i in {-1} {
      \foreach \j in {0,...,4} {
        \filldraw[red,rounded corners,opacity=.2] (\i*\sep-0.4*\sep,\j*\sep-0.4*\sep) rectangle (\i*\sep+.4*\sep,\j*\sep+0.4*\sep);
      }
    }
  \end{scope}
}

\newcommand{\drawlatticeya}[2]{
  \begin{scope}[shift={(#1,#2)}]
    \foreach \i in {0,...,4} {
      \foreach \j in {0,...,4} {
        \shade[ball color=blue!80,opacity=1] (\i*\sep,\j*\sep) circle (\rad);
      }
    }
      \foreach \j in {0,2} {
      \foreach \i in {0,...,4} {
        \filldraw[red,rounded corners,opacity=.2] (\i*\sep-0.4*\sep,\j*\sep-0.4*\sep) rectangle (\i*\sep+.4*\sep,\j*\sep+1.4*\sep);
      }
    }
    \foreach \j in {4} {
      \foreach \i in {0,...,4} {
        \filldraw[red,rounded corners,opacity=.2] (\i*\sep-0.4*\sep,\j*\sep-0.4*\sep) rectangle (\i*\sep+.4*\sep,\j*\sep+.4*\sep);
      }
    }
  \end{scope}
}

\newcommand{\drawlatticeyb}[2]{
  \begin{scope}[shift={(#1,#2)}]
    \foreach \i in {0,...,4} {
      \foreach \j in {-1,...,3} {
        \shade[ball color=blue!80,opacity=1] (\i*\sep,\j*\sep) circle (\rad);
      }
    }
      \foreach \j in {0,2} {
      \foreach \i in {0,...,4} {
        \filldraw[red,rounded corners,opacity=.2] (\i*\sep-0.4*\sep,\j*\sep-0.4*\sep) rectangle (\i*\sep+0.4*\sep,\j*\sep+1.4*\sep);
      }
    }
    \foreach \j in {-1} {
      \foreach \i in {0,...,4} {
        \filldraw[red,rounded corners,opacity=.2] (\i*\sep-0.4*\sep,\j*\sep-0.4*\sep) rectangle (\i*\sep+0.4*\sep,\j*\sep+0.4*\sep);
      }
    }
  \end{scope}
}

\drawlatticexb{2.5}{-0.8};
\drawlatticexa{-3.7}{-0.8};
\drawlatticeyb{-0.8}{2.5};
\drawlatticeya{-0.8}{-3.7};
\drawafm{-0.62}{-0.6};

\end{tikzpicture}}
		\subfloat[$-1 < \Delta \le -1/\sqrt{2}$]{\begin{tikzpicture}[scale=1]
\draw[->] (-4, -4) -- (4, -4) node[right] {$\delta_x$};
\draw[->] (-4,-4) -- (-4,4) node[above] {$\delta_y$};
\fill[blue,opacity=0.1] (-3.9,-3.9) rectangle (3.9,3.9);
\fill[blue,opacity=0.1] (-2.5,3.9)   to[bend right] (-1,0) to[out=90,in=180] (0,1) to[out=0,in=90] (1,0) to[bend right] (2.5,3.9);
\draw[blue,dashed] (-2.5,3.9) to[bend right] (-1,0) to[out=90,in=180] (0,1) to[out=0,in=90] (1,0) to[bend right] (2.5,3.9);
\fill[red,opacity=0.2] (0,0) circle (1.);
\draw[thick,red] (0,0) circle (1.);
draw[dashed] ()
\def\sep{.5}
\newcommand{\drawafm}[2]{
  \begin{scope}[shift={(#1,#2)}]
    \foreach \i in {0,2} {
      \foreach \j in {0,2} {
        \draw[-{Stealth[open]}] (\i*\sep-0.25*\sep,\j*\sep-0.25*\sep) -- (\i*\sep+0.25*\sep,\j*\sep+0.25*\sep);
      }
    }
    \foreach \i in {1,3} {
      \foreach \j in {1,3} {
        \draw[-{Stealth[open]}] (\i*\sep-0.25*\sep,\j*\sep-0.25*\sep) -- (\i*\sep+0.25*\sep,\j*\sep+0.25*\sep);
      }
    }
        \foreach \i in {0,2} {
      \foreach \j in {1,3} {
        \draw[-{Stealth[open]}]  (\i*\sep+0.25*\sep,\j*\sep+0.25*\sep) -- (\i*\sep-0.25*\sep,\j*\sep-0.25*\sep) ;
      }
    }
      \foreach \i in {1,3} {
      \foreach \j in {0,2} {
        \draw[-{Stealth[open]}]  (\i*\sep+0.25*\sep,\j*\sep+0.25*\sep) -- (\i*\sep-0.25*\sep,\j*\sep-0.25*\sep) ;
      }
    }
  \end{scope}
}

\def\sep{0.4}      
\def\rad{0.1}     

\newcommand{\drawlatticexa}[2]{
  \begin{scope}[shift={(#1,#2)}]
    \foreach \i in {0,...,4} {
      \foreach \j in {0,...,4} {
        \shade[ball color=blue!80,opacity=1] (\i*\sep,\j*\sep) circle (\rad);
      }
    }
      \foreach \i in {0,2} {
      \foreach \j in {0,...,4} {
        \filldraw[red,rounded corners,opacity=.2] (\i*\sep-0.4*\sep,\j*\sep-0.4*\sep) rectangle (\i*\sep+1.4*\sep,\j*\sep+0.4*\sep);
      }
    }
    \foreach \i in {4} {
      \foreach \j in {0,...,4} {
        \filldraw[red,rounded corners,opacity=.2] (\i*\sep-0.4*\sep,\j*\sep-0.4*\sep) rectangle (\i*\sep+0.4*\sep,\j*\sep+0.4*\sep);
      }
    }
  \end{scope}
}
\newcommand{\drawlatticexb}[2]{
  \begin{scope}[shift={(#1,#2)}]
    \foreach \i in {-1,...,3} {
      \foreach \j in {0,...,4} {
        \shade[ball color=blue!80,opacity=1] (\i*\sep,\j*\sep) circle (\rad);
      }
    }
      \foreach \i in {0,2} {
      \foreach \j in {0,...,4} {
        \filldraw[red,rounded corners,opacity=.2] (\i*\sep-0.4*\sep,\j*\sep-0.4*\sep) rectangle (\i*\sep+1.4*\sep,\j*\sep+0.4*\sep);
      }
    }
    \foreach \i in {-1} {
      \foreach \j in {0,...,4} {
        \filldraw[red,rounded corners,opacity=.2] (\i*\sep-0.4*\sep,\j*\sep-0.4*\sep) rectangle (\i*\sep+.4*\sep,\j*\sep+0.4*\sep);
      }
    }
  \end{scope}
}

\newcommand{\drawlatticeya}[2]{
  \begin{scope}[shift={(#1,#2)}]
    \foreach \i in {0,...,4} {
      \foreach \j in {0,...,4} {
        \shade[ball color=blue!80,opacity=1] (\i*\sep,\j*\sep) circle (\rad);
      }
    }
      \foreach \j in {0,2} {
      \foreach \i in {0,...,4} {
        \filldraw[red,rounded corners,opacity=.2] (\i*\sep-0.4*\sep,\j*\sep-0.4*\sep) rectangle (\i*\sep+.4*\sep,\j*\sep+1.4*\sep);
      }
    }
    \foreach \j in {4} {
      \foreach \i in {0,...,4} {
        \filldraw[red,rounded corners,opacity=.2] (\i*\sep-0.4*\sep,\j*\sep-0.4*\sep) rectangle (\i*\sep+.4*\sep,\j*\sep+.4*\sep);
      }
    }
  \end{scope}
}

\newcommand{\drawlatticeyb}[2]{
  \begin{scope}[shift={(#1,#2)}]
    \foreach \i in {0,...,4} {
      \foreach \j in {-1,...,3} {
        \shade[ball color=blue!80,opacity=1] (\i*\sep,\j*\sep) circle (\rad);
      }
    }
      \foreach \j in {0,2} {
      \foreach \i in {0,...,4} {
        \filldraw[red,rounded corners,opacity=.2] (\i*\sep-0.4*\sep,\j*\sep-0.4*\sep) rectangle (\i*\sep+0.4*\sep,\j*\sep+1.4*\sep);
      }
    }
    \foreach \j in {-1} {
      \foreach \i in {0,...,4} {
        \filldraw[red,rounded corners,opacity=.2] (\i*\sep-0.4*\sep,\j*\sep-0.4*\sep) rectangle (\i*\sep+0.4*\sep,\j*\sep+0.4*\sep);
      }
    }
  \end{scope}
}

\drawlatticexb{2.5}{-0.8};
\drawlatticexa{-3.7}{-0.8};
\drawlatticeyb{-0.8}{2.5};
\drawlatticeya{-0.8}{-3.7};
\drawafm{-0.62}{-0.6};

\end{tikzpicture}}
		\subfloat[Possible proximate phase diagram]{\begin{tikzpicture}[scale=1]
\draw[->] (-4, -4) -- (4, -4) node[right] {$\delta_x$};
\draw[->] (-4,-4) -- (-4,4) node[above] {$\delta_y$};
\fill[blue,opacity=0.1] (-3.9,-3.9) rectangle (3.9,3.9);
\draw[red,thick](-1.15,0) -- (1.15,0);
\fill[blue,opacity=0.1] (-2.4,3.9) to[bend right] (-1.15,0) -- (1.15,0) to[bend right] (2.4,3.9);
\draw[dashed] (-2.4,3.9) to[bend right] (-1.15,0);
\draw[dashed](2.4,3.9) to[bend left]  (1.15,0);
\node[] at (0,-0.2) {?};
\def\sep{.5}

\def\sep{0.4}      
\def\rad{0.1}     

\newcommand{\drawlatticexa}[2]{
  \begin{scope}[shift={(#1,#2)}]
    \foreach \i in {0,...,4} {
      \foreach \j in {0,...,4} {
        \shade[ball color=blue!80,opacity=1] (\i*\sep,\j*\sep) circle (\rad);
      }
    }
      \foreach \i in {0,2} {
      \foreach \j in {0,...,4} {
        \filldraw[red,rounded corners,opacity=.2] (\i*\sep-0.4*\sep,\j*\sep-0.4*\sep) rectangle (\i*\sep+1.4*\sep,\j*\sep+0.4*\sep);
      }
    }
    \foreach \i in {4} {
      \foreach \j in {0,...,4} {
        \filldraw[red,rounded corners,opacity=.2] (\i*\sep-0.4*\sep,\j*\sep-0.4*\sep) rectangle (\i*\sep+0.4*\sep,\j*\sep+0.4*\sep);
      }
    }
  \end{scope}
}
\newcommand{\drawlatticexb}[2]{
  \begin{scope}[shift={(#1,#2)}]
    \foreach \i in {-1,...,3} {
      \foreach \j in {0,...,4} {
        \shade[ball color=blue!80,opacity=1] (\i*\sep,\j*\sep) circle (\rad);
      }
    }
      \foreach \i in {0,2} {
      \foreach \j in {0,...,4} {
        \filldraw[red,rounded corners,opacity=.2] (\i*\sep-0.4*\sep,\j*\sep-0.4*\sep) rectangle (\i*\sep+1.4*\sep,\j*\sep+0.4*\sep);
      }
    }
    \foreach \i in {-1} {
      \foreach \j in {0,...,4} {
        \filldraw[red,rounded corners,opacity=.2] (\i*\sep-0.4*\sep,\j*\sep-0.4*\sep) rectangle (\i*\sep+.4*\sep,\j*\sep+0.4*\sep);
      }
    }
  \end{scope}
}

\newcommand{\drawlatticeya}[2]{
  \begin{scope}[shift={(#1,#2)}]
    \foreach \i in {0,...,4} {
      \foreach \j in {0,...,4} {
        \shade[ball color=blue!80,opacity=1] (\i*\sep,\j*\sep) circle (\rad);
      }
    }
      \foreach \j in {0,2} {
      \foreach \i in {0,...,4} {
        \filldraw[red,rounded corners,opacity=.2] (\i*\sep-0.4*\sep,\j*\sep-0.4*\sep) rectangle (\i*\sep+.4*\sep,\j*\sep+1.4*\sep);
      }
    }
    \foreach \j in {4} {
      \foreach \i in {0,...,4} {
        \filldraw[red,rounded corners,opacity=.2] (\i*\sep-0.4*\sep,\j*\sep-0.4*\sep) rectangle (\i*\sep+.4*\sep,\j*\sep+.4*\sep);
      }
    }
  \end{scope}
}

\newcommand{\drawlatticeyb}[2]{
  \begin{scope}[shift={(#1,#2)}]
    \foreach \i in {0,...,4} {
      \foreach \j in {-1,...,3} {
        \shade[ball color=blue!80,opacity=1] (\i*\sep,\j*\sep) circle (\rad);
      }
    }
      \foreach \j in {0,2} {
      \foreach \i in {0,...,4} {
        \filldraw[red,rounded corners,opacity=.2] (\i*\sep-0.4*\sep,\j*\sep-0.4*\sep) rectangle (\i*\sep+0.4*\sep,\j*\sep+1.4*\sep);
      }
    }
    \foreach \j in {-1} {
      \foreach \i in {0,...,4} {
        \filldraw[red,rounded corners,opacity=.2] (\i*\sep-0.4*\sep,\j*\sep-0.4*\sep) rectangle (\i*\sep+0.4*\sep,\j*\sep+0.4*\sep);
      }
    }
  \end{scope}
}

\drawlatticexb{2.5}{-0.8};
\drawlatticexa{-3.7}{-0.8};
\drawlatticeyb{-0.8}{2.5};
\drawlatticeya{-0.8}{-3.7};

\end{tikzpicture}}
		\caption{ (a,b) The phase diagram of \cref{appeq:XXZ} in the easy-plane limit $|\Delta|\le 1$. The region near the origin contains a planar N\'{e}el antiferromagnet separated from the peripheral states by an Ising universality class. (a) For $-1/\sqrt{2} < \Delta < 1$, the boundary is gapless only along a one-dimensional subspace. (b) For $-1<\Delta < -1/\sqrt{2}$, the boundary Luttinger liquid becomes stable and we get an extended region with edge modes. (c) We conjecture that in the vicinity of the parameter space with stable edge modes, an unnecessary critical line should be present.}
		\label{appfig:2d_XXZ}
	\end{figure}
	Let us demonstrate the construction of models with topological families in 2d  using suspension isomorphism~\cite{HermeleGappedfamiliesPhysRevB.108.125147}. We consider a pump of the Haldane phase in a   model of qubits on a square lattice with Hamiltonian
	\begin{multline}
		H=  \sum_{\vec{r}} \left[\frac{(1+ (-1)^{r_x} \delta_x)}{2 (1-\delta_x^2)} \left(S^x_{\vec{r}} S^x_{\vec{r} + \hat{x}} + S^y_{\vec{r}} S^y_{\vec{r} + \hat{x}} + \Delta S^z_{\vec{r}} S^z_{\vec{r} + \hat{x}}\right) + \frac{(1+ (-1)^{r_y} \delta_y)}{2 (1-\delta_y^2)} \left(S^x_{\vec{r}} S^x_{\vec{r} + \hat{y}} + S^y_{\vec{r}} S^y_{\vec{r} + \hat{y}} + \Delta S^z_{\vec{r}} S^z_{\vec{r} + \hat{y}}\right) \right]. \label{appeq:XXZ}
	\end{multline}
	We will study two-dimensional phase diagrams by varying $-1< \delta_{x,y} < 1$ while keeping all other parameters fixed and focusing on the  `easy-plane' regime $|\Delta|\le 1$. In the absence of spatial boundaries, along the periphery of the phase diagram, the ground state family consists of singlets oriented along different directions as shown in \cref{appfig:2d_XXZ}. This is the non-contractible family corresponding to a `Haldane phase pump' by construction. {The circular family of states also appears in the so-called chiral Floquet phase of many-body localized systems~\cite{ChiralFloquet_PhysRevX.6.041070}.}} .

Let us begin with a discussion of the boundary. We consider periodic boundary conditions along the $x$ direction and open boundary conditions on the $y$ direction and focus on only one of the ends. The $\delta_x=0$ line consists of stacks of Haldane chains for $\delta_y\sim 1$ and trivial chains for $\delta_y \sim -1$. For $\delta_y \sim 1$, each Haldane chain supplies an edge qubit. The effective theory for these qubits can be determined easily from the remaining terms in the Hamiltonian. For small $\delta_x$, this is
\begin{equation}
	H_{\text{edge}} \approx  \sum_{j} \left[(1+ (-1)^{j} \delta_x) \left(S^x_{j} S^x_{j+1} + S^y_{j} S^y_{j+1} + \Delta S^z_{j} S^z_{j+1}\right) \right]. \label{appeq:Edge_XXZ}
\end{equation}
In other words, we have the bond-dimerized XXZ spin chain, whose phase diagram can be reproduced in the bosonized description,
\begin{equation}
	H \approx \frac{v}{2 \pi} \int dx \left(\frac{1}{4K} \left(\partial_x \phi\right)^2 + K \left(\partial_x \theta\right)^2\right) +2\mathcal{AC} \delta  \int dx~    \cos \phi + \ldots 
\end{equation}
with $K$, $v$ determined from the Bethe ansatz solution as shown in \cref{appeq:Scaling_dim_perturbative}. For $\delta_x = 0$, we get a gapless state on the edge described by a compact boson CFT.  For $-\frac{1}{\sqrt{2}}<\Delta\le 1$, the bond dimerization is relevant and the edge theory gaps out giving us the phase diagram of the kind shown in \cref{appfig:2d_XXZ}(a). For a range $-1<\Delta<-\frac{1}{\sqrt{2}}$ however, bond dimerization (and all symmetry allowed operators) become irrelevant and the gapless edge becomes parametrically stable as shown in \cref{appfig:2d_XXZ}(b). In this regime, we can expect some modification of \cref{appeq:XXZ} to exhibit unnecessary criticality near the origin. 

However, at its origin $\delta_x = \delta_y =0$, \cref{appeq:XXZ} is the two-dimensional XXZ model whose ground state on the square lattice in the easy-plane regime $|\Delta|<1$ is the easy-plane N\'{e}el antiferromagnet~\cite{BISHOP2017178_XXZ} which spontaneously breaks the planar spin rotation $U(1)$ symmetry. Thus the obstruction to contracting the non-trivial family appears as a transition (in this case belonging to the Ising universality class) to a new phase, not an unnecessary critical line. This can change with suitable perturbations such as next-nearest neighbour interactions and $x$-$y$ anisotropic couplings and we can reasonably hope to find unnecessary criticality. Indeed, it is known that in the proximity of the N\'{e}el phase, there exists a critical Dirac spin liquid~\cite{Song2019unifying} which forms an unnecessary critical surface in appropriate contexts~\cite{ZhangSenthilUC2024Dirac}. This would be a promising candidate for an unnecessary critical theory in our model. We leave such a systematic search to future work.

We end with a comment on symmetries. The extended region of edge modes found near the unnecessary critical line of the $d=1$ model studied in this paper required non-Abelian symmetries. From our conjecture that stable edge modes are a necessary condition for unnecessary criticality, it follows that non-Abelian symmetries are also necessary for unnecessary criticality in $d=1$. In higher dimensions, the symmetry requirements are less restrictive. The Haldane pump model in \cref{appeq:XXZ} has an on-site non-Abelian $O(2)$ symmetry. However, even if we break the $O(2)$ symmetry down to $U(1)$, for example by adding a magnetic field in the $z$ direction
\begin{equation}
	\delta H = h \sum_{\vec{r}}  S^z_{r},
\end{equation}
the non-trivial Haldane phase pump is still stable as long as we also preserve a time-reversal symmetry $\ztwo^T$ generated by complex conjugation. The magnetic field term, in fact, further stabilizes the edge modes and potentially helps drive the system closer to an unnecessary critical line.
\end{appendices}

\newpage
	\bibliography{references}
    
\end{document}


\begin{tikzpicture}[scale=1.5]
		\draw[thick,->] (-0.35,-0.1) -- ++(2.5,0.0) node[align=right,below] {$J_U$};
		\draw[thick,->] (-0.35,-0.1) -- ++(0.0,3.) node[align=left,above] {$\delta$};
		\fill[blue,opacity=0.05] (-0.25,0.) rectangle ++(2.25,2.8);
            \draw[dashed,blue] (1.8/2,2.8) -- (1.8/2,1.4);

  \draw[dotted] (-0.1,2.8) to[out=-90,in=140] (.4,1.8);
  \draw[dotted] (-0.1,0) to[out=90,in=-140] (.4,1);

  \draw[dotted] (1.85,2.8) to[out=-90,in=40] (1.35,1.8);
  \draw[dotted] (1.85,0) to[out=90,in=-40] (1.35,1);

\node[] at(0.16,2.6) {\scriptsize T$_1$};
\node[] at(1.67,2.6) {\scriptsize T$_1$};
\node[] at(0.0,1.8) {\scriptsize T$_2$};
\node[] at(1.8,1.8) {\scriptsize T$_4$};
\node[] at(0.95,0.67) {\scriptsize T$_3$};

            \def\ya{2.25};
              \def\yb{2.38};
              \foreach   \x in {1} {\fill[rounded corners,gray,opacity=0.2] (\x,\ya) rectangle (\x+0.5,\ya+.5);\draw[blue, snake it]  (\x+0.15,\yb)--(\x+0.15,\yb+0.23); \draw[thick] (\x+0.35,\yb+0.23)-- (\x+0.35,\yb)--(\x+0.55,\yb);\draw[thick]  (\x+0.35,\yb+0.23)--(\x+0.55,\yb+0.23);
              \shade[inner color=white, outer color = red] (\x+0.15,\yb) circle (0.06);\shade[inner color=white, outer color = red] (\x+0.35,\yb) circle (0.06);\shade[inner color=white, outer color = red] (\x+0.15,\yb+0.23) circle (0.06);\shade[inner color=white, outer color = red] (\x+0.35,\yb+0.23) circle (0.06);};

              \foreach   \x in {.3} {\fill[rounded corners,gray,opacity=0.2] (\x,\ya) rectangle (\x+0.5,\ya+.5);\draw[thick]  (\x+0.15,\yb)--(\x+0.15,\yb+0.23); \draw[thick] (\x+0.35,\yb+0.23)--   (\x+0.35,\yb)--(\x+0.55,\yb);\draw[thick]  (\x+0.35,\yb+0.23)--(\x+0.55,\yb+0.23);
              \shade[inner color=white, outer color = red] (\x+0.15,\yb) circle (0.06);\shade[inner color=white, outer color = red] (\x+0.35,\yb) circle (0.06);\shade[inner color=white, outer color = red] (\x+0.15,\yb+0.23) circle (0.06);\shade[inner color=white, outer color = red] (\x+0.35,\yb+0.23) circle (0.06);};
            
            \def\ya{1.15};
              \def\yb{1.28};
           \foreach   \x in {-.2} {\fill[rounded corners,gray,opacity=0.2] (\x,\ya) rectangle (\x+0.5,\ya+.5); \draw[thick]  (\x+0.15,\yb)--(\x+0.15,\yb+0.23);\draw[thick]  (\x+0.35,\yb)--(\x+0.35,\yb+0.23); 
              \shade[inner color=white, outer color = red] (\x+0.15,\yb) circle (0.06);\shade[inner color=white, outer color = red] (\x+0.35,\yb) circle (0.06);\shade[inner color=white, outer color = red] (\x+0.15,\yb+0.23) circle (0.06);\shade[inner color=white, outer color = red] (\x+0.35,\yb+0.23) circle (0.06);};

           \foreach   \x in {1.45} {\fill[rounded corners,gray,opacity=0.2] (\x,\ya) rectangle (\x+0.5,\ya+.5); \draw[blue, snake it]  (\x+0.15,\yb)--(\x+0.15,\yb+0.23);\draw[blue, snake it]  (\x+0.35,\yb)--(\x+0.35,\yb+0.23); 
              \shade[inner color=white, outer color = red] (\x+0.15,\yb) circle (0.06);\shade[inner color=white, outer color = red] (\x+0.35,\yb) circle (0.06);\shade[inner color=white, outer color = red] (\x+0.15,\yb+0.23) circle (0.06);\shade[inner color=white, outer color = red] (\x+0.35,\yb+0.23) circle (0.06);};              

              \def\ya{.05};
              \def\yb{.18};
              \foreach   \x in {1.35/2} {\fill[rounded corners,gray,opacity=0.2] (\x,\ya) rectangle (\x+0.5,\ya+.5); \draw[thick]  (\x+0.15,\yb)rectangle (\x+0.35,\yb+0.23);
              \shade[inner color=white, outer color = red] (\x+0.15,\yb) circle (0.06);\shade[inner color=white, outer color = red] (\x+0.35,\yb) circle (0.06);\shade[inner color=white, outer color = red] (\x+0.15,\yb+0.23) circle (0.06);\shade[inner color=white, outer color = red] (\x+0.35,\yb+0.23) circle (0.06);};

            \fill[] (1.8/2,1.4) circle (.05);
            \node[] at(1.8/2,1.24) {\scriptsize XY$_3$};
            \node[] at(1.8/2,-0.28) {0};
            \node[] at(-0.55,0.25) {-1};
            \node[] at(-0.55,1.4) {0};
            \node[] at(-0.55,2.7) {1};
            \node[] at(0.9,3.0) {\scriptsize $(b)~J_M \approx 5.2 ,\Delta\approx -0.05$};
		\end{tikzpicture}